\documentclass[%
 aps,
twocolumn,preprintnumbers,floatfix,
prev,
nofootinbib,
 amsmath,amssymb,
aps,prd,longbibliography]{revtex4-1}

\newcommand{\be}{\begin{equation}}
\newcommand{\ee}{\end{equation}}

\newcommand{\apjl}{ApJLett}		
\newcommand{\mnras}{MNRAS}		
\newcommand{\jcap}{JCAP}		

\newcommand{\kahlerij}{\mathcal{K}_{ij}}

\usepackage{graphicx}
\usepackage{dcolumn}
\usepackage{bm}
\usepackage{xcolor}
\usepackage{dsfont}
\usepackage{graphicx}
\usepackage{tabularx}
\usepackage{color}
\usepackage[table]{colortbl}
\usepackage{empheq}
\usepackage{lipsum}
\usepackage{yfonts}
\usepackage{float}
\usepackage{mathtools}
\usepackage{enumerate}
\usepackage{hyperref}
\usepackage{bbold}
\usepackage{cleveref}
\usepackage{mathrsfs}
\usepackage[caption=false]{subfig}
\newcolumntype{K}[1]{>{\centering\arraybackslash}p{#1}}
\usepackage{cancel}
\usepackage{nth}

\usepackage{xfrac}
\newcommand{\beq}{\begin{equation}}
\newcommand{\eeq}{\end{equation}}
\newcommand{\beqa}{\begin{eqnarray}}
\newcommand{\eeqa}{\end{eqnarray}}
\usepackage{amsmath}

\usepackage{setspace}
\usepackage[T1]{fontenc}

\begin{document}
\newcommand{\tkDM}[1]{\textcolor{red}{#1}}  

\begin{flushleft}
KCL-PH-TH/2018-17
\end{flushleft}


\title{Black Hole Spin Constraints on the Mass Spectrum and Number of Axion-like Fields}


\author{Matthew J. Stott$^a$}
\email{matthew.stott@kcl.ac.uk}
\author{David J. E. Marsh$^{b}$}
\email{david.marsh@uni-goettingen.de}

\affiliation{
$^a$ Theoretical Particle Physics and Cosmology Group, Department of Physics, King's College London, University of London, Strand, London, WC2R 2LS, United Kingdom \\
$^b$ Instit\"{u}t f\"{u}r Astrophysik, Georg-August Universit\"{a}t, Friedrich-Hund-Platz 1, D-37077 G\"{o}ttingen, Germany}

\date{\today}


\begin{abstract}

Astrophysical observations of spinning BHs, which span $ 5M_\odot\lesssim M_{\rm BH}\lesssim 5\times 10^8 M_\odot$, can be used to exclude the existence of certain massive bosons via the superradiance phenomenon. In this work, we explore for the first time how these measurements can be used to constrain properties of statistical distributions for the masses of multiple bosonic fields. Quite generally, our methodology excludes $N_{\rm ax}\gtrsim 30$ scalar fields with a range of mass distribution widths and central values spanning many orders of magnitude. We demonstrate this for the specific example of axions in string theory and M-theory, where the mass distributions in certain cases take universal forms. We place upper bounds on $N_{\rm ax}$ for certain scenarios of interest realised approximately as mass distributions in M-theory, including the QCD axion, grand unified theories, and fuzzy dark matter.
\end{abstract}


\maketitle
\begin{spacing}{0.95}
\tableofcontents
\end{spacing}
\section{Introduction}
\label{sec:intro}

The Penrose process~\cite{1969NCimR...1..252P} allows bosonic waves infalling into a Kerr black hole (BH) to emerge with more energy than incident upon entry at the horizon, in exact analogy to other superradiant processes in physics, such as Cherenkov radiation. If the bosons can be confined around the BH by a mirror, then this amplification process continues without limit leading to Press and Teukolsky's ``black hole bomb'' scenario~\cite{1972Natur.238..211P,1973ApJ...185..649P}. Massive bosonic fields on a Kerr spacetime possess hydrogenic bound states. In this case the potential barrier provided by the particle mass can play the role of the mirror, leading to a natural realisation of the BH superradiance process for massive bosons in orbits around astrophysical BHs (see Ref.~\cite{Brito:2015oca} for a review). 
\begin{figure*}
\centering
\includegraphics[width=\textwidth]{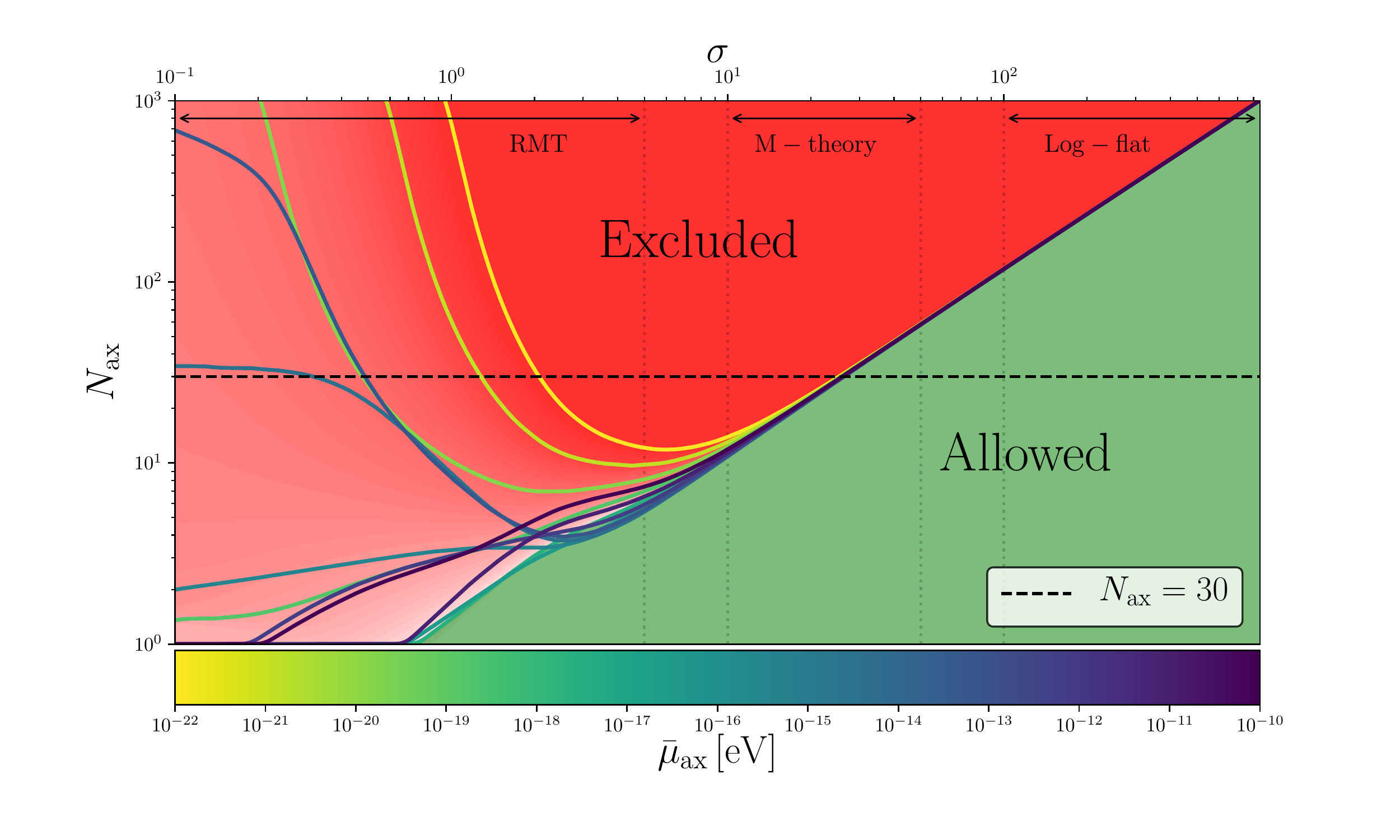}
\caption{Summary of results displaying contours for the 95\% exclusion regions for log-normal axion mass distributions as a function of the width, $\sigma$, and number of fields, $N_{\rm ax}$, for various central masses, $\bar{\mu}_{\rm ax}$. Regions above the contours are excluded. Certain ranges of $\sigma$ correspond closely to RMT and M-theory mass spectra, and can also be used to approximate the log-flat spectrum. For $1\lesssim\sigma\lesssim 20$, $N_{\rm ax}\geq 30$ is excluded for an extremely wide range of central masses. Constraints neglect axion self-interactions and apply approximately in the limit of large decay constants, $f_a\gtrsim 10^{14}\text{ GeV}$.}
\label{fig:summary}
\end{figure*}
The historic Laser Interferometry Gravitational-Wave Observatory (LIGO) observations of gravitational waves from the binary coalescence of astrophysical BHs has ushered in a new era of interest in BH physics~\cite{2016PhRvL.116f1102A}. Gravitational wave data can be used to infer the mass and spin of the two BHs in the binary. LIGO has the prospects to detect the existence of many hundreds of such events, accurately determining the mass and spin distribution of BHs. The future of BH superradiance constraints derived from LIGO, the growing global network of GW observatories, and future space-based missions, is extremely promising as a probe of fundamental physics~\cite{2017PhRvD..95d3001A,2017PhRvD..96c5019B,2017PhRvD..96f4050B,2018arXiv180101420C,Hannuksela:2018izj}.

The ability to constrain ultralight bosonic fields from BH-scalar condensate systems come in the form of two phenomena. It may be possible to identify the presence of scalar clouds in the vicinity of BHs as emission sources of monochromatic gravitational waves (GWs). The signal frequency, $f \sim \sfrac{\mu_{\rm ax}}{\pi}$, with boson mass, $\mu_{\rm ax}$ could potentially be detected by either ground or space-based GW observatories and proposes to be an exciting methodology to enhance constraints on the mass bounds for bosonic fields. This subject has been extensively discussed in Refs.~\cite{Arvanitaki:2010sy,Baryakhtar:2017ngi,Brito:2017zvb,Baumann:2018vus}. The second phenomenon of interest, and the subject of this work, is the spin down of astrophysical BHs. If the superradiance rate is faster than any other astrophysical process affecting the BH mass, $M_{\rm BH}$, and dimensionless spin, $a_*$, then the BH superradiance process can efficiently reduce these quantities. This occurs when the boson Compton wavelength is of the order of the gravitational radius of the BH. Thus, if a massive boson exists, then astrophysical BHs of particular values in the $(M_{\rm BH},a_*)$ ``Regge plane'' (which, according to the no-hair theorems, gives a complete description of spinning BHs) should be absent in observations. The masses and spins of a large number of astrophysical BHs have been measured, often incorporating either X-ray reflection spectroscopy or continuum-fitting methods (see Table~\ref{tab:cosmopar} for BH parameter measurements and corresponding references). These measurements can be used to probe the possible existence of massive bosons~\cite{2015CQGra..32m4001B,2014PTEP.2014d3E02Y}. BH superradiance constraints apply to a range of particle physics models, including a possible mass for the graviton or the photon~\cite{2012PhRvL.109m1102P} (and indeed to the photon plasma mass near the BH), as well as to exotic particles, such as massive vector (Proca) fields~\cite{2017PhRvD..96c5019B}, massive spin-two fields~\cite{2008PThPS.172...11K}, and axion-like particles and other massive scalars~\cite{axiverse,Arvanitaki:2010sy,2015PhRvD..91h4011A}. 

BH superradiance excludes two separate ranges of axion masses, $\mu_{\rm ax}$. Stellar mass BHs exclude $7\times 10^{-14}\text{ eV}<\mu_{\rm ax}/{\rm eV}<2\times 10^{-11}$ at the 95\% C.L., while supermassive BHs (SMBHs) exclude $7\times 10^{-20}\text{ eV}<\mu_{\rm ax}/{\rm eV}<1\times 10^{-16}$ at the 95\% C.L.. These limits apply strictly in the regime of zero self-coupling. Assuming a self-coupling derived from a standard instanton potential, they apply for axions with decay constants $f_a\gtrsim 10^{14}\text{ GeV}$~\cite{2015PhRvD..91h4011A}, a limit we assume throughout the remainder of this work.

These are powerful and generic exclusions, but they leave many axion models of interest unconstrained. Stellar BHs are too heavy to place constraints on the QCD axion~\cite{pecceiquinn1977,weinberg1978,wilczek1978} possessing a decay constant far below the Planck scale~\cite{2015PhRvD..91h4011A}. ``Fuzzy dark matter (DM)'' with $\mu_{\rm ax}\approx 10^{-22}\text{ eV}$~\cite{1990PhRvL..64.1084P,hu2000,Marsh:2013ywa,2014NatPh..10..496S,2017PhRvD..95d3541H}, which has novel effects on the formation of galaxies, is too light to make predictions about the spin distribution of SMBHs with $M_{\rm BH}<10^9 M_\odot$ that inhabit the centres of galaxies. Finally, the axion mass scale associated to grand unification (GUTs) in M-theory, $\mu_{\rm ax}\approx 10^{-15}\text{ eV}$~\cite{Acharya:2010zx} is in the ``desert'' of intermediate mass BHs (IMBHs) which so far have not been observed. There is hope, however, since each of these models is only a small logarithmic distance from the BH superradiance constrained regions, while axion models typically have a spectrum spanning many orders of magnitude~\cite{axiverse,2017PhRvD..96h3510S}. All previous studies of BH superradiance constraints on bosons have focused on the range of excluded masses assuming the existence of a single new bosonic field. In the present work we assess, for the first time, what constraints can be drawn on the properties of axion mass distributions from BH superradiance.  

String theory and M-theory predict that there should be a large number of as-yet-undiscovered light bosonic degrees of freedom, including hidden $U(1)$ gauge fields, moduli, and axions~\cite{axiverse,1984PhLB..149..351W,Svrcek:2006yi,2006JHEP...05..078C,Acharya:2010zx,2012JHEP...10..146C}. The number of axion fields depends on details of the compactification of the 6/7 extra-dimensional space determining the required 3+1 spacetime dimensions. Typical numbers of axions are of order 30 in Calabi-Yau compactifications~\cite{Kreuzer:2000xy,Altman:2014bfa}, with a similar expectation for $G_2$ manifolds~\cite{Corti:2012kd,Halverson:2014tya,Halverson:2015vta,Braun:2016igl,Braun:2017uku,Braun:2017ryx,Braun:2017csz}, although certain flux compactifications could contain upwards of $10^5$ axions~\cite{2007RvMP...79..733D}. Significant progress can also be made towards general predictions since the mass distributions of large numbers of axions possess universal properties thanks to results from random matrix theory (RMT)~\cite{mehta,Easther:2005zr,2013JHEP...01..136B,Long:2014fba,Brodie:2015kza,Bachlechner:2017hsj,2017PhRvD..96h3510S}. 

Any string or M-theory model that realises one of the models of interest (QCD axion, fuzzy DM, or GUTs) will likely contain a distribution of masses around this value. Even a small spread on a logarithmic scale could lead to strong constraints on the model. The central observation of the present work is that, simply from the statistical overlap between a mass distribution and the BH superradiance bounds, it is possible to place constraints on the allowed \emph{mass distributions} of axions. Furthermore, these constraints get increasingly more stringent as the number of axions increases, placing \emph{upper bounds on the number of axion-like fields}. 
\begin{figure}
\centering
\includegraphics[width=0.5\textwidth]{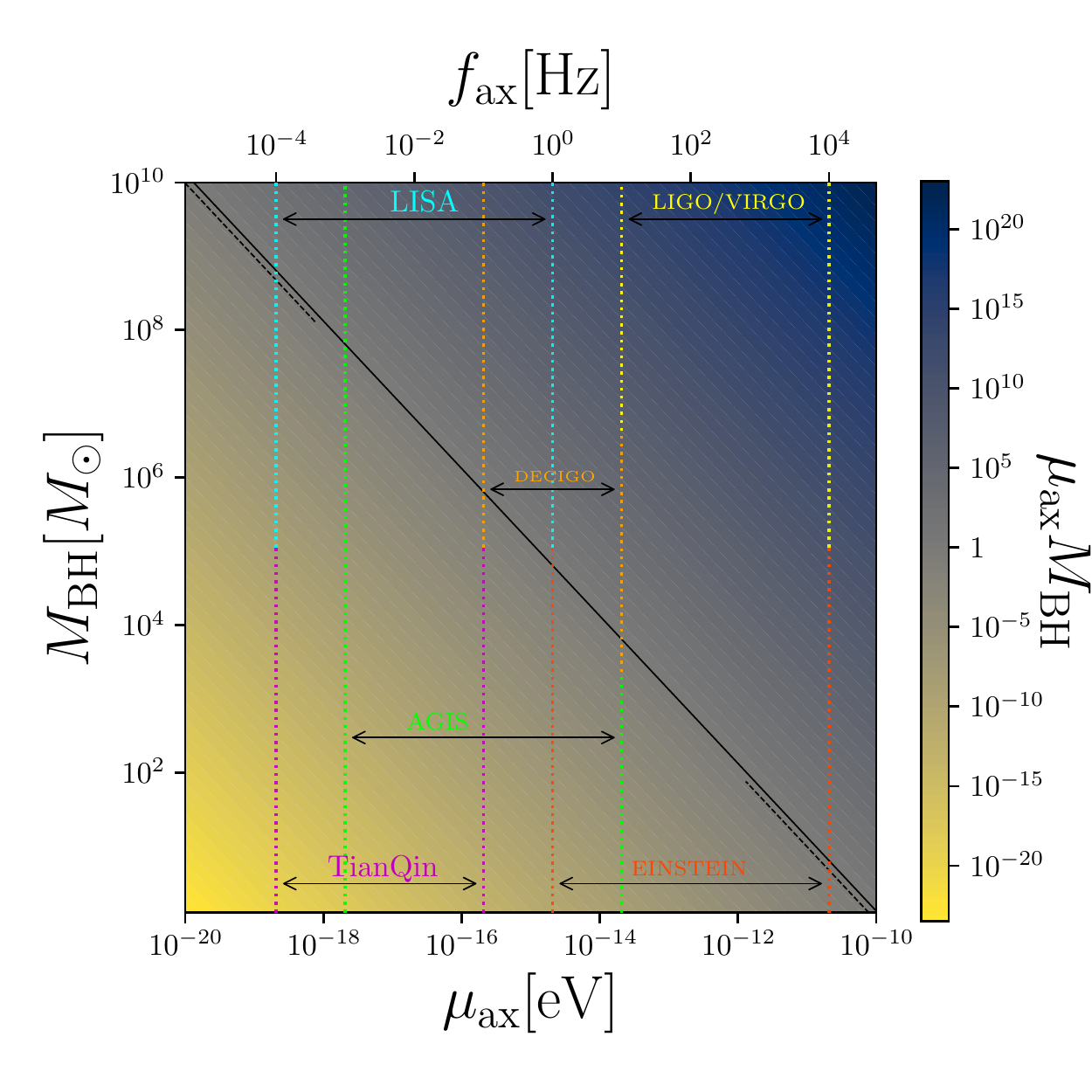}	
\caption{The BH-scalar condensate coupling, $\alpha = \mu_{\rm ax}M_{\rm BH}$. The solid black line represents the unity limit for non-relativistic and relativistic regimes. The dashed line corresponds to $\alpha = 0.5$, the approximate limit in which the analytical approximation for the instability rate is valid. Dotted lines correspond to frequency ranges for monochromatic gravitational wave emission from the scalar cloud accessible to current and future GW observatories \cite{TheLIGOScientific:2014jea,AmaroSeoane:2012je,AmaroSeoane:2012km,Dimopoulos:2008sv,Luo:2015ght,Sathyaprakash:2012jk,1742-6596-120-3-032004}.}
\label{fig:alpha}
\end{figure}
Consider the following toy model. In Ref.~\cite{axiverse} it was suggested that axion masses have a log-flat distribution from the Planck scale to the Hubble scale, covering approximately sixty orders of magnitude. The BH superradiance constraints cover approximately four orders of magnitude. Assuming independent and identically distributed draws from the log-flat distribution, this naive model of the axiverse is excluded with probability $P=1-(56/60)^N$, which is greater than 95\% C.L. if $N_{\rm ax}\geq 44$. Clearly, \emph{the model with a log-flat prior on the axion mass is excluded by BH superradiance for large numbers of fields}. The exclusion is a function of the upper and lower bounds on the mass spectrum. The constraint gets considerably stronger if the upper bound is below the Planck scale, and vanishes if the distribution does not extend below about $10^{-11}\text{ eV}$. Such a truncated spectrum, on the other hand, cannot realise many of the models of interest discussed above. 

Fortunately for phenomenologists, the mass distributions arising from RMT models are not log flat from the Hubble scale to the Planck scale. The log-normal distribution, centred on a particular mean mass, $\mu_{\rm ax}$, and with a variance $\sigma^2$, provides a useful benchmark, covering different types of models. For small $\sigma$, it resembles a degenerate spectrum, large $\sigma$ is approximately log-flat, and intermediate values of $\sigma$ are statistically similar to eigenvalue distributions found in RMT and M-theory. Fig.~\ref{fig:summary} summarises our conclusions, showing the allowed number of axionic fields drawn from log-normal distributions as a function of the width and central value. 
 
The structure of this work is as follows: Section~\ref{sec:bhsr_main} contains a brief review of BH superradiance along with the Regge plane and BH spin measurements, while Section~\ref{sec:axion_models} overviews our models for the axion mass matrix and collects our BH data. In Section~\ref{sec:results} we present constraints on axion mass spectra from BH mass and spin measurements under a frequentist framework. We first reproduce the known single-field results and then move on to considering mass distributions. We conclude our work in Section~\ref{sec:conclusions}. Further details of our BH superradiance calculations are given in Appendix~\ref{appendix:bhsr}. Appendix~\ref{appendix:stats}  describes our statistical methods, which we believe are somewhat novel in this context. Appendix~\ref{appendix:axiverse} collects results from Ref.~\cite{2017PhRvD..96h3510S} on the axion mass matrix and RMT.

\section{Black Hole Superradiance}
\label{sec:bhsr_main}

\subsection{Scalar Fields on Kerr Background}

The action for $N$ real scalar fields $\Psi_i$ with masses $\mu_i$ takes the form
\begin{equation}
S = \int d^4x \sqrt{-g}\sum_i\left(-\frac{1}{2}\nabla_\mu \Psi_i\nabla^{\mu}\Psi_i - \frac{1}{2}\mu_i\Psi_i^2\right) \, ,	
\end{equation}
where $\nabla_\mu$ is the covariant derivative on the spacetime with metric $g$. The metric is assumed to be the Kerr metric for a spinning BH. A review of the Kerr geometry is given in Appendix~\ref{appendix:geometry}. This geometry is taken as a background. The superradiant process leads to time dependence of the BH mass and spin, but the structure of the metric does not change due to backreaction. It is known for single field superradiance that the backreaction of the scalar condensate on the Kerr geometry is small. This is because, although the cloud can obtain a large mass, it is distributed over a large volume compared to the BH, leading to low scalar energy density (and thus a low source of curvature) in the cloud~\cite{Brito:2014wla}.

Concerns that backreaction is a more severe problem with large numbers of fields as opposed to dealing with a single field can be alleviated considering the properties of the scalar cloud. The gravitational backreaction is a function of $M_S/M_{\rm BH}$, where $M_S$ is the total mass in the scalar cloud. There is a maximum value of $M_S$ independent of the number of axion fields, which is determined by the BH mass at the initial spin, $M_{\rm BH}(a_*)$, and the irreducible mass after all the spin has been extracted, $M_{\rm BH}(a_*=0)$. $N_{\rm ax}$ fields cannot extract any more total mass than a single field, and for resonant modes the cloud size is of the same order of magnitude for all the fields, therefore gravitational backreaction is not enhanced to a greater severity than the single field case. Non-linearities coming from axion interactions, on the other hand, can increase with the number of fields. We discuss this briefly later.

Thus, neglecting the self-interactions, each field $\Psi_i$ evolves independently on the fixed background. In this separable limit, the total rate of the superradiant process is given simply by the sum of the single field rates:
\be
\Gamma_{\rm tot} = \sum_i\Gamma_i \, .
\label{eqn:rate_sum}
\ee
Solutions of the single field Klein-Gordon equation are discussed in detail in Appendix~\ref{sec:klein}, and the superradiance phenomenon for multiple fields is described in terms of these.

\subsection{Superradiance}

Astrophysical BHs with a mass $M_{\rm BH}$ and spin $J=aM_{\rm BH}$ will spin down via superradiant instabilities extracting energy and angular momentum \cite{Arvanitaki:2010sy,Brito:2015oca}, forming very large gravitationally bound states comprising of a scalar cloud containing exponentially large axion population numbers. Axions bound in this way with a BH form a \emph{gravitational atom}, where superradiant instabilities are found to be strongest when the Compton wavelength of the field, $\lambda_{\rm ax}=\sfrac{\bar{h}}{\mu_{\rm ax}c}$ is comparable to the Schwarzschild radius of the BH, $r_{\rm s} = \sfrac{2GM_{\rm BH}}{c^2}$. 

\begin{figure}
\centering
\includegraphics[width=0.5\textwidth]{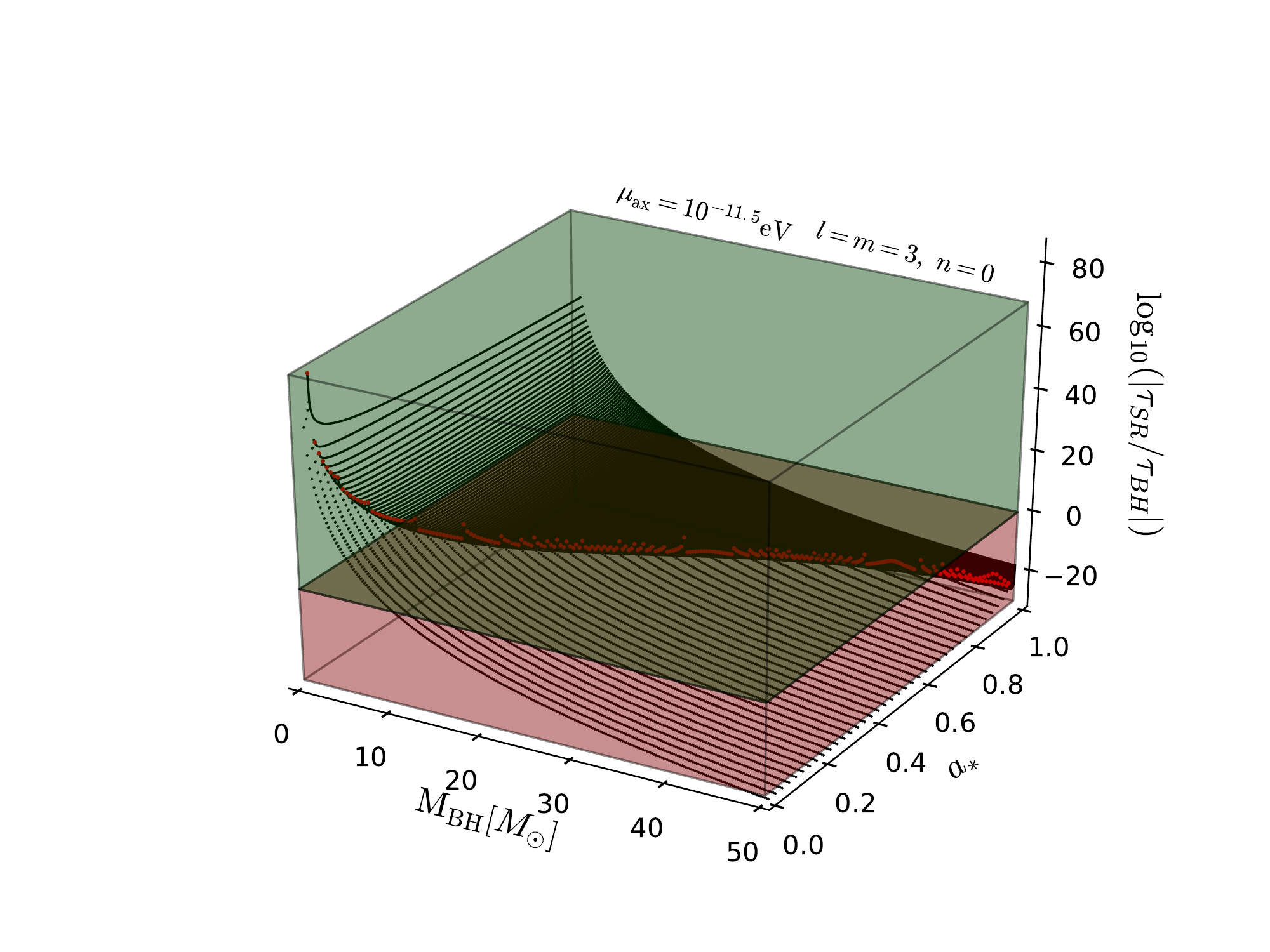}
\caption{Timescale ratios for the superradiance rates for an axion with mass $\mu_{\rm ax}=10^{-11.5}{\rm\ eV}$ compared with a typical BH astrophysical timescale, here taken to be $\tau_{\rm Salpeter}$ (Eq.~(\ref{eq:salpeter})). Each cusp represents the analytical limit beyond which Eq.~(\ref{eq:condition}) is satisfied. The limit to the right of the cusp (sold line) represents the ratio defining the nature of the timescales where superradiance is apparent. The red volume defines the limit in the two dimensional BH mass/spin parameter space where superradiance occurs within the defined astrophysical timescale used to map the Regge plane isocontour limits.}
\label{fig:timescale}
\end{figure}

The condition for mode amplification of the scalar field requires the angular velocity of the BH horizon to exceed the angular phase velocity of the wave mode, defining the superradiance condition (see Fig.~\ref{fig:timescale}) 
\beq
\label{eq:condition}
\frac{\omega}{m}<\omega_{+}\,,
\eeq
where $m$ is the spherical harmonic quantum number. The effective angular velocity of the BH as a function of the dimensionless rotation spin parameter is 
\beq
\omega_{+} = \frac{a_{*}}{2r_{g}(1+\sqrt{1-a_*^2})}\,,
\eeq 
where $a_*$ is defined in region $0 \leq |a_*| < 1$ as\,,
\beq
 a_* = \frac{a}{r_{g}}\,,
\eeq 
in Boyer-Lindquist coordinates. The gravitational radius of the BH is, 
\beq
r_{g} \equiv G_{\rm N}{M_{\rm BH}}\,.
\eeq 
In parts of the following we shall work in units $c=\hbar=G=1$ such that $r_{g} \equiv M_{\rm BH}$.
The Kerr-Klein-Gordon system admits quasi-bound states with complex eigenfrequencies
\begin{equation}
\omega_{nlm} = \omega_{R} + i\omega_{I}\,,	
\label{eq:eigenfrequencies}
\end{equation}
where $\{\omega_R,\omega_I\} \in \mathbb{R}$. Kerr BHs present a critical frequency for superradiant scattering 
\begin{equation}
\omega_{c} \equiv m\Omega_{H}\,,	
\end{equation}
 with $m$ representing the angular momentum about the BH spin axis. This defines the stability thresholds for the scalar modes:
 \begin{align}
 \omega_{nlm} &> 	m\Omega_{H} \ \rightarrow \ {\rm Stable} \,, \\
 \omega_{nlm} &< 	m\Omega_{H} \ \rightarrow \ {\rm Unstable}\,.
 \end{align}
For values of $\omega_{nlm}$ satisfying $0<\omega_{nlm}<\omega_c$ the imaginary component is positive defining the superradiant regime. 
Scalar modes in the presence of the Kerr BH spacetime with scalar mass, $\mu_{\rm ax }$ contain a natural confinement mechanism in the limit 
 \begin{equation}
 0<\omega_{nlm}<\mu_{\rm ax}	\,,
 \end{equation}
where they are bounded from escaping via their potential (Eq.~(\ref{eq:effectivepot})). Modes satisfying these conditions will grow exponentially over time identifying the presence of an instability in the Kerr spacetime. When $\omega_{nlm} = \omega_{c}$ the imaginary component of the frequency drops out allowing for the formation of bound states or scalar clouds. 

Aside from regions within a significant proximity to the BH the gravitational potential is $\propto \sfrac{1}{r}$ where the spherically symmetric properties of the potential to leading order allow for a separation of variables of the field evolution in the background reproducing a Schr\"{o}dinger type wave-equation (see Section~\ref{sec:klein}). The equation for the separated radial wave function (Eq.~(\ref{eq:radial})) is the equivalent to that of the Scalar Coulomb, thereby presenting hydrogenic wavefunctions. To leading order the energy levels for the bound states are well approximated by the spectrum of the hydrogen atom in the non-relativistic limit. When the superradaiance condition is saturated the eigenfrequencies take the approximate form, 
\begin{equation}
\omega_{nlm} \equiv \omega_{R} \approx \mu_{\rm ax} \left(1 - \frac{\alpha^2}{2(n+l+1)^2}\right) \approx \mu_{\rm ax}.	
\end{equation}
The orbitals around the BH are indexed by the overtone ($n$), orbital multi-pole ($l$) and azumutal ($m$) quantum numbers satisfy $l \leq n-1$ and $|m|\leq l$ forming discrete sets, \{n,l,m\} used to quantise the superradiant behaviour. Superradiance requires evolving modes to co-rotate with the BH which satisfy, $m>0$. Details of the methodology used to determine the approximated eigenspectrum are given in Appendix~\ref{appendix:bhsr}. The dimensionless coupling of the gravitational BH-scalar condensate system is, 
\beq
\alpha = r_{g}\mu_{\rm ax} \equiv \mu_{\rm ax}M_{\rm BH}.
\eeq  
in our choice of units. Fig.~\ref{fig:alpha} presents the coupling strength for potential regions of the axion mass parameter space open to investigation for BH masses spanning the stellar and supermassive limits. 

\subsection{Superradiance Rates} 
\label{sec:rates}
The evolution of the axion field is defined by the characteristic eigenfrequencies corresponding to the instability timescales for the unstable modes of the system. The nature of scalar instabilities is well researched covering both the frequency \cite{PhysRevD.22.2323,ZOUROS1979139,Dolan:2007mj} and time domains \cite{Dolan:2012yt}. In the frequency regime in order to extract valid quasibound state instability rates, $\Gamma_{nlm}$, which depend on the wavefunction near the horizon, either one of two approaches can be implemented. The superradiance rates are defined as the small imaginary component of the energy of the free field solution on the Kerr background. Analysing the region of the parameter space where $\alpha \sim 1$, solutions for the unstable modes can be found using a numerical analysis of the wave equation (see Appendix~\ref{app:numerics}) \cite{Dolan:2007mj,Furuhashi:2004jk,Cardoso:2005vk}. When $\alpha$ surpasses unity WKB methods are formulated to evaluate the rate, presenting an exponential suppression proportional to $\alpha$ where $\Gamma_{nlm} \propto e^{-3.7\alpha}$ \cite{ZOUROS1979139,Arvanitaki:2010sy}.

It has been shown it is possible to find analytical solutions to approximate the instability rate, incorporating matching techniques between different regimes of validity as a a function of $\alpha$. For a particular bound state if the superradiance condition is satisfied then providing that the instability rate is quicker than relevant astrophysical timescales, wave modes will extract energy and angular momentum from the BH. It has been shown in the ${\alpha}\ll 1$ regime known as the ``small mass approximation'' the evolution of the superradiant instability can be analytically described via a matched asymptotic expansion. This solution was initially derived by Detweiler to solve the Klein-Gordon equation of the scalar field perturbation \cite{PhysRevD.22.2323}. Comparing the large $r$ behaviour of the near-region solution with the small $r$ behaviour of the far-region solution yields the allowed values of the small imaginary component of the frequency $\omega_{I}$. The instability rate in the small mass approximation is defined as 
\begin{equation}
\Gamma_{nlm} = 2\mu_{\rm ax}r_{+}\left( m \Omega_{H} - \mu_{\rm ax}\right)\left(\mu_{\rm ax}M_{\rm BH}\right)^{4l +4}\mathcal{C}_{nlm},
\label{eq:det}	
\end{equation}
where, 
\begin{widetext}
\begin{equation}
\mathcal{C}_{nlm} = \frac{2^{4l+2}(2l+n+1)!}{n!(n+l+1)^{2l+4}}\left[ \frac{l!}{(2l+1)! (2l)! }\right]^2 \\ \times \prod_{j=1}^{l}\left[ j^2\left(1-\frac{a^2}{M_{\rm BH}^2}\right) +4r^2_{+}\left(\mu_{\rm ax} - m\Omega_h \right)^2   \right].
\end{equation}
	
\end{widetext}

\begin{figure*}
\centering
\begin{tabular}{cc}
    \includegraphics[width=0.49\linewidth]{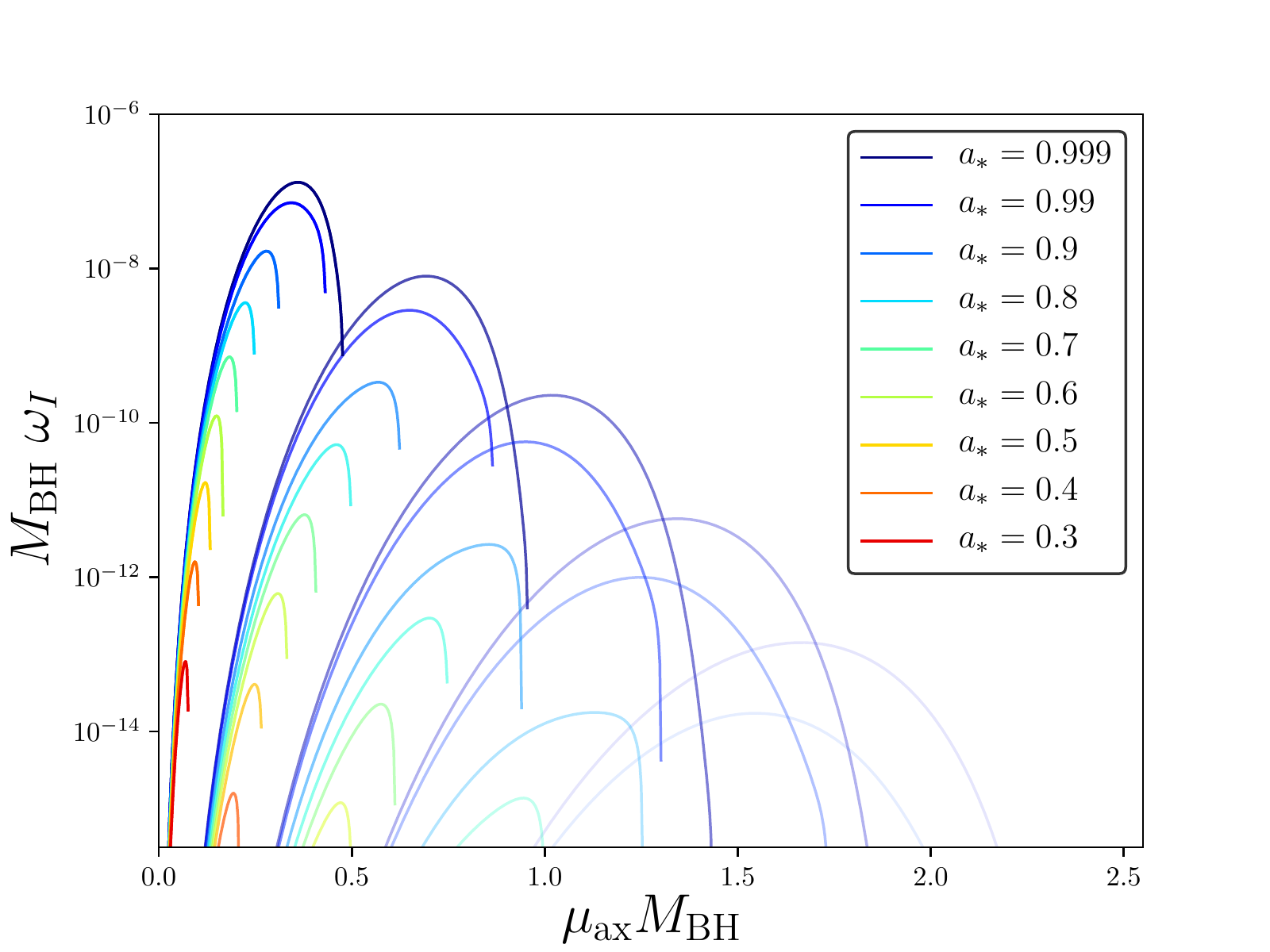}&
    \includegraphics[width=0.49\linewidth]{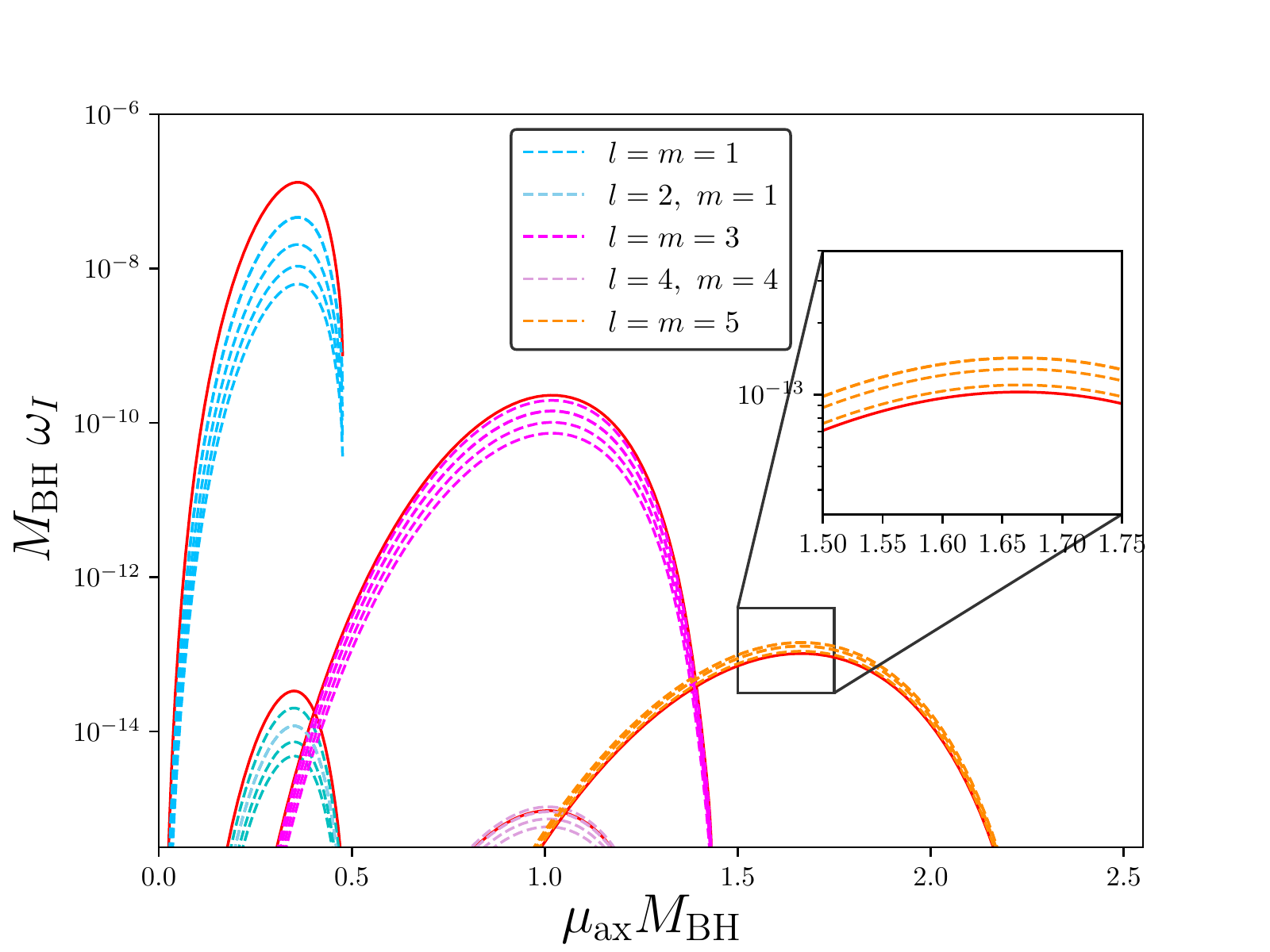}\\
\end{tabular}
\caption{Imaginary component of the bound-state frequency, $M_{\rm BH}\omega_{I}$ representing the superradiance instability rate, $\Gamma_{\rm nlm} $ as a function of the dimensionless coupling, $\alpha = \mu_{\rm ax} M_{\rm BH}$. \emph{Left panel:} Superradiance rates for each orbital/azimuthal quantum numbers, l = m = 1 to 5 for various values of the dimensionless BH spin $a_{*}$, approaching the extremal limit. \emph{Right panel:} Superradiance rates for the fundamental and higher order overtone modes n = 0 to 4 for configurations satisfying $l = m$ and $l>m$. The red lines correspond to the fundamental overtone modes, $n=0$ which become subdominant for values of $l = m \geq 4$.}
\label{fig:superradiance_rates}
\end{figure*}

 It can be seen from Eq.~(\ref{eq:det}) the superradiance rates for scalar fields scale approximately as 
 \begin{equation}
 \Gamma_{nlm} \propto \alpha^{4l+4} \mu_{\rm ax}, 
 \end{equation}
which is maximised close to the superradiance boundary. In Fig.~\ref{fig:superradiance_rates} we present the superradiance rates for a range of modes and spins as a function of the axion/BH coupling, $\mu_{\rm ax}M_{\rm BH}$. The fastest growing mode occurs for $\Gamma_{011}$ with the superradiance rates exponentially suppressed for higher values of $l$. The maximum superradiance rates are found by fixing the values of $l$ and $m$ such that, $l=m$ where $m$ determines the ability to satisfy the superradiance condition in Eq.~(\ref{eq:condition}) (See \emph{right panel} of Fig.~\ref{fig:superradiance_rates}). The value of $\Gamma_{nlm}$ has a limited dependance on the overtone mode, $n$. When the BH possesses significant spin higher order overtone modes for larger values of $l=m$ can present greater superradiance rates as compared to the fundamental overtone mode. Analytically this is apparent for $l=m=4$ (see \emph{right panel} of Fig.~\ref{fig:superradiance_rates}) where it has also been shown to occur for $l=m=3$ considering numerical solutions \cite{Yoshino:2015nsa}.

\subsection{Superradiant Evolution}

Sequential to the formational phase of a BH, superradiant evolution can begin via quantum fluctuations in the vacuum where each of the quantised superradiant levels begin to grow exponentially with their corresponding superradiance rates. The fastest-growing level which satisfies the superradiance condition always dominates the initial superradiant evolution until it has extracted enough spin so that the superradiance condition is no longer satisfied. Once the scalar cloud has extracted the maximal spin for the dominant mode the system can be be considered as a (quasi)-stationary hairy BH for astrophysical purposes. The BH energy loss through mass reduction is minimal compared to the shift in angular momentum due to the extend of the scalar cloud. Once the growth of the dominant level has stopped the BH will spend a significant portion of its lifetime on a Regge trajectory (dashed lines in Fig.~\ref{fig:solarregge}) separating higher mode instability bounds. This can be seen from the basic intuition that as the higher modes of the BH begin to spin down the BH perturbing it from the Regge trajectory the negative component of the eigenfrequency for the previous mode dominates the evolution, spinning up the BH. This process is apparent until a significant portion of the scalar density in the cloud is reduced from the previously dominant level.  At this point the BH traverses the Regge plane towards the successive superradiant boundary, repeating the process until the timescales considered are to large for superradiance to occur. 
 
If non-linearities are taken into account level mixing can increase the time spent on the superradiance condition boundary via perturbations of the gravitational potential around the BH. Dissipation of the scalar cloud can occur through processes such as the annihilation of axions into gravitons or unbound axions~\cite{Arvanitaki:2010sy,2015PhRvD..91h4011A}. In general the scalar cloud becomes maximally occupied before annihilation processes begin in the non-relativistic limit. Further complications to the trajectory evolution of the BH could come from the bosenova phenomena, introducing intermediate stages comprising of bursts of GWs and phases spinning down the BH before the superradiance condition is finally saturated. Given the hierarchy of timescales between the superradiant instability and the GW emission from non-linearities when compared to the dynamical time scale of the BH it is possible to study the systems evolution in the quasi-adiabatic approximation for $N_{\rm ax}$ fields \cite{Brito:2014wla,Brito:2015oca,Brito:2017zvb}. The total scalar energy flux from the superradiance process through the horizon is, 
\begin{equation}
\dot{E} = 2M_S \sum_{g=1}^{N_{\rm ax}}  \omega_{I_{,g}}\,.
\end{equation}
With a disregard for accretion the evolution of the system is described by the following equations  
\begin{align}
 -\dot{E}_S &= \dot{M}_{\rm BH}  \label{eq:M}\,, \\
 -\dot{E} &= \dot{M}_{\rm BH} + \dot{M}_S\,, \\
\sfrac{-m \dot{E_S}}{\sum_{g=1}^{N_{\rm ax}}\omega_{R,g}} &= \dot{J}_{\rm BH} \label{eq:J}\,,\\ 
 -m\sfrac{\dot{E}}{\sum_{g=1}^{N_{\rm ax}}\omega_{R,g}} &= \dot{J}_{\rm BH} + \dot{J}_S  \, ,
\end{align}
where $E_S$ is the energy of the scalar cloud. The scalar cloud extracts mass and spin until reaching the saturation point. The final BH spin is, 
\begin{equation}
J_{\rm BH,F} = \frac{4mM_{\rm BH,F}^3\sum_{g=1}^{N_{\rm ax}}\omega_{R,g}}{m^2+4M_{\rm BH,F}^2 \sum_{g=1}^{N_{\rm ax}}\omega^2_{R,g}}\,. 	
\end{equation}
The final mass of the BH after the phase of superradiant evolution is defined by Eq.~(\ref{eq:J}) were the variations in the defining BH parameters are related by,
\begin{equation}
\delta J_{\rm BH} = \frac{m}{\sum_{g=1}^{N_{\rm ax}}\omega_{R,g}}\delta M_{\rm BH}\,.	
\end{equation}
This defines the final mass of the BH:
\begin{equation}
M_{\rm BH,F} = M_{\rm BH,I} - \frac{\sum_{g=1}^{N_{\rm ax}}\omega_{R,g}}{m}(J_{\rm BH,I} - J_{\rm BH,F}) \, .
\end{equation}

The true evolution of course is a complicated picture where non-linearities must be accounted for along with the properties of each system. In particular for SMBHs their mass are generally accumulated via accretion which requires very significant perturbations in order to match the evolutionary traits a stellar BH may follow for example in terms of traversing the mass-spin Regge plane.  
\subsection{The Regge Plane}

\begin{figure*}[t]
\centering
\includegraphics[width=0.85\textwidth]{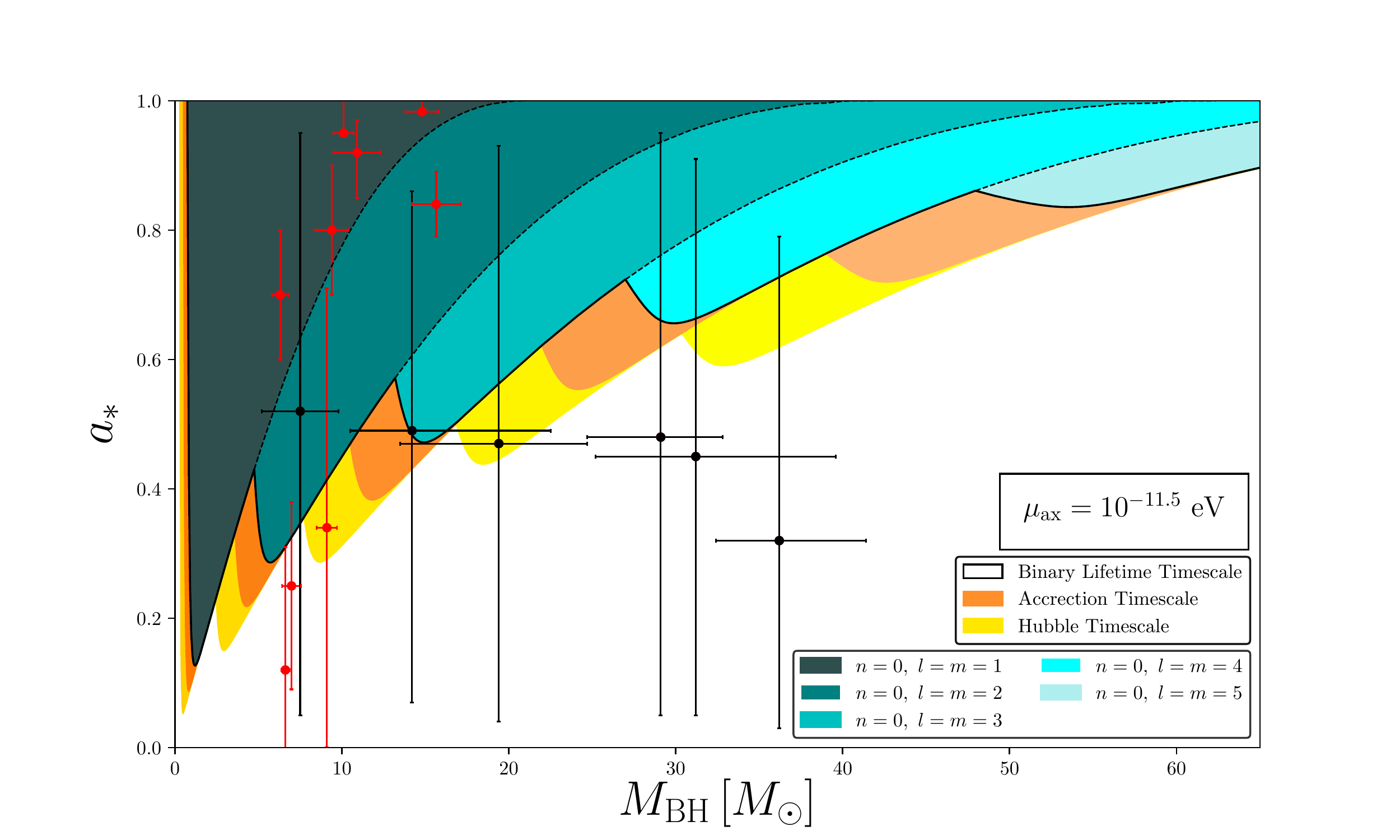}
\caption{Isocontour exclusion bounds in the BH mass-spin Regge plane for an axion mass, $\mu_{\rm ax} = 10^{-11.5} \ {\rm eV}$ probing the stellar BH parameter space. The limits (black outline) for the instability threshold are obtained by fixing the superradiant instability time scales for each value of the orbital/azimuthal quantum numbers, l = m = 1 to 5 equal to the timescale of a typical BBH system shown in Eq.~(\ref{eq:stellar_life}). The extended limits come from considering  superradiant instability timescales shorter than $\tau_{\rm Salpeter}$ (orange, Eq.~({\ref{eq:salpeter}})) and $\tau_{\rm Hubble}$ (Yellow, Eq.~(\ref{eq:hubble_time})). The red/black data points denote mass and spin estimates of the stellar BHs from X-ray/BBH sources presented in Tabel~\ref{tab:cosmopar}.}
\label{fig:solarregge}
\end{figure*}

\begin{figure*}[t]
\centering
\includegraphics[width=0.85\textwidth]{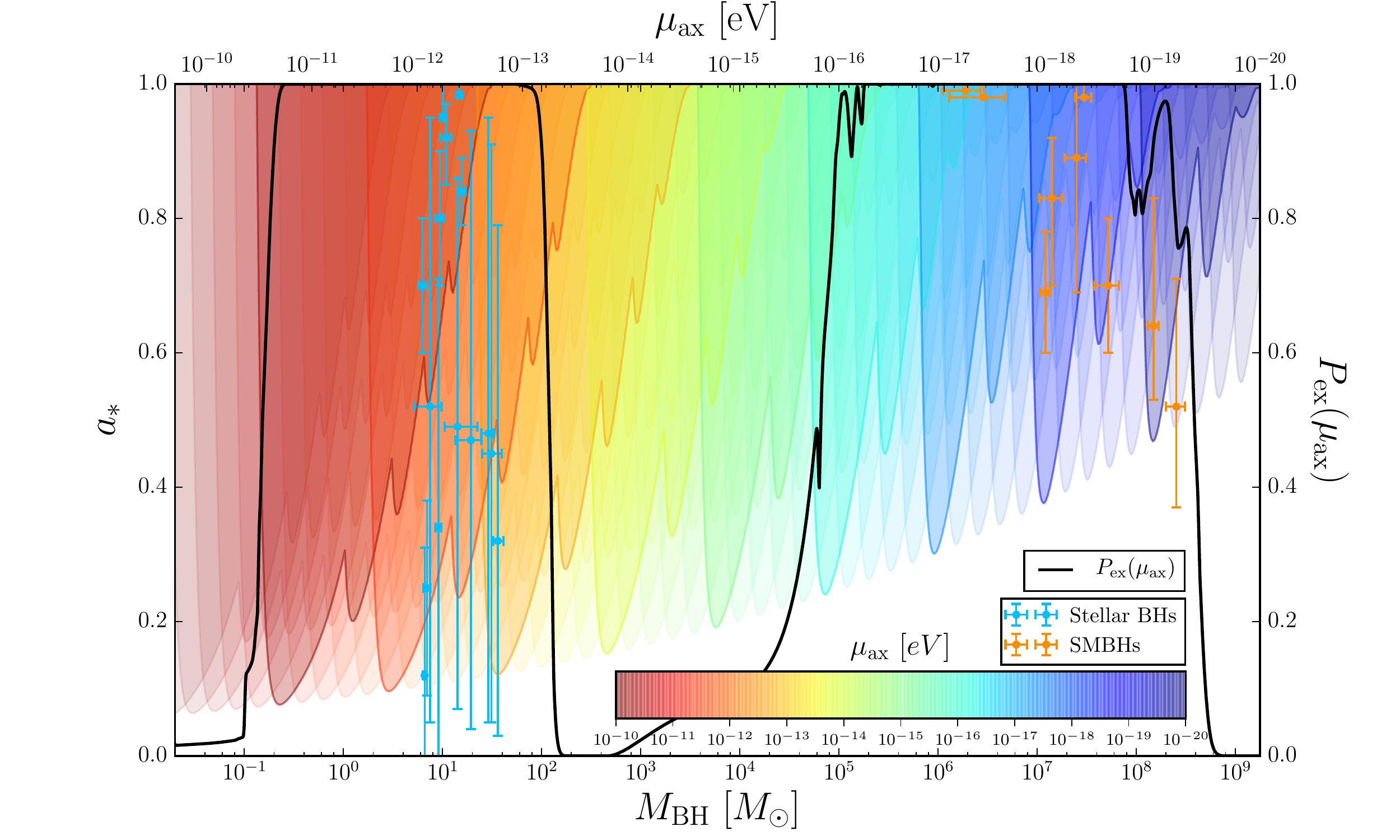}
\caption{Isocontour exclusion bounds with calculated total exclusion probabilities in the BH mass-spin Regge plane from superradiant instabilities with a single axion field with mass, $\mu_{\rm ax}$ spanning the limits in Eq.~(\ref{eq:alp_mass_window}). The shaded regions represent instability thresholds shorter than the time scale $\tau_{\rm Salpeter}$ in Eq.~(\ref{eq:salpeter}) for each value of the dominant orbital/azimuthal quantum numbers, l = m = 1 to 5. The blue data points are mass/spin estimates of stellar X-ray and BBH systems. The orange points correspond to mass/spin estimates of SMBHs from X-ray reflection spectroscopy. The exclusion probability function (black line) is calculated using the statistical model in Appendix~\ref{appendix:stats} using the BHs compiled in Table~\ref{tab:cosmopar} and is given as a function of the axion mass spanning both the stellar and supermassive regimes.}
\label{fig:regge_spectrum}
\end{figure*}
A fundamental prediction stemming from superradiant instabilities of bosonic fields is the existence of exclusion regions in the BH Regge plane. Estimates of the instability time scale, $\tau_{\rm SR}$ partnered with reliable spin measurements for BHs, can be used to impose stringent constraints on the allowed masses of ultralight bosons. These bounds on the parameters of ultralight bosons follow from the requirement that in principle an astrophysical spinning BH should be stable over its lifetime. A superradiant instability time scale which acts faster than core processes such as accretion form observational thresholds on the expected regions of the two-dimensional mass-spin parameter space BHs should fall in. Following the process of superradiant evolution a large number of BH observations should trace out the superradiance condition boundaries, mapping the Regge trajectories given the existence of as yet unidentified fields. For axions the shape of the gaps in the Regge plane are extremely sensitive to variations in the superradiant growth rate with the scalar mass. A BH therefore should be excluded from observational measurements given the existence of an ultralight boson if it's spin is measured above the relevant level curves for different orbital states of the quantised modes for the field. The bounds for bosonic fields with spin are wider than those for axion-like particles and so the potentially large systematic errors in BH spin measurements could act as a current restriction to this approach for spin-0 fields. The axion mass window which can be probed is fixed by the heaviest supermassive BHs with accurate recorded spin measurements along with a lower bound defined by the lightest measured stellar mass BHs. 

The current lower and upper bounds on BH masses from X-ray spectroscopy and emission data covers the approximate region 
\begin{equation}
 5M_{\odot}\lesssim M_{\rm BH} \lesssim 5\times10^8 M_{\odot}\,,	
\label{eq:massrange}
\end{equation}
which defines the relevant axion mass window as, 
\begin{equation}
10^{-20}{\rm eV}\lesssim \mu_{\rm ax} \lesssim 10^{-11}{\rm eV}\,.
\label{eq:alp_mass_window}	
\end{equation}
The isocontours defining the exclusion bounds are a function of the instability timescale and the boson mass. As the axion mass decreases the instability exclusion contours reduce in size. This corresponds to tighter instability regions which require larger spins for more massive BHs. Taking into account accretion and GW emissions can also slightly reduce the bounds in the Regge plane \cite{Brito:2014wla}. The timescales associated to the astrophysical processes of relevance alter when considering different compact object systems. 

For rapidly spinning BH candidates in X-ray binary systems or binary BH (BBH) mergers identified as detectable GW sources by LIGO more accurate constraints can be imposed when considering the typical timescales associated to a binary systems lifetime as other astrophysical processes such as accretion are sub-leading in this regard. A typical lower bound approximation for the lifetime of the binary system is given as
\begin{equation}
\tau_{\rm BH} \sim 10^{6} \ {\rm yrs}\,,
\label{eq:stellar_life}
\end{equation}
for the most accurate constraints. The most conservative limits come from exclusion regions constructed using the Hubble time, 
 \begin{equation}
 \tau_{\rm H} \sim 10^{10} \ {\rm yrs}\,.
 \label{eq:hubble_time}
 \end{equation}
 
 As opposed to stellar binary objects the relevant timescales for AGN in order for superradiance to maximally grow the scalar cloud for each quantised level come from accretion models. A statistical analysis of the exclusion limits over the whole BH mass region defined in Eq.~(\ref{eq:massrange}) requires us to use a characteristic timescale derived from accretion considerations. The time scale for mass growth increases exponentially with an e-folding time given by a fraction $1/f_{\rm Edd}$ of the Salpeter time scale, where $f_{\rm Edd}$ is the Eddington ratio for mass accretion. The accretion time scale is estimated using the Salpeter time for a BH radiating at it Eddington limit
\begin{equation}
\label{eq:salpeter}
\tau_{\rm Salpeter} = \frac{\sigma_{\rm T}}{4 \pi m_{\rm P}} \sim 4.5 \times 10^7 \  {\rm yrs}\,,
\end{equation}
where $\sigma_{\rm T}$ is the Thompson cross section and $m_{\rm P}$ is the proton mass \cite{Shankar:2007zg}. In order to model the accretion time the following parameters can be introduced \cite{Brito:2017zvb}
\begin{equation} 
\tau_{\rm Salpeter} =  4.5 \times 10^8 \ { \rm yrs} \ \frac{\eta}{f_{\rm Edd}(1-\eta)}\,,
\end{equation}
where $\eta$, the thin-disk radiative efficiency is a function of the spin related to a specific energy at the innermost stable circular orbit (ISCO). We select a typical value for the efficiency, $\eta = 0.1$ and the most conservative value of $f_{\rm Edd} = 1$ to model the effects of accretion. This fixes the superradiant instability timescale as $\tau_{\rm SR} = 45 \ {\rm Myrs}$. Increasing the bounds on $f_{\rm Edd}$ allows for more optimistic models incorporating potential periods of super-Eddington accretion. A redefinition of $f_{\rm Edd}$ holds the same equivalence as considering a subpopulation of degenerate mass fields (see Section.~\ref{sec:degenerate}) or considering different astrophysical processes to define the superradiance timescale. Such considerations are a limitation in the ``logistics'' of encapsulating the behaviour of the total BH spectrum and as such we follow the most conservative limit defined above. 

An individual treatment of the instability timescales derived from the properties of the accretion disc stability for each BH candidate can be used to tighten constraints of the field mass exclusions \cite{Cardoso:2018tly}. As the timescale limits for the superradiant instability are increased the limits for each mode, $m$ will begin to saturate to the limits set by the boundaries of the superradiance condition. This effect is most prominent for higher order modes in the spin axis of the Regge plane allowing for enhancements in the potential to constrain ultralight bosons using observations of BHs with spins a moderate fraction of the extremal limit.              
 
In Fig.~\ref{fig:solarregge} this is shown in the example exclusion window for a fixed axion mass of $\mu_{\rm ax} = 10^{-11.5}\ {\rm eV}$ in the stellar BH parameter space for each of the instability timescales in Eq.~(\ref{eq:stellar_life}), Eq.~(\ref{eq:hubble_time}) and Eq.~(\ref{eq:salpeter}). As the considered timescale increases the saturation of the mode bounds in the limit of the superradiance condition sees the greatest enhancement for $l=m=5$. The red data points are the X-ray binary system BHs from Table~\ref{tab:cosmopar}. The black data points are the \emph{primary} and \emph{secondary} sources involved in the BBH coalescence events (GW150914,GW151226 and GW170104) for several LIGO detections. Extremal BHs such as NGC 4051 impose constraints on each of the $l=m=1,2$ and $3$ modes demonstrating the ability of well defined rapidly spinning BHs to constrain significant portions of the axion mass parameter space. An axion mass of $\mu_{\rm ax} \approx 10^{-11.5}\ {\rm eV}$ is therefore tightly constrained by known X-ray binary sources as shown in both Fig.~\ref{fig:solarregge} with the poor measurements from LIGO data open to a far greater uncertainty if treated separately.
      
Fig.~\ref{fig:regge_spectrum} details the exclusion bounds for the treatment of a single axion covering the full region of the axion mass window in Eq.~(\ref{eq:alp_mass_window}), along with the full stellar BH and SMBH data presented in Table~\ref{tab:cosmopar}. The \emph{primary} axis presents the Regge exclusion bounds for an instability time scale $\tau_{\rm SR} = 45 \ {\rm Myrs}$ as a function of the axion mass, $\mu_{\rm ax}$. The \emph{blue/orange} data points are the stellar/SM BHs in Table~\ref{tab:cosmopar}. The \emph{secondary} axis displays the probability exclusion function formulated from the statistical model in Appendix~\ref{appendix:stats} across the total BH mass range. The function ``well'' corresponds to the absence of any well defined IMBH candidates. Well defined mass and spin measurements for BHs covering the approximate region $10^{2}M_{\odot}-10^{6}M_{\odot}$ could fill the currently inaccessible portion of the parameter space and probe interesting masses for axions associated to GUT and supersymmetric models in string/M-theory. The most promising realisation of detecting BHs in this space comes from the proposed space based gravitational wave observatories such as the Laser Interferometer Space Antenna (LISA) (see Fig.~\ref{fig:alpha}).      
   
\subsection{Black Hole Spin Measurements from Binary Systems and Active Galactic Nuclei}

The identification of compact systems has seen a steady increase over the past decades with a number of X-ray binary sources and active galactic nuclei (AGN) now providing well defined measurements for the masses and spins of these systems. Currently the main sources of error for catalogued BHs comes from the systematic errors when modelling the emission of the accreting disc of the system. Both stellar BH and SMBH measurements come from analysing the X-ray spectrum of the accretion disk for identified compact sources. Assuming that General Relativity holds true as a valid description of the spacetime region outside the BH horizon and the ISCO of the accretion disk possesses a monotonic function potential then estimates on the spin of BHs can be made. In principle most BH candidates with well defined parameter estimates come from either thermal continuum fitting of the inner accretion disk or inner disk reflection modelling in order to determine the size of the ISCO. Further to this BH spin data has recently been collected via the observations made in several BBH mergers by LIGO \cite{TheLIGOScientific:2016pea,2041-8205-851-2-L35,PhysRevLett.119.141101}. Currently such observations contain large errors on both the mass and spin of the BHs when compared to existing X-ray binary system records. The resultant BHs formed from such astrophysical events cannot be included in considerations of constraining the masses of bosons given their timescale for observation is less than typical instability timescales by definition in the process of identification. Generally though future generation ground based detectors are still expected to produce large error measurements on BHs identified in this way and so impose a strong limitation on the accuracy of measurements used for constraints. Improvements in observatory sensitivity with space operated missions such as LISA \cite{Klein:2015hvg} will open up the potential for a large catalogue of accurate BH measurements capable of probing a large potion of the cosmologically significant sector for axion-like fields. A large exclusion in the fully accessible space could also lead to tight constraints on how the axion population or sub-populations may be distributed when seeking realisations of desirable models in the context of cosmology.  

We restrict ourselves to considering only BHs with detailed mass and spin errors. Each BH chosen for our analysis therefore has upper and lower bounds on both their mass and spin with well defined quoted uncertainties. In Table~\ref{tab:cosmopar} we present all the stellar BHs and SMBHs used to constrain our axion distributions in Section~\ref{sec:results} along with their associated references. For a review of compiled stellar BH data see Refs.~\cite{McClintock:2013vwa,Middleton:2015osa} and for SMBHs see Refs.~\cite{2011ApJ...736..103B,Reynolds:2013qqa}.        

{
\renewcommand{\arraystretch}{1.0}
\begin{table*}
\begin{center}
\caption{Stellar BH and SMBH systems used to apply constraints on axion masses and values of $N_{\rm ax}$ for various model mass spectra. BHs are selected with reliable mass and spin measurements and associated errors are quoted with their confidence limits and corresponding references. Stellar BH measurements come from both X-ray binary systems via X-ray continuum-fitting methods and BBH mergers from detected coalescence events at LIGO. SMBHs are measured AGN using X-ray reflection spectroscopy. Where two methods have been stated we use averaged posterior values for each. For review material and collections of stellar BHs see Refs.~\cite{Middleton:2015osa,Miller:2014aaa}. Compiled AGN data can be found in Refs.~\cite{Reynolds:2013qqa,Reynolds:2013rva,2011ApJ...736..103B}. }
\label{tab:cosmopar}
\begin{tabular*}{\textwidth}{K{0.24\textwidth}K{0.22\textwidth}K{0.10\textwidth}K{0.10\textwidth}K{0.09\textwidth}K{0.09\textwidth}K{0.09\textwidth}}
\toprule
    \text{$Object$}   & \text{$Method$} &\text{$Mass\ (M_{\rm BH})$}       & \text{$Spin \ (a_{*})$}& \text{$Mass \ CL$}& \text{$Spin \ CL$} & \text{$Ref.$}  \\ 
    \hline
    \rule{0pt}{2.6ex}
    {\bf Stellar }   & &$[M\odot]$
      \rule[-1.2ex]{0pt}{0pt}
\\ 
   \hline
   GW150914 (Primary) &EOBNR+IMRPhenom &$36.2^{+5.20}_{-3.80}$ & $0.32^{+0.47}_{-0.29}$&90\%&90\%&\cite{TheLIGOScientific:2016pea}
\\
GW150914 (Secondary)&EOBNR+IMRPhenom &$29.1^{+3.70}_{-4.40}$ & $0.48^{+0.47}_{-0.43}$&90\%&90\%&\cite{TheLIGOScientific:2016pea}
\\
GW151226 (Primary)&EOBNR+IMRPhenom &$14.2^{+8.30}_{-3.70}$ & $0.49^{+0.37}_{-0.42}$&90\%&90\%&\cite{TheLIGOScientific:2016pea}
\\
GW151226 (Secondary) & EOBNR+IMRPhenom& $7.5^{+2.30}_{-2.30}$ & $0.52^{+0.43}_{-0.47}$&90\%&90\%&\cite{TheLIGOScientific:2016pea}
\\
GW170104 (Primary) & Eff+Full precession &$31.2^{+8.40}_{-6.00}$ & $0.45^{+0.46}_{-0.40}$&90\%&90\%&\cite{Abbott:2017vtc}
\\
GW170104  (Secondary) & Eff+Full precession& $19.4^{+5.30}_{-5.90}$ & $0.47^{+0.46}_{-0.43}$&90\%&90\%&\cite{Abbott:2017vtc}
\\
Cygnus X-1 & Continuum (KERRBB2) &$14.8^{+1.00}_{-1.00}$ & $\geq 0.983$ &$1\sigma$&$3\sigma$&\cite{2011ApJ...742...84O}/\cite{Gou:2013dna}
\\
XTE J1550-564 & Continuum (KERRBB2) & $9.10^{+0.61}_{-0.61}$ & $0.34^{+0.37}_{-0.34}$&$1\sigma$&90\%&\cite{2011ApJ...730...75O}/\cite{2011MNRAS.416..941S}
\\
A 0620-00&Continuum (KERRBB2) &$6.61^{+0.25}_{-0.25}$ & $0.12^{+0.19}_{-0.19}$&$1\sigma$&$1\sigma$&\cite{2010ApJ...710.1127C}/\cite{2010ApJ...718L.122G}
\\
4U 1543-475&Continuum (KERRBB)& $9.4^{+1.00}_{-1.00}$ & $0.8^{+0.10}_{-0.10}$ &$1\sigma$&$1\sigma$& \cite{Orosz67}/\cite{Shafee:2005ef} 
\\

GRO J1655-40&Continuum (KERRBB)& $6.30^{+0.50}_{-0.50}$ & $0.7^{+0.10}_{-0.10}$&95\%&$1\sigma$&\cite{1538-4357-636-2-L113}/\cite{Greene:2001wd}
\\
GRS 1915+105&Continuum (KERRBB2)& $10.1^{+0.60}_{-0.60}$ & $\geq 0.95$&$1\sigma$&$1\sigma$&\cite{Steeghs:2013ksa}/\cite{McClintock:2006xd}
\\
LMC X-1&Continuum (KERRBB2)& $10.91^{+1.41}_{-1.41}$ & $0.92^{+0.05}_{-0.07}$&$1\sigma$&$1\sigma$&\cite{Orosz:2008kk}/\cite{2009ApJ...701.1076G}
\\
LMC X-3&Continuum (KERRBB2)& $6.98^{+0.56}_{-0.56}$ & $0.25^{+0.13}_{-0.16}$&$1\sigma$&$1\sigma$& \cite{2014ApJ...794..154O}/\cite{Steiner:2014zha}
\\
M33 X-7&Continuum (KERRBB2)& $15.65^{+1.45}_{-1.45}$ & $0.84^{+0.05}_{-0.05}$&$1\sigma$&$1\sigma$&\cite{Orosz:2007ng}/\cite{1538-4357-679-1-L37}
\\

    \hline
    \rule{0pt}{2.6ex}
    {\bf Supermassive}   &&$\times10^{6}[M\odot] $&
      \rule[-0.0ex]{0pt}{0pt}
      \\
      \hline
Mrk 335   & Reflection (Suzaku)&$14.20^{+3.70}_{-3.70}$&$0.83^{+0.09}_{-0.13}$&$1\sigma$&90\%& \cite{Peterson:2004nu}/\cite{2013MNRAS.428.2901W}     \\
Fairall 9  &Reflection (Suzaku)&$255.0^{+56.0}_{-56.0}$ &$0.52^{+0.19}_{-0.15}$ &$1\sigma$&90\%&\cite{Peterson:2004nu}/\cite{2012ApJ...758...67L}     \\
Mrk 79 &Reflection (Suzaku)&$52.40^{+14.40}_{-14.40}$  & $0.70^{+0.10}_{-0.10}$&$1\sigma$&90\%& \cite{Peterson:2004nu}/\cite{2011MNRAS.411..607G}     \\
NGC 3783 &Reflection (Suzaku)&$29.80^{+5.40}_{-5.40}$  & $\geq 0.98$&$1\sigma$&90\%& \cite{Peterson:2004nu}/\cite{2011Brenneman}     \\
MCG-6-30-15  &Reflection (Suzaku)&$2.90^{+1.80}_{-1.60}$  & $\geq 0.98$&$1\sigma$&90\%&\cite{McHardy:2005ut}/\cite{Brenneman:2006hw}       \\
NGC 7469 &Reflection (Suzaku)&$12.20^{+1.40}_{-1.40}$  & $0.69^{+0.09}_{-0.09}$&$1\sigma$&90\%&\cite{Peterson:2004nu}/\cite{doi:10.1111/j.1365-2966.2011.19224.x}      \\
Ark 120 &Reflection (Suzaku)&$150.0^{+19.0}_{-19.0}$  & $0.64^{+0.19}_{-0.11}$&$1\sigma$&90\%&\cite{Peterson:2004nu}/\cite{2013MNRAS.428.2901W}      \\
 Mrk 110 &Reflection (Suzaku)&$25.10^{+6.10}_{-6.10}$  & $\geq 0.89$&$1\sigma$&90\%&\cite{Peterson:2004nu}/\cite{2013MNRAS.428.2901W}      \\
NGC 4051 &Reflection (Suzaku)&$1.91^{+0.78}_{-0.78}$  & $\geq 0.99$&$1\sigma$&90\%&\cite{Peterson:2004nu}/\cite{8175999}      \\
\botrule  
\end{tabular*}
\end{center}
\end{table*}
}

\section{The Axion Mass Spectrum}
\label{sec:axion_models}
\begin{figure*}
\centering
\begin{tabular}{cc}
\includegraphics[width=0.49\linewidth]{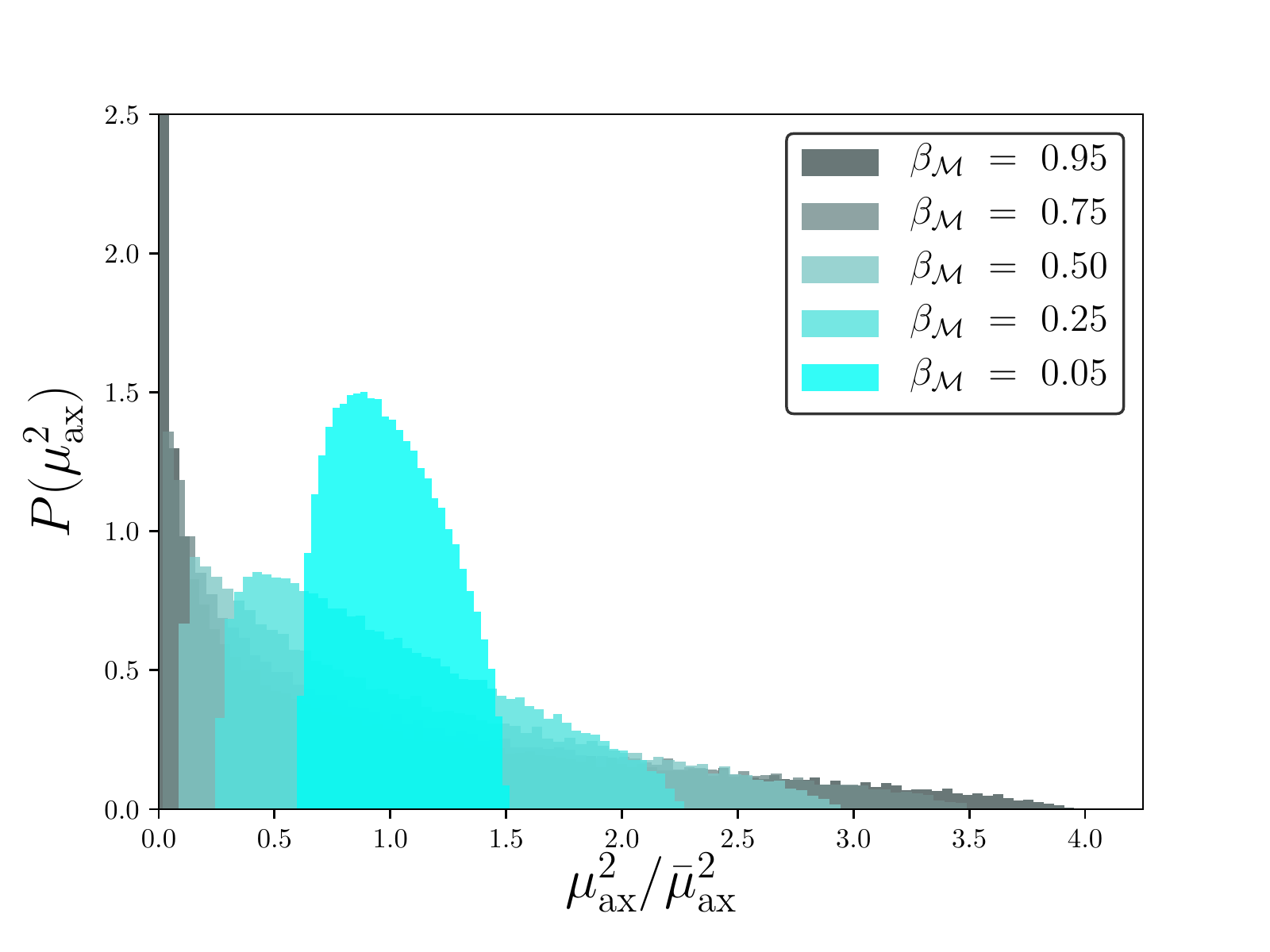}&
    \includegraphics[width=0.49\linewidth]{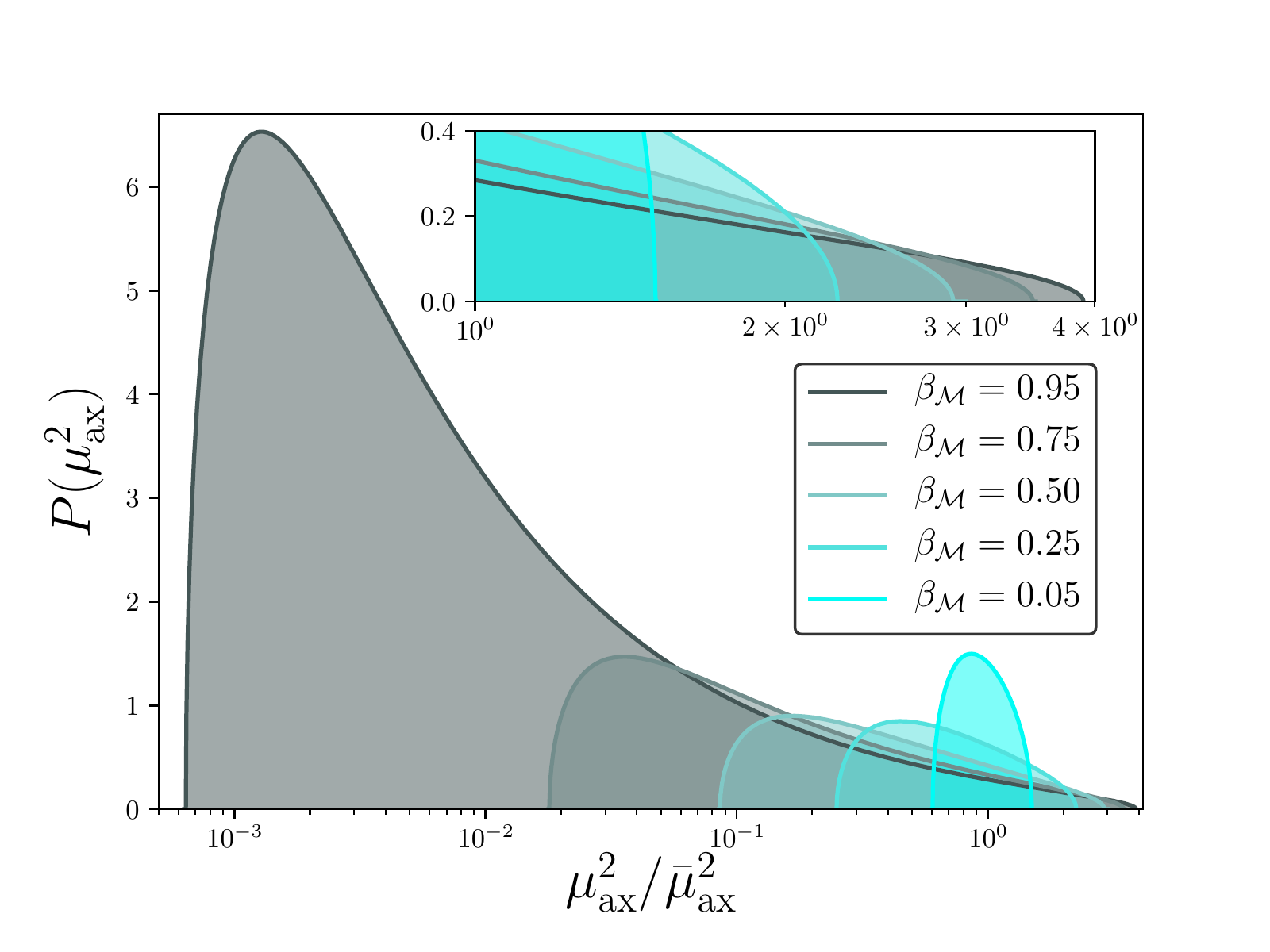}\\
\end{tabular}
\caption{Mar\v{c}henko-Pastur model normalised eigenvalue spectra and probability density functions for axion masses, $\mu_{\rm ax}^2$ with linear and logarithmic scales respectively. Each panel represents five selected  values of the spectrum shaping parameter $\beta_{\mathcal{M}}$ approximately covering its defining interval $\beta_{\mathcal{M}} \in (0,1]$. \emph{Left panel:} The mass distribution converges to the Mar\v{c}henko-Pastur limiting law as $N_{\rm ax}\rightarrow \infty$. Asymptotically the largest eigenvalue fluctuations outside its defined compact interval are determined by the Tracy-Widom law \cite{johnstone2001}. \emph{Right panel:} Probability density functions for each of the associated distributions in the \emph{left panel} displayed on a logarithmic mass scale. \emph{Inset:} As $\beta_{\mathcal{M}}$ increases the positive logarithmic displacement of the upper bound (Eq.~(\ref{eq:mp_upper_bound})) is limited compared to the negative displacement of the lower bound (Eq.~(\ref{eq:mp_lower_bound})) away from the mean scale of the distribution, $\bar{\mu}^2_{\rm ax}$. }
\label{fig:MP_spectra}
\end{figure*}
\begin{figure} 
\centering
\includegraphics[width=0.5\textwidth]{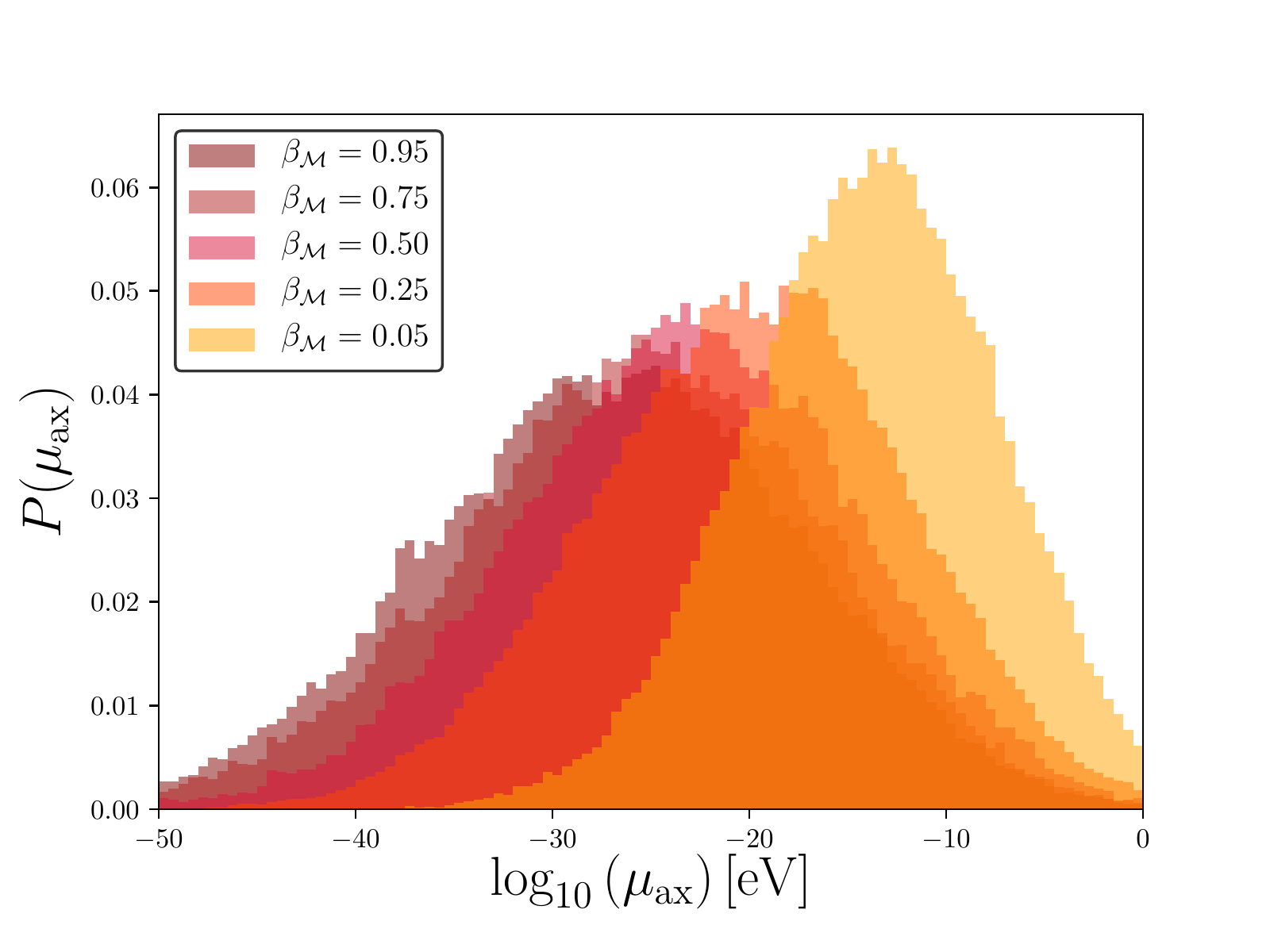}	
\caption{ M-theory model eigenvalue spectra for axion masses $\mu_{\rm ax}$ for different values of the spectrum shaping parameter $\beta_{\mathcal{M}}$. The mass spectra converge to an approximate log-normal distribution in the mass eigenstate basis. Each spectrum is constructed using a fixed value of the average three-cycle volume, $\langle V_{X}\rangle = 25$ required for GUT scale unification.}
\label{fig:m-theory_mass}
\end{figure}

The generic multi axion Lagrangian is:
\begin{equation}
\mathcal{L} = -\sum_{i,j = 1}^{N_{\rm ax}}\mathcal{K}_{ij} \partial_{\mu} \theta_i \partial^{\mu} \theta_j - \sum_{\alpha = 1}^{n_{\rm inst}}\sum_{j = 1}^{N_{\rm ax}}\Lambda_\alpha U_\alpha(\mathcal{Q}_{j,\alpha}\theta_j + \delta_\alpha),	
\label{eq:multiaxion}
\end{equation}
where $\theta_i$ are the dimensionless axion fields, ${\mathcal{K}_{ij}}$ is the kinetic matrix with mass dimension two, $U$ is a general periodic instanton potential with charge matrix $\mathcal{Q}$ and phases, $\delta$. Expanding the potential to the mass term only and diagonalising Eq.~(\ref{eq:multiaxion}) can be reduced to the simple form: 
\begin{equation}
\mathcal{L} = -\frac{1}{2}\partial_{\mu}\phi_i\partial^{\mu}\phi_i - \frac{1}{2}{\rm diag}(\mu^2_{\rm ax})\phi_i \phi_i\,.
\label{eq:finall}
\end{equation}

The spectrum of the model is given by the mass eigenvalues, $\{\mu_i\}$, which can be determined after expanding the instanton potential to quadratic order and obtaining a mass matrix, $\mathcal{M}_{ij}$. Diagonalising these matrices following the methodology in Appendix~\ref{sec:multi} gives the mass eigenstates of a spectrum of fields. The canonically normalised dimensionful mass eigenstate fields, $\phi_i$, are defined in Eq.~\eqref{eqn:mass_eigenstates} from the \emph{eigenvalues} of the kinetic matrix. Adopting random matrix models for $\mathcal{K}_{ij}$ and $\mathcal{M}_{ij}$ it is possible to determine the distribution of $\{\mu_i\}$ for various models. This process is reviewed in Appendix~\ref{appendix:axiverse} and covered extensively in Ref.~\cite{2017PhRvD..96h3510S}. In the following we will consider just two simple models for the mass eigenvalues.

The first follows the celebrated Mar\v{c}henko-Pastur law for the eigenvalues of white Wishart matrices~\cite{mehta}. The spectrum is thought to describe Type-IIB string theory models of inflation with large numbers of axions~\cite{Easther:2005zr}. The limiting distribution as the matrix size goes to infinity is given by
\begin{equation}
    P\left(\mu^2_{\rm ax}\right) = 
\begin{cases}
    \frac{1}{2 \pi \mu^2_{\rm ax} \beta_{\mathcal{M}} \bar{\mu}^2_{\rm ax}}\sqrt{\left(\gamma_+ - \mu^2_{\rm ax}\right)\left(\mu^2_{\rm ax} - \gamma_-\right)}\\
    0 
\end{cases} \,,
\label{eq:MPprobden}
\end{equation}
on the compact interval 
\begin{equation}
\gamma_- \leq \mu^2_{\rm ax} \leq \gamma_+\,,	
\end{equation}
where $\gamma_+$ and $\gamma_-$ are defined as,
\begin{align}
\gamma_+ = \bar{\mu}^2_{\rm ax} \left(1+\sqrt{\beta_{\mathcal{M}}}\right)^2\,,\label{eq:mp_upper_bound}\\
\gamma_- = \bar{\mu}^2_{\rm ax} \left(1-\sqrt{\beta_{\mathcal{M}}}\right)^2\,.	
\label{eq:mp_lower_bound}
\end{align}
The expectation value of $\mu^2_{\rm ax}$ is $\bar{\mu}^2_{\rm ax}$ and the shape parameter $0<\beta_{\mathcal{M}}\leq 1$ determines the spread of the distribution, with large $\beta_{\mathcal{M}}$ giving larger spreads as shown in the \emph{right panel} of Fig.~\ref{fig:MP_spectra}. The distribution for random realisations with finite $N_{\rm ax}$ is shown in the \emph{left panel} of Fig.~\ref{fig:MP_spectra}, and is well fit by the limiting law.

Our second model for the mass eigenvalues follows from the ``M-theory axiverse''~\cite{Acharya:2010zx}. In this case the mass eigenvalues follow an approximately log-normal distribution~\cite{2017PhRvD..96h3510S}, as shown in Fig.~\ref{fig:m-theory_mass}.\footnote{The naive expectation of log-flat eigenvalues turns out not to be realised for large numbers of fields after applying rotations to the canonical basis.} In this example, the spread is controlled by the shaping parameter $\beta_{\mathcal{M}}=N_{\rm ax}/N_{\rm inst}$ which takes values $0<\beta_{\mathcal{M}}\leq1$. Increasing $\beta_{\mathcal{M}}$ leads to a larger mean and smaller variance. 

There are five parameters in the model of Ref.~\cite{2017PhRvD..96h3510S} in total, but here we use a two-parameter approximate fit:
\be
P(\mu^2_{\rm ax}) = \frac{1}{\sqrt{2\pi \sigma^2}}\exp \left[ \frac{-\log_{10}(\mu_{\rm ax}/\bar{\mu}_{\rm ax})^2}{2\sigma^2}\right] \, .
\ee 
The mean of the log-normal distribution can be related to the expectation value of the 3-cycle volumes in the $G_2$ manifold, $\langle V_X\rangle$ (see Eq.~(\ref{eqn:cycle_volume})), and as in the above example, the variance, $\sigma^2$, can be controlled by the number of instantons in the potential sum. The variance of the log-normal distribution is dimensionless, and so should take on some $\mathcal{O}(1)$ value. In Ref.~\cite{2017PhRvD..96h3510S} we typically found $\sigma\gg 1$.

Axion self-interactions can also play an important role in BH superradiance. In principle, by expanding the instanton potential to higher orders our RMT approach could lead to a distribution for the quartic interaction tensor:
\be
\mathcal{L}_{\rm int} = \lambda_{ijkl}\phi_i\phi_j\phi_k\phi_l \, .
\ee
We are unaware of any study of the distribution of $\lambda_{ijkl}$ in RMT, and thus the treatment of interactions is beyond the scope of the present work. For sparse charge matrices the flavour changing, non-diagonal, entries in $\lambda_{ijkl}$ will be rare.

If the attractive self-interactions are too strong then the superradiant cloud collapses via a bosenova before it can extract large amounts of spin from the BH. Superradiance can also be shut off by non-linear level mixing, or affected by axion emission due to annihilations~\cite{Arvanitaki:2010sy,2014PTEP.2014d3E02Y,2015PhRvD..91h4011A}. The interaction tensor can used to calculate these rates, e.g. for axion emission via the $\phi\phi\phi\rightarrow\phi$ process. 
\begin{figure*}
\centering
\begin{tabular}{cc}
\includegraphics[width=0.49\linewidth]{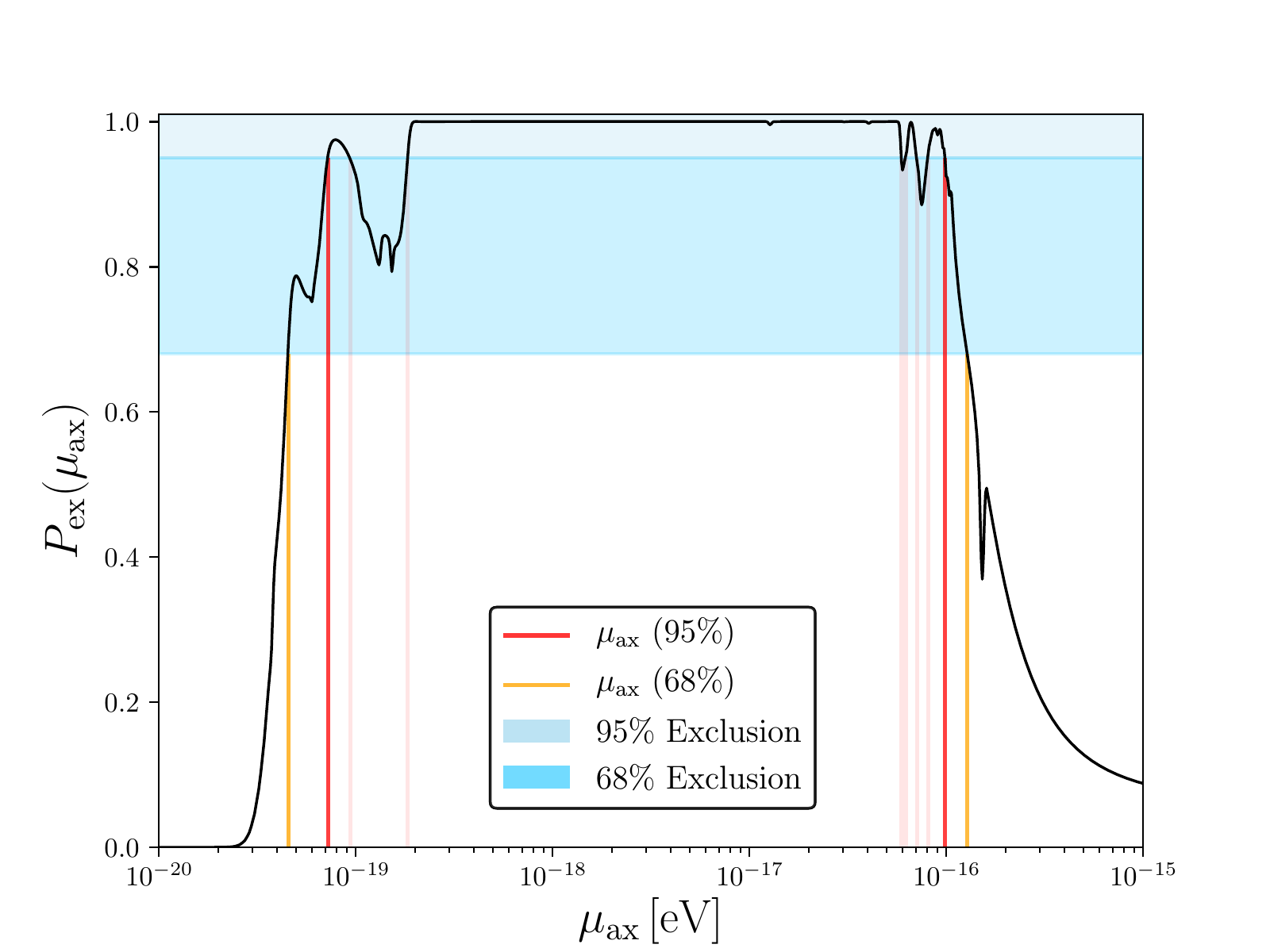}&
    \includegraphics[width=0.49\linewidth]{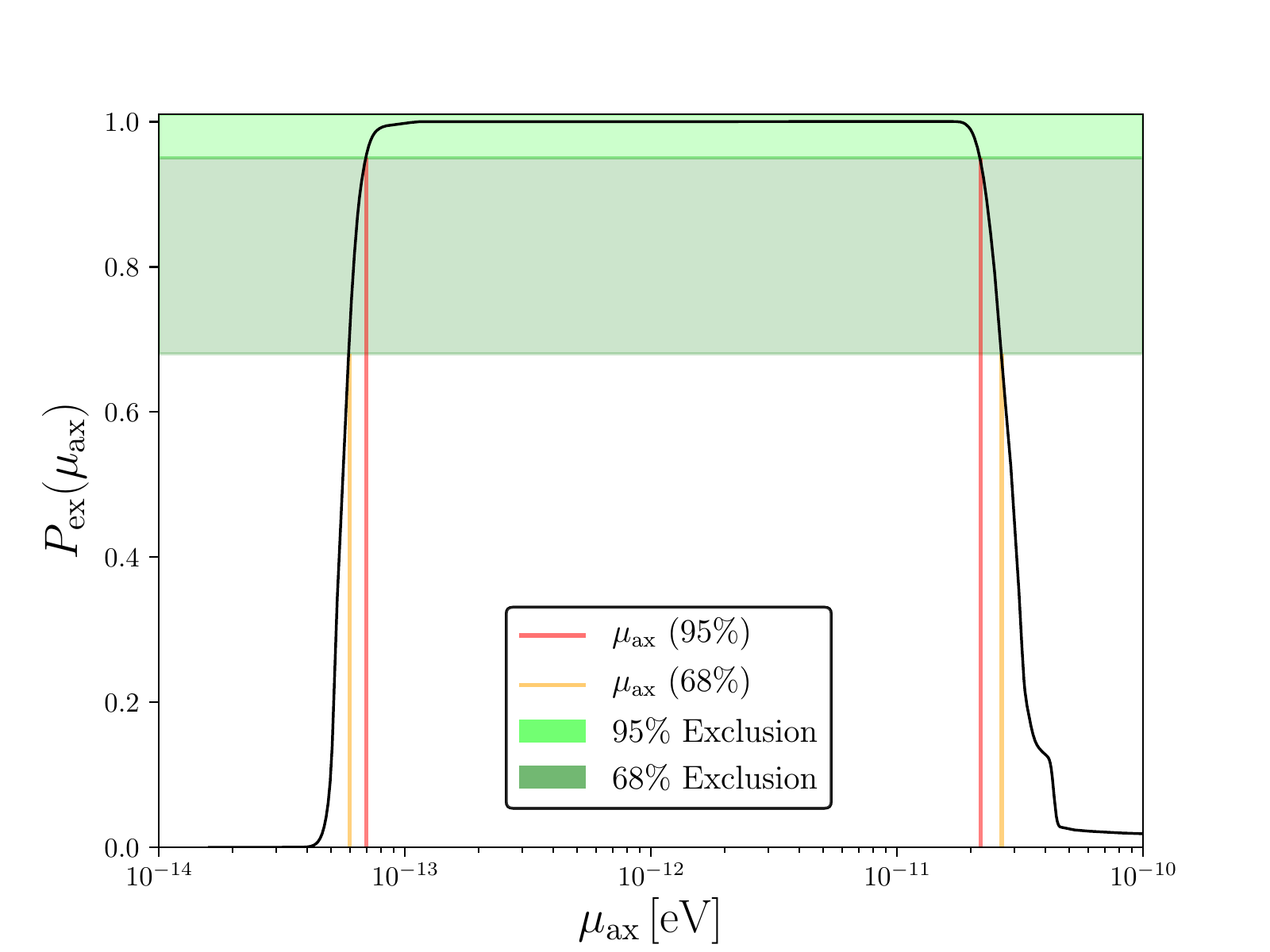}\\
\end{tabular}
\caption{Constraints on the masses of ultralight axions, $\mu_{\rm ax}$ for singular fields determined by the total probability of exclusion calculated using the methodology in Appendix~\ref{appendix:stats} via Eq.~(\ref{eq:probex}). Exclusion bounds are presented in the $68\%$ and $95\%$ confidence intervals as a function of $\mu_{\rm ax}$ with orange/red lines representing the upper and lower limits of the 68\%/95\% interval. \emph{Left panel:} Limits determined using the SMBHs given in Table~\ref{tab:cosmopar}. \emph{Right panel:} Limits Determined using stellar mass BHs given in Table~\ref{tab:cosmopar}.} 
\label{fig:stellar_single}
\end{figure*}
The level-mixing will be enhanced if the $\lambda_{ijkl}$ are non-diagonal and allow scattering of axions of different flavours. Decays from one flavour into another will have a similar effect of additional cooling of the cloud as the axion photon coupling considered in Ref.~\cite{Arvanitaki:2010sy}. The Bosenova critical size, $N_{\rm Bosenova}$, could also become smaller in such a case due to the increased phase space for the scattering. How these and other non-linear effects compete with the basic increase of the BH superradiance rate and increased probability of mass outliers at large $N_{\rm ax}$ is unclear.

Using the single instanton, dilute gas potential, $V(\phi)=\mu^2f_a^2[1- \cos (\phi/f_a)]$ for a single field, it can be shown that the ratio of emission via the quartic interaction compared to graviton emission due to annihilations is given by~\cite{Arvanitaki:2010sy}:
\be
\frac{P_\lambda}{P_{\rm grav}}\approx 10^{-2}\alpha^4 \frac{M_a}{M_{\rm BH}}\left(\frac{M_{pl}}{f_a}\right)^4 \, .
\ee

The overall strength of the interactions, and their importance relative to gravity, is controlled by the axion decay constants, $f_a$. The $f_a$ distributions for multiple fields derived from RMT can be computed (see e.g. Ref.~\cite{2017PhRvD..96h3510S}). Distributions with a high probability of small decay constants will have non-linearities dominated by self-interactions, while a for high probability of large decay constants the pure-gravity results can be used. Since we consider BH superradiance dominated by gravity, our results should be understood to apply strictly to distributions dominated by large $f_a$. Taking the single-field results of Ref.~\cite{2015PhRvD..91h4011A} as a guide, this should be for $f_a\gtrsim 10^{14-16}\text{ GeV}$. In the context of string models, our results should apply well to small volume compactifications~\cite{2003PhRvD..68d6005K}, whereas self-interactions will play an important role in the Large Volume Scenario~\cite{2006JHEP...05..078C}.

\section{Results}
\label{sec:results}

\subsection{Single Field}
\label{sec:single_field_results}

In this short section, we begin the presentation of our results by computing single field limits to check our methodology is consistent with other results in the literature. Our statistical methods are described in Appendix~\ref{appendix:stats}, and we calculate the exclusion probability, $P_{\rm ex}(\mu_{\rm ax})$.

Treating the stellar BHs and SMBHs as a single data set, our results for a single axion field with mass $\mu_{\rm ax}$ are shown in Fig.~\ref{fig:regge_spectrum}, superimposed on the Regge plane with the data. In this combined data set the exclusion probability remains finite over a range of intermediate axion masses due to the large mass errors on the lightest SMBHs. The absence of IMBHs means that the regions with $P_{\rm ex}(\mu_{\rm ax})>0.68$ (``$1\sigma$ exclusion'') do not overlap between the two datasets and they can be considered separately. 

The exclusion probability for the stellar BH data set is shown in the \emph{right panel} of Fig.~\ref{fig:stellar_single}. The high quality of these measurements, and the large number of them, leads to a smooth exclusion probability. At the 95\% C.L. the stellar BHs exclude:
\be
7\times 10^{-14}\text{ eV}<\mu_{\rm ax}<2\times 10^{-11}\text{ eV}\, .
\ee
\begin{figure*}
\centering
\begin{tabular}{cc}
\includegraphics[width=0.49\linewidth]{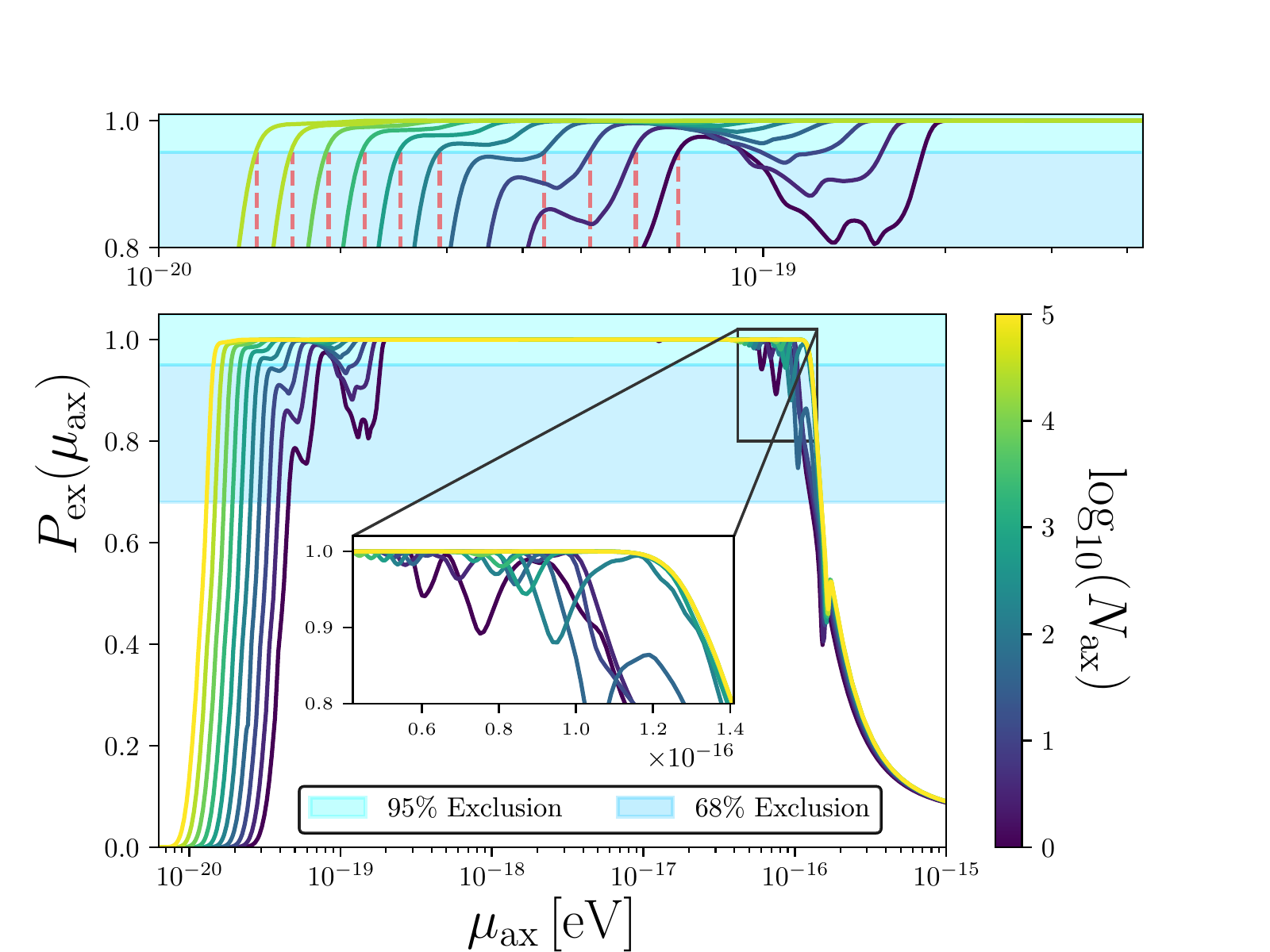}&
    \includegraphics[width=0.49\linewidth]{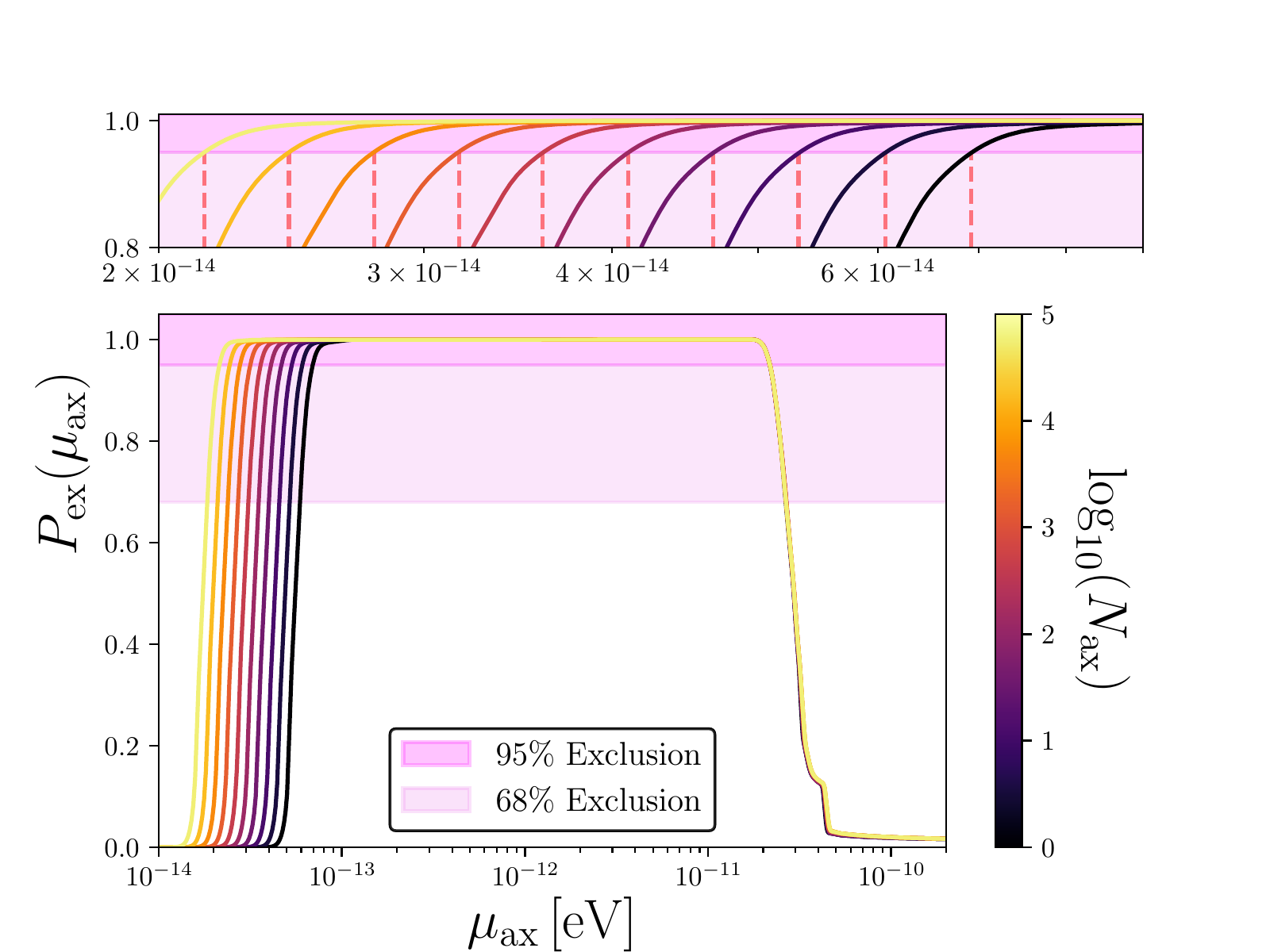}\\
\end{tabular}
\caption{Constraints on masses of ultralight axions, $\mu_{\rm ax}$, via the total exclusion probability in the $68\%$ and $95\%$ confidence limits for large numbers, $N_{\rm ax}$, of degenerate fields. \emph{Upper panels:} Dashed red lines represent the shift of the lower bound in the 95\% confidence limit, which decreases as $N_{\rm ax}$ increases. \emph{Left panel:} Exclusion probability for the SMBH data set. \emph{Inset:} Oscillatory behaviour of the exclusion probability due to higher values of the orbital/azimuthal quantum numbers passing over low mass SMBHs. \emph{Right panel:} Exclusion probability for the stellar BH data set.}
\label{fig:constrains_N}
\end{figure*}
The exclusion probability for the SMBH data set is shown in the \emph{left panel} of Fig.~\ref{fig:stellar_single}. The data is of general poorer quality than the stellar data, with certain systems containing significantly large mass errors. It is also much sparser, with fewer SMBHs in the set. The sparseness of the data leads to oscillatory features in the exclusion probability, driven by the shape of the BH superradiance contours for each of the modes, with the exclusions being driven by individual BHs. This causes the probability of exclusion to oscillate between the 95\% C.L when transitioning between certain BHs (faded red lines in the \emph{left panel} of Fig.~\ref{fig:stellar_single}). The largest candidate, \emph{Fairall 9} drives the non-monotonic nature of the function at low axion masses. The large mass errors lead to non-zero exclusion probability extending to large axion masses. Taking the outer edge of the 95\% C.L. region, the SMBHs exclude:
\be
7\times 10^{-20}\text{ eV}<\mu_{\rm ax}<1\times 10^{-16}\text{ eV}\, .
\ee

Our exclusions for the stellar BH and SMBH datasets are consistent with the results of Refs.~\cite{Cardoso:2018tly,2015PhRvD..91h4011A}, after accounting for the differences in the data sets and methodology used. In particular comparing to Ref.~\cite{2015PhRvD..91h4011A} our choice to include BBH coalescence events with large masses when partnered with their large uncertainties push the constraints to incorporate lower masses, increasing the lower bound on the axion mass exclusion.  

\subsection{Degenerate Masses}
\label{sec:degenerate}
\begin{figure}
\centering
\includegraphics[width=0.5\textwidth]{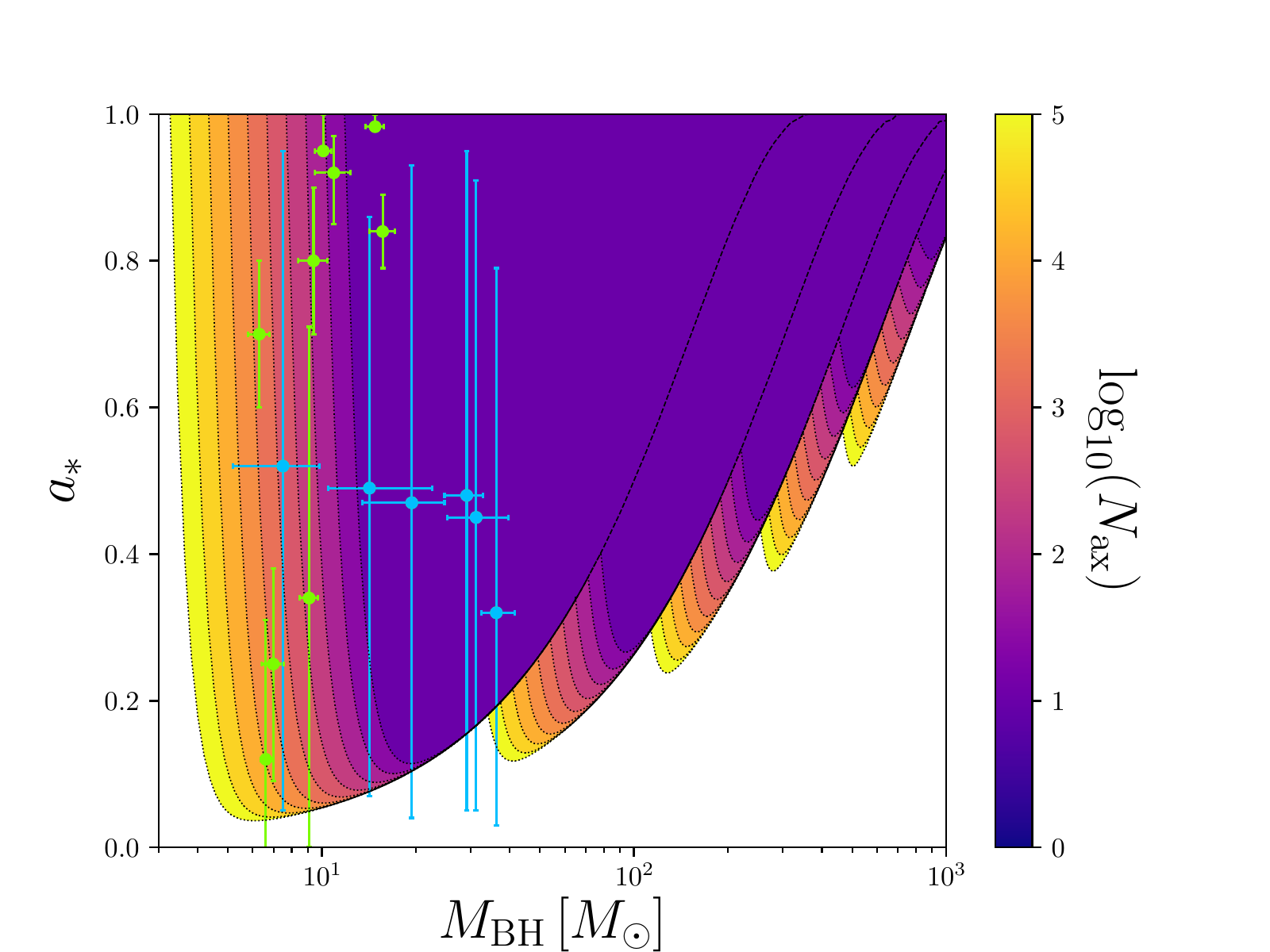}
\caption{Isocontour exclusion regions for degenerate mass axion populations with $N_{\rm ax}=\mathcal{O}(1) \rightarrow \mathcal{O}(10^5)$ in the stellar BH parameter-space. The limits for the instability threshold are obtained by fixing the superradiant instability time scales for each value of the orbital/azimuthal quantum numbers, l = m = 1 to 5 equal to $\tau_{\rm Salpeter}$ (Eq.~(\ref{eq:salpeter})) for an axion mass $\mu_{\rm ax} = 10^{-12.75}\ {\rm eV}$. Large values of $N_{\rm ax}$ effectively correspond to greater superradiance instability timescales considering a single field. Green data points are mass/spin estimates of X-ray binary stellar BH candidates. Blue data points are primary and secondary sources from BBH coalescence detections at LIGO.}
\label{fig:degen}
\end{figure}
We now begin to consider cases with multiple axion masses. The degenerate case is trivial to treat for any number of $N_{\rm ax}$ axions with identical masses, $\mu_{\rm ax}$. Since the rate is additive in $N_{\rm ax}$ we have:
\be
\Gamma_{\rm tot}=N_{\rm ax}\Gamma \, .
\ee
Therefore, setting $\tau_{\rm BH}\Gamma_{\rm tot}=1$ is equivalent to the single field case with the timescale rescaled as $\tau_N = N_{\rm ax}\tau_{\rm BH}$. Thus, for the degenerate case the exclusion probabilities are trivial to compute for any $N_{\rm ax}$, and they will simply grow wider for increasing $N_{\rm ax}$ corresponding to larger rates such as those shown in Fig.~\ref{fig:solarregge}. 
\begin{figure*}
\centering
\begin{tabular}{cc}
\includegraphics[width=0.49\linewidth]{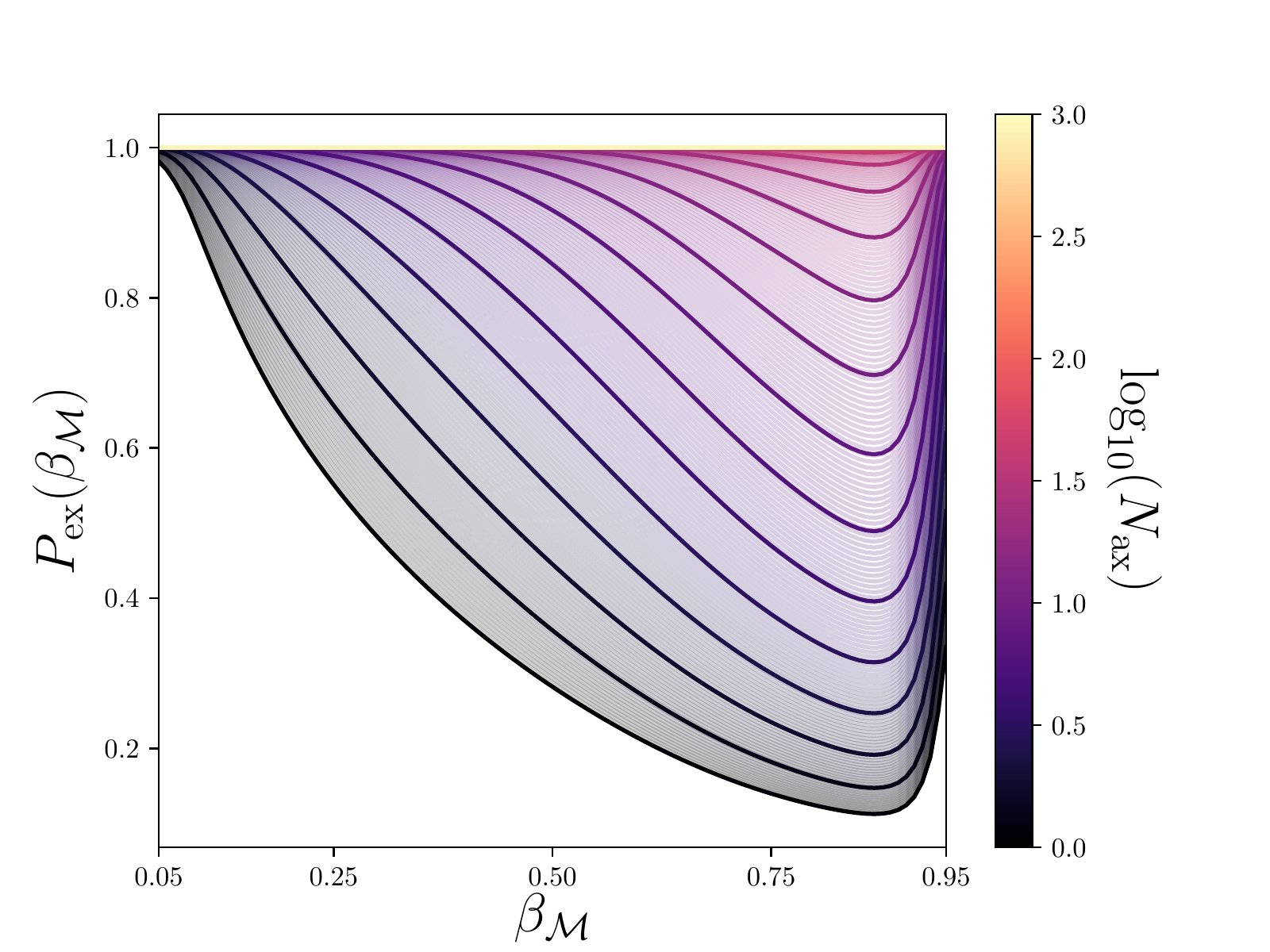}&
    \includegraphics[width=0.49\linewidth]{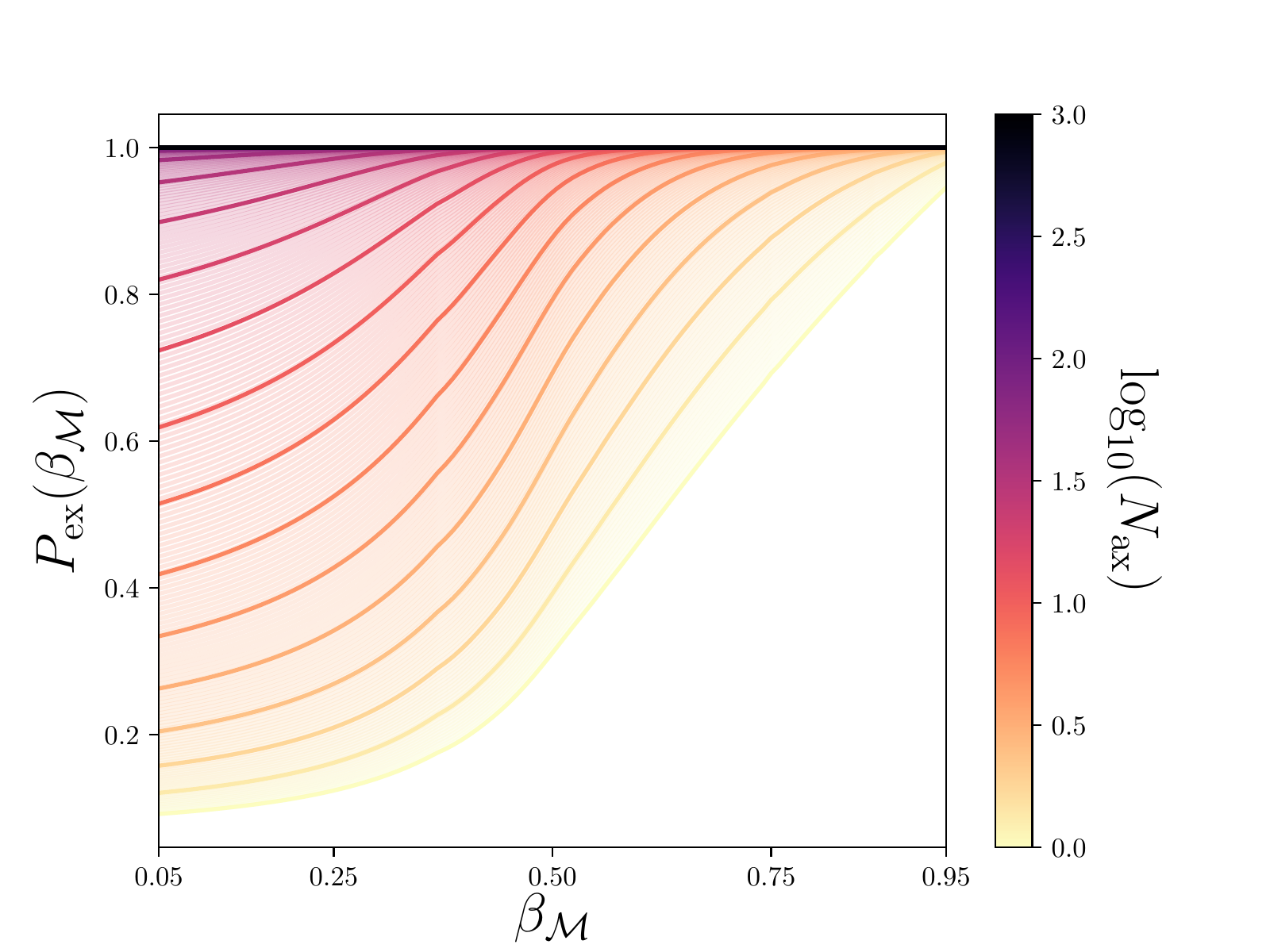}\\
\end{tabular}
\caption{Probability of exclusion as a function of the dimensionless shaping parameter, $\beta_{\mathcal{M}}$ defining the mass spectra in the Mar\v{c}henko-Pastur model for $N_{\rm ax} = 1 \rightarrow 1000$. In general large populations or sub-populations are heavily constrained as the field masses saturate the limiting spectrum of the model. \emph{Left panel:} The mean of the distribution is fixed to $\bar{\mu}_{\rm ax} = 10^{-13}\ {\rm eV}$ on the edge of the constrained region from stellar BHs. \emph{Right panel:} The mean of the distribution is fixed to $\bar{\mu}_{\rm ax} = 10^{-15}\ {\rm eV}$ inside the ``well'' of the constrained region from both stellar BHs and SMBHs.}
\label{fig:mp_exclusions}
\end{figure*}
In Fig.~\ref{fig:degen} we show the effect on the Regge plane with a degenerate population of axions with masses $\mu_{\rm ax} = 10^{-12.75}\ {\rm eV}$. It is clear that an increase in $N_{\rm ax}$ can lead to an exclusion on $\mu_{\rm ax}$ where there was not one in the single field case (\emph{purple} limits). As the instability thresh-holds sweep through the Regge plane as a function of the axion mass, the wider instability limits possess the ability to ``catch'' lighter BHs in their exclusion bounds.   

We present the exclusion probabilities $P_{\rm ex}(\mu_{\rm ax})$ for various values of $\log_{10}N_{\rm ax}$ for each regime in the \emph{left} and \emph{right panels} of Fig.~\ref{fig:constrains_N}. The contours in Fig.~\ref{fig:degen} always increase in the direction of smaller $M_{\rm BH}$, and so the constraints in Fig.~\ref{fig:constrains_N} only broaden relative to the single field case for smaller axion masses. For SMBHs, where the higher harmonics play a role in the exclusion, the oscillations in the exclusion probability at high mass are also mildly affected. This is shown in the \emph{inset} of the \emph{left panel} of Fig.~\ref{fig:constrains_N}. Extremely large values of $N_{\rm ax}$ quench the oscillations from the instability bounds of the higher order modes, saturating the upper bounds on the constraints.

The 95\% excluded regions for $\mu_{\rm ax}$ for the degenerate case change by less than an order of magnitude compared to the single field case for $N_{\rm ax}\lesssim 10^5$. This shows that the increase in the superradiance rate for multiple fields (i.e. the rate sum in Eq.~(\ref{eqn:rate_sum})) can be virtually neglected when computing the exclusion probability, even in the most extreme case of a very large number of degenerate superradiant fields. 


\subsection{Mass Distributions}

There are two effects on BH superradiance constraints for mass distributions. The first is the effect of rate addition, the second is the effect of an overlap between the mass distribution and the exclusion probability. The results of the previous section show that even for the extreme case of degenerate masses, this effect is virtually negligible in the the exclusion probability for $\mu_{\rm ax}$. Rate addition will be even more negligible for mass distributions with finite width, where off-resonant superradiance rates are exponentially suppressed. This leaves probability overlap as the dominant effect for mass distributions of finite width.
\begin{figure} 
\centering
\includegraphics[width=0.5\textwidth]{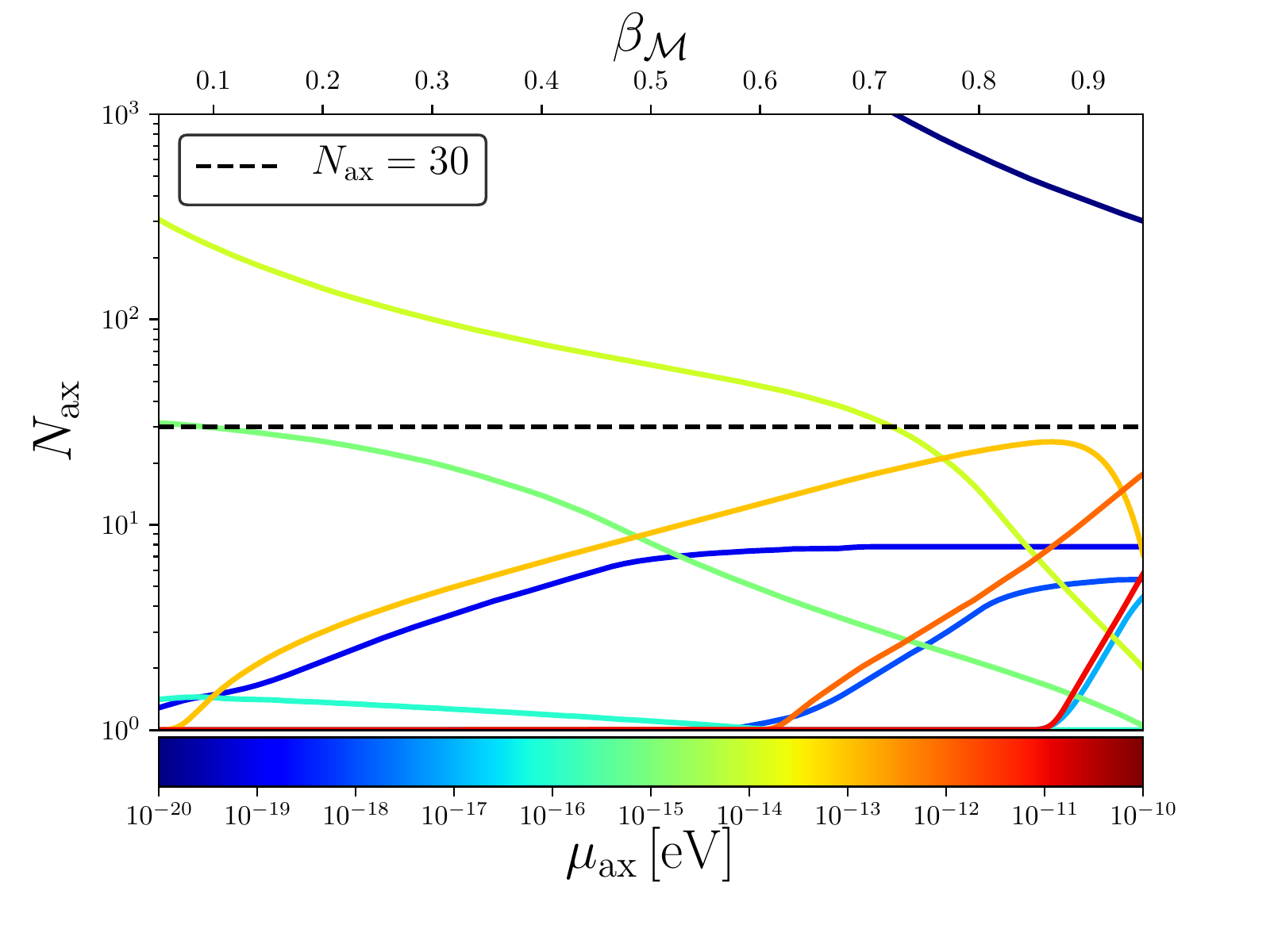}	
\caption{Contours representing the 95\% exclusion for Mar\v{c}henko-Pastur axion mass distributions as a function of the distribution shape, $\beta_{\mathcal{M}}$, and number of fields, $N_{\rm ax}$, for various distribution mean scales, $\bar{\mu}_{\rm ax}$. Regions above the contours are excluded. Large numbers of fields are constrained for a significant region of the probable axion mass space, with $N_{\rm ax} \geq 30$ constrained for a wide range of $\beta_{\mathcal{M}}$ over the considered scales.}
\label{fig:MP_95}
\end{figure}
With the effect of rate addition neglected, the exclusion probability for a mass distribution is trivial to construct from the exclusion probability for a single mass from the overlap integral. We use the probability that a model is allowed, since this trivially accounts for the combinatorics, and the excluded probability is in turn found trivially from this. Let $P_{\rm al}(\mu_{\rm ax}|N_{\rm ax}=1)=1-P_{\rm ex}(\mu_{\rm ax}|N_{\rm ax}=1)$ be the probability that a given axion mass is allowed, assuming just one axion field. We then have that in a given model $\mathcal{M}$ with one axion, the probability that some parameters $\theta$ are allowed is
\be
P_{\rm al}(\theta,N_{\rm ax}=1|\mathcal{M}) = \int {\rm d}\mu_{\rm ax} p(\mu_{\rm ax} |\theta,\mathcal{M}) P_{\rm al}(\mu_{\rm ax}|N_{\rm ax}=1) \, ,
\label{eqn:p_allowed_dist_n1}
\ee 
where ${\rm d}\mu_{\rm ax} p(\mu_{\rm ax} |\theta,\mathcal{M})$ is the probability distribution for $\mu_{\rm ax}$ in the model. The single axion allowed regions were evaluated numerically in Section~\ref{sec:single_field_results} and the integral in Eq.~\eqref{eqn:p_allowed_dist_n1} can be evaluated numerically given $p(\mu_{\rm ax} |\theta,\mathcal{M})$. The above trivially generalises to the case of $N_{\rm ax}$ fields:
\be
P_{\rm al}(\theta,N_{\rm ax}|\mathcal{M}) = \left[\int {\rm d}\mu_{\rm ax}\,\, p(\mu_{\rm ax} |\theta,\mathcal{M}) P_{\rm al}(\mu_{\rm ax}|N_{\rm ax}=1) \right]^{N_{\rm ax}}\, .
\label{eqn:p_allowed_dist_N}
\ee
The exclusion probability for $N_{\rm ax}$ fields is then given by $P_{\rm ex}(\theta,N_{\rm ax}|\mathcal{M}) = 1-P_{\rm al}(\theta,N_{\rm ax}|\mathcal{M})$.

\subsubsection{The Mar\v{c}henko-Pastur Distribution}

The Mar\v{c}henko-Pastur (MP) distribution depends on two parameters: a mean mass, $\bar{\mu}_{\rm ax}$, and a shape parameter, $\beta_{\mathcal{M}}$. In order to probe the potential of a spectrum of fields scanning the Regge plane analogous to our single field constraints we highlight several interesting configurations. Consider the case $\bar{\mu}_{\rm ax}=10^{-13}\text{ eV}$, shown in Fig.~\ref{fig:mp_exclusions}, \emph{left panel}. A single axion at this mass is excluded by the stellar BH data. However, for large spreads, i.e. $\beta_\mathcal{M}\rightarrow 1$, the mode of the distribution moves to smaller values of the mass (shown in the \emph{right panel} of Fig.~\ref{fig:MP_spectra}), which are not constrained. Eventually at still larger $\beta_\mathcal{M}$ the mode moves down to masses excluded by the SMBH data. In Fig.~\ref{fig:MP_95} we present the axion mass window open to superradiance as distribution mean scales $\bar{\mu}_{\rm ax}$. Increasing $N_{\rm ax}$ makes the exclusion probability grow, and for all $\beta_\mathcal{M}$ there is a maximum $N_{\rm ax}\approx 20$ above which the model is excluded at better than the 95\% C.L for all $\beta_\mathcal{M}$ (see Fig.~\ref{fig:MP_95}). The maximum $N_{\rm ax}$ allowed grows with $\beta_\mathcal{M}$.

Now consider the case $\bar{\mu}_{\rm ax}=10^{-15}\text{ eV}$, shown in Fig.~\ref{fig:mp_exclusions}, right panel. In this case, the mean mass is in between the stellar and SMBH exclusions, and is allowed by the data. Thus, increasing $\beta_\mathcal{M}$ now increases the exclusion probability. The non-zero exclusion probability at $\mu_{\rm ax}=10^{-15}\text{ eV}$ coming from the SMBH data lowest mass points with large error causes the exclusion probability to grow as $N_{\rm ax}$ increases even for small $\beta_\mathcal{M}$. Once again, there is a maximum $N_{\rm ax}\approx 50$ above which the model is excluded at better than the 95\% C.L for all $\beta_\mathcal{M}$ (see Fig.~\ref{fig:MP_95}). The maximum $N_{\rm ax}$ allowed decreases with $\beta_\mathcal{M}$.

\begin{figure} 
\centering
\includegraphics[width=0.5\textwidth]{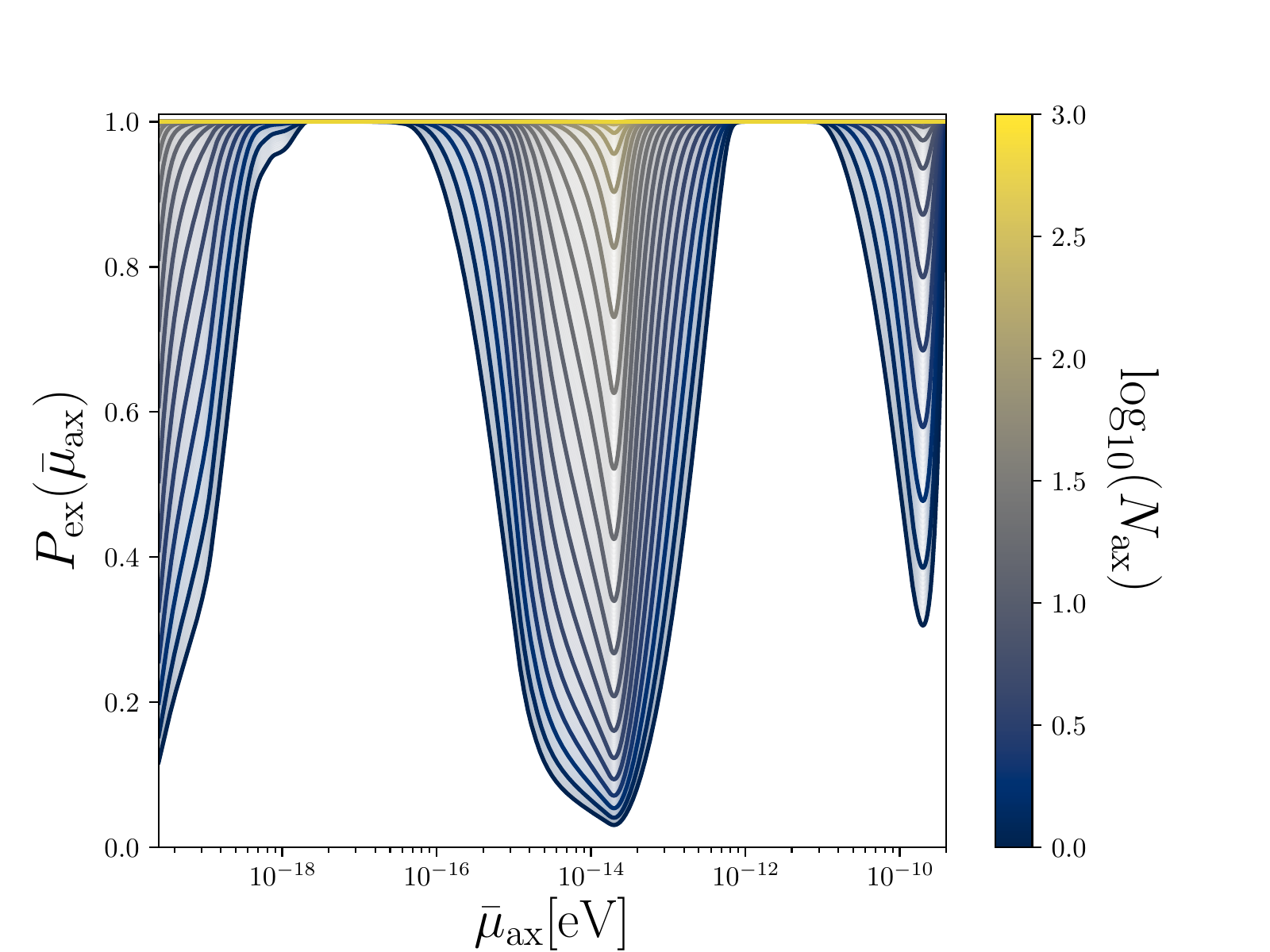}	
\caption{Probability of exclusion as a function of the Mar\v{c}henko-Pastur distribution mean scale, $ \bar{\mu}_{\rm ax}$ for $N_{\rm ax} = 1 \rightarrow 1000$. Each probability function is determined using a fixed shape parameter, $\beta_{\mathcal{M}} = 0.5$. In general large populations or sub-populations are heavily constrained as the field masses saturate the limiting spectrum of the model. The singular field bounds trace the limits in Fig.~\ref{fig:regge_spectrum} defining the white region within statistical fluctuations of the mean.}
\label{fig:MP_total}
\end{figure}

Motivated by the peak in the Calabi-Yau distribution along the self-mirror manifold line, Fig.~\ref{fig:MP_total} shows constraints on $\bar{\mu}_{\rm ax}$ at fixed $\beta_\mathcal{M}=0.5$. The excluded region has the same approximate shape as the single field exclusions for small $N_{\rm ax}$. As $N_{\rm ax} \rightarrow 1$ the exclusion limits trace out the constraints for the single field case up to statistical fluctuations about the mean scale. The softer edges to the untouched regions when compared with the single field exclusion bounds come from the non-equidistant logarithmic spread of the mass spectrum about the mean scale when $\beta_\mathcal{M}=0.5$. Reducing $\beta_\mathcal{M}$ relaxes the limits to fully match the single field case in the low $N_{\rm ax}$ limit. Increasing the number of fields, the model is excluded at better than the 95\% C.L. for the range of mean masses shown for all $N_{\rm ax}\gtrsim 100$.

\subsubsection{The M-theory Axiverse: the QCD axion, GUTs, and Fuzzy DM}
\label{sec:mtheory_axiverse}
The axion mass spectrum of the M-theory axiverse~\cite{Acharya:2010zx} was computed from RMT models in Ref.~\cite{2017PhRvD..96h3510S} and is well described by a log-normal distribution. The 95\% excluded region in $(\sigma,N_{\rm ax})$ for the log-normal distribution across a range of central values is shown in Fig.~\ref{fig:summary}. We now derive BH superradiance constraints on three scenarios of interest realised approximately from this simple model for the M-theory axiverse. 
\begin{figure*}
  \subfloat[GUT.]{\includegraphics[width=0.49\textwidth]{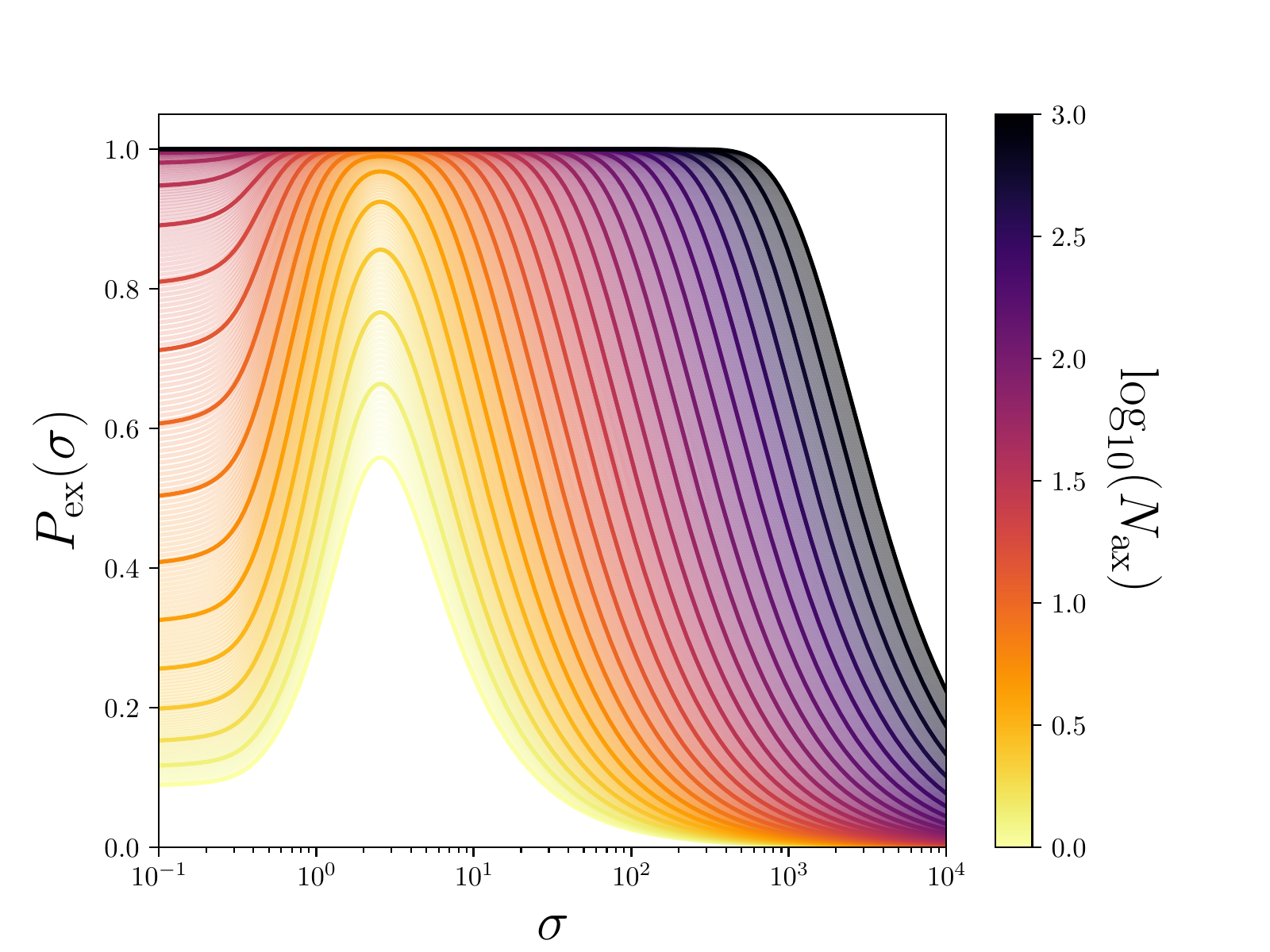}\label{fig:fig_a}}
  \hfill
  \subfloat[Fuzzy DM.]{\includegraphics[width=0.49\textwidth]{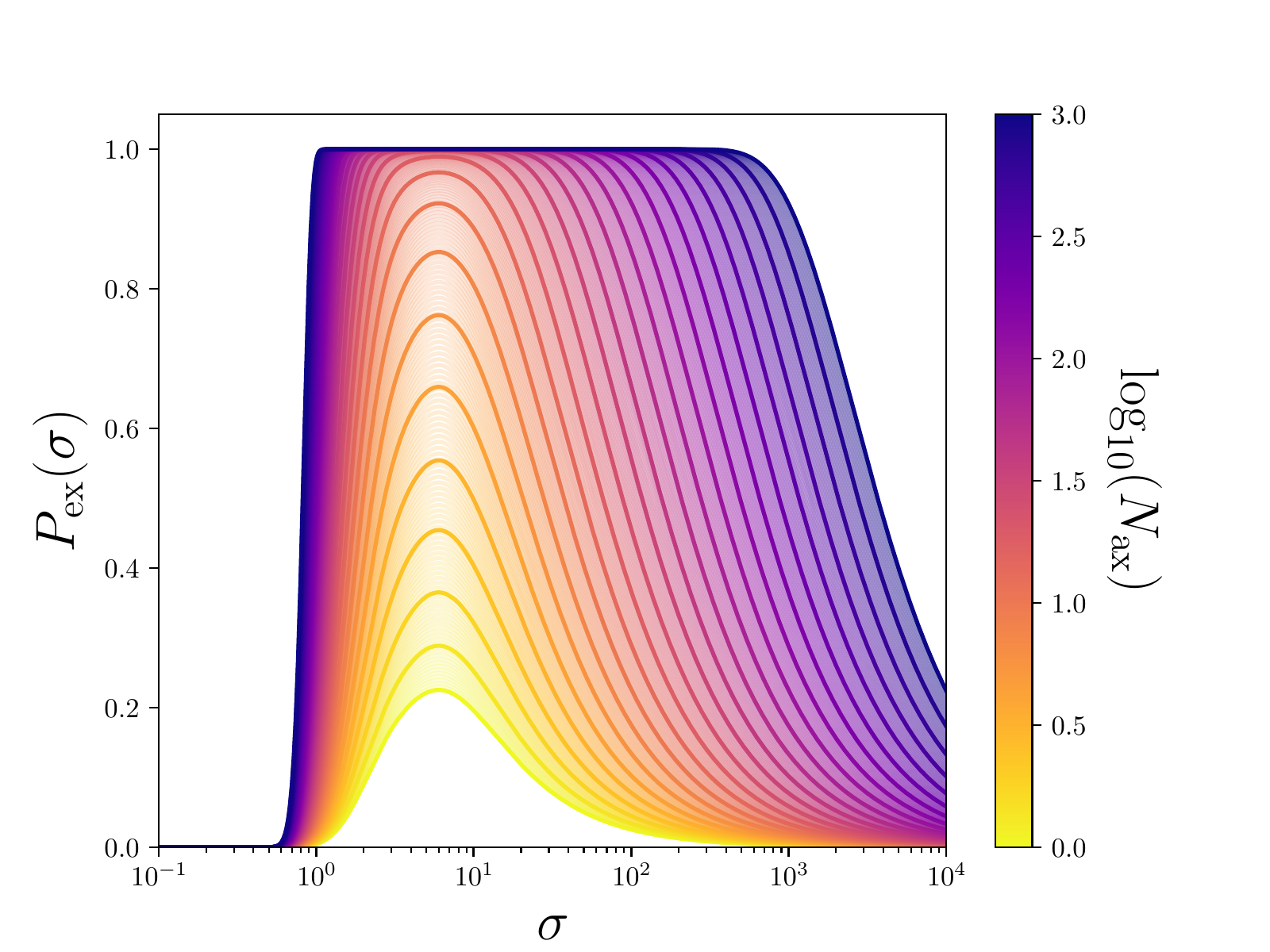}\label{fig:fig_b}}\\
  \subfloat[QCD.]{\includegraphics[width=0.49\textwidth]{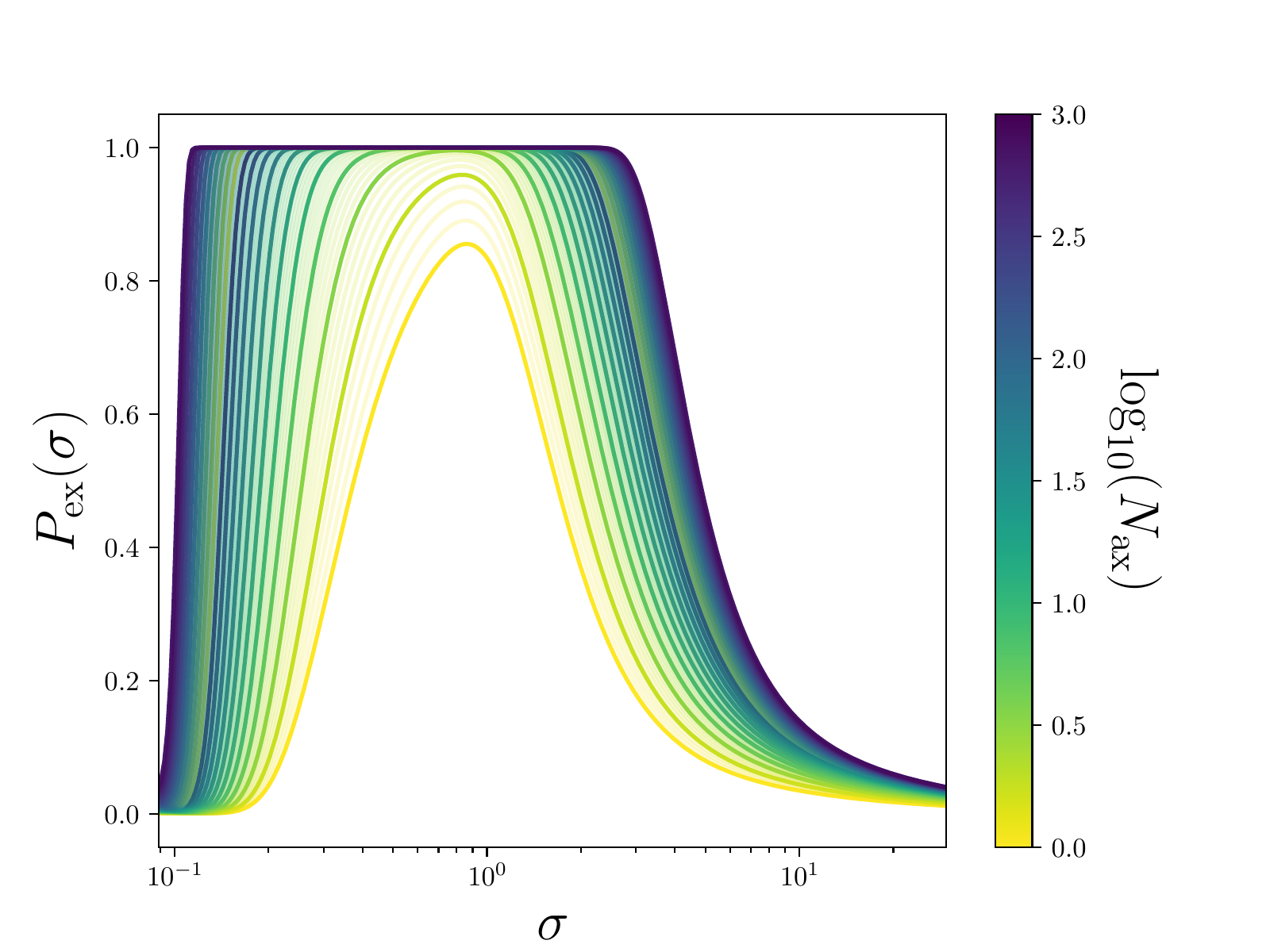}\label{fig:fig_c}} 
  \hfill
  \label{fig:mtheory_models_results}
  \begin{minipage}[b][7cm][c]{0.5\textwidth}         
\caption{Probability of exclusion as a function of the dimensionless spread, $\sigma$ determined by the model variance defining the mass spectra in the M-theory axiverse for $N_{\rm ax} = 1 \rightarrow 1000$. Each panel corresponds to three unique scenarios which determine the mean of the mass spectrum required to maximise the probability of drawing the desired masses detailed in Section~\ref{sec:mtheory_axiverse}. In the limit $\sigma \gg 1$ the total probability for $N_{\rm ax} = 1 \rightarrow \infty$ converges to zero as the spread crosses the bounds probable by BH spin measurements. The behaviour in the limit $\sigma \ll 1$ is determined by the accuracy of the available BH mass/spin measurements.}
\label{fig:mtheory_models_results}
\end{minipage}
\end{figure*}
The M-theory axiverse with GUT scale unification predicts the existence of an axion with
\be
\mu_{\rm GUT}\approx 10^{-15}\text{ eV}\,,
\ee
which arises from fixing a single modulus to give the correct GUT scale coupling, $\alpha_{\rm GUT}=1/25$ arising from a 3-cycle with volume $V_X=25$ in string units (see Appendix~\ref{app:m_theory}). We model this by fixing the log-normal mean to $\log_{10}\bar{\mu}_{\rm ax}=-15$.

The fuzzy DM model~\cite{1990PhRvL..64.1084P,hu2000,Marsh:2013ywa,2014NatPh..10..496S,2017PhRvD..95d3541H} posits that DM composed of axions with mass
\be
\mu_{\rm FDM} \approx 10^{-22}\text{ eV}\, ,
\ee
has certain desirable properties that could lead to its being favoured over standard cold DM by observations of galactic structure. We model this by fixing the log-normal mean to $\log_{10}\bar{\mu}_{\rm ax}=-22$. The QCD axion~\cite{pecceiquinn1977,weinberg1978,wilczek1978} mass is given by:
\be
\mu_{{\rm QCD}}\approx 6\times 10^{-10}\text{ eV}\left(\frac{10^{16}\text{ GeV}}{f_a} \right)\, .
\label{eqn:qcd_mass_zero_T}
\ee
In order to realise the QCD axion in M-theory, some light eigenstate in the ``pure M-theory'' spectrum should receive its mass dominantly from QCD instantons. Furthermore, the VEV of this field should be not far displaced from $\theta=0$ to solve the strong-CP problem. These two conditions together require that there is at least one eigenstate in the pure M-theory spectrum with~\cite{Acharya:2010zx}:
\be
\mu_{\rm ax}\lesssim \mu_{\rm ax, low}\approx 10^{-14}\text{ eV}\, .
\ee
We model this by fixing $\bar{\mu_{\rm ax}}$ and $\sigma$ such that $\mu_{\rm ax, low}$ is within 95\% of the probability at the lower end of the distribution after $N_{\rm ax}$ draws. This fixes $\bar{\mu}_{\rm ax}(\sigma,N_{\rm ax})$ in terms of standard error functions:
\be
N_{\rm ax}{\rm erfc}\left[ - \frac{\log_{10}(\mu_{\rm ax, low}/\bar{\mu}_{\rm ax})}{\sqrt{2\sigma^2}}  \right]=0.1 \, .
\ee
With the above fixed, one linear combination of axions receives its mass from QCD instantons. Therefore, we remove one axion from the M-theory distribution and replace it with the QCD axion. The probability that the QCD axion in M-theory is allowed based on BH superradiance data is thus:
\begin{widetext}
\be
P_{\rm al}(\sigma,N_{\rm ax}) = P_{\rm al}(\mu_{\rm ax, QCD}|N_{\rm ax}=1)\left\{\int {\rm d}\mu_{\rm ax}\,\, p[\mu_{\rm ax} |\sigma,\bar{\mu}_{\rm ax}(\sigma,N_{\rm ax})] P_{\rm al}(\mu_{\rm ax}|N_{\rm ax}=1) \right\}^{N_{\rm ax}-1}\, .
\ee
\end{widetext}
Constraints on the distribution parameters of each of these benchmark models are shown in Fig.~\ref{fig:mtheory_models_results}. While none of these models are ruled out for a single axion, in all cases the exclusion probability starts to become significant for non-zero distribution widths and large numbers of fields. In all cases, the maximum allowed value of $N_{\rm ax}$ increases for very large $\sigma$. For large $\sigma$ the distribution is approximately log-flat with respect to the data exclusions, and increasing the width simply reduces the probability of overlap.

The GUT model has a small exclusion probability at zero width due to the large mass errors on the lightest SMBHs (\emph{NGC 4051} and \emph{MCG-6-30-15}). The GUT model is excluded at better than the 95\% C.L. for all widths $\sigma<\mathcal{O}(100)$ for $N_{\rm ax}\gtrsim 100$. The fuzzy DM model is excluded at better than the 95\% C.L. for all widths $1\lesssim \sigma\lesssim 10^3$ if $N_{\rm ax}\gtrsim 100$. 

The QCD axion model is the least constrained by the data. The mass of the QCD axion with $f_a\gtrsim 10^{17}\text{ GeV}$ is not excluded itself by BH superradiance, nor is the light mass $\mu_{\rm ax, low}$ required from the M-theory part of the spectrum. There is a small range of intermediate widths where the distribution does overlap the excluded region, excluding $0.2\lesssim \sigma\lesssim 4$ if $N_{\rm ax}\gtrsim 100$ at 95\% C.L. while $\sigma\gtrsim 4$ is allowed for all $N_{\rm ax}<1000$ considered.

\subsubsection{Comment on Fuzzy DM and BH Superradiance}

Recently it has been claimed that the global 21cm signal~\cite{2018Natur.555...67B}, which is strong evidence that the Universe was undergoing reionization at redshift $z_{\rm re}\approx 17$, places a lower bound on the fuzzy DM mass of $\mu_{\rm ax}\geq 5{\text -}8\times 10^{-21}\text{ eV}$~\cite{Lidz:2018fqo,Schneider:2018xba}. This result is extremely interesting since, if it is to be believed in its accuracy, it significantly shrinks the gap between fuzzy DM bounds from BH superradiance and structure formation. In the context of the present work, if fuzzy DM is realised from a mass distribution, then respecting the reionization bound and BH superradiance demands an extremely narrow distribution with a small number of light fields. If the gap between fuzzy DM constraints from BH superradiance and reionization is closed, either by the measurement of spins of the most massive SMBHs, or improvements on the lower limit to $z_{\rm re}$, then fuzzy DM with no self-interactions will be completely excluded. Rescuing fuzzy DM from BH superradiance constraints in such a case would require self interaction strengths corresponding to decay constants $f_a\lesssim 10^{16}\text{ GeV}$. Low decay constants open the door to new fuzzy DM phenomenology~\cite{2017JCAP...03..055H,2017PhRvL.118a1301L,2018PhRvD..97b3529D}, but may become increasingly hard to realise in small-volume string compactifications.

\section{Discussion and Conclusions}
\label{sec:conclusions}

BH superradiance places strong constraints on the possible existence of light bosonic fields with small self-interactions, in particular on axion-like fields. Many authors have considered these constraints for the case of a single new light field. The excluded ranges of axion mass are:
\begin{align}
7\times 10^{-14}\text{ eV}&<\mu_{\rm ax}<2\times 10^{-11}\text{ eV}\, , \nonumber\\
7\times 10^{-20}\text{ eV}&<\mu_{\rm ax}<1\times 10^{-16}\text{ eV}\, .\nonumber
\end{align}
A model with multiple axions is excluded if just one field lies in these ranges. We have studied this possibility, and used BH superradiance to exclude certain distributions of axion masses. The constraints become more severe with larger numbers of axion-like fields due to the increased probability of drawing an outlier. This allows us to place constraints on the number of axion-like fields, $N_{\rm ax}$.

Models for axions coming from string theory and M-theory typically involve many axion-like fields. These fields have their masses determined by microscopic quantities related to the geometry of the compact space. Their masses, however, are expected to follow particular statistical distributions independently of the microscopic details. We have considered various different distributions, log-flat, log-normal, and Mar\v{c}henko-Pastur, using BH superradiance to bound both the parameters of the distribution, and, more significantly, the number of light axions within that distribution.

Constraints on $N_{\rm ax}$ from a process such as BH superradiance, which relies only on the existence of the vacuum fluctuations of the given field, are extremely powerful, and could be used in this context to bound the dimensionality of phenomenologically consistent moduli spaces in string/M-theory. Indeed we have seen that the benchmark value of $N_{\rm ax}\approx 30$ found in the majority of known Calabi-Yau manifolds can be excluded for a wide range of distribution parameters. Only a small number of fields should obtain masses anywhere in the BH superradiance region from $10^{-10}\text{ eV}\lesssim \mu_{\rm ax}\lesssim 10^{-20}\text{ eV}$, which can be accommodated with a single very wide distribution $\sigma\gtrsim 30$, or bimodal distributions containing only very light or relatively heavy axions.

Our analysis has neglected axion self-interactions, which shut off BH superradiance if they are strong, and other constraints, for example coming from the relic abundance. It would be interesting in this regard to combine our previous analysis in Ref.~\cite{2017PhRvD..96h3510S} with the current analysis and compute, in addition to axion masses, the axion decay constants, relic density, and self-interaction potential. The present work is more model-independent, since it does not rely on any cosmological assumptions, and applies to any model for light scalars with sufficiently small self-interactions. The extended and combined analysis will be the subject of future work.

\section*{Acknowledgments}
We acknowledge useful conversations with Bobby Acharya, Katy Clough and Chakrit Pongkitivanichkul. The work of MJS is supported by funding from the UK Science and Technology Facilities Council (STFC). DJEM is supported by the Alexander von Humboldt Foundation and the German Federal Ministry of Education and Research.

\appendix

\section{Computations of Black Hole Superradiance}
\label{appendix:bhsr}

\subsection{The Geometry of the Kerr Spacetime}
\label{appendix:geometry}
The 3+1 dimensional spacetime region outside the horizon of a rotating Kerr BH is described by the invariant line element, $ds^2 = g_{\alpha \beta }dx^{\alpha}dx^\beta$ which, using the standard Boyer-Lindquist coordinates $(t,r,\theta,\phi)$ and metric signature $[-,+,+,+]$, takes the form 

\begin{widetext}
\begin{equation}
ds^2_{\rm Kerr} = - \left( 1 - \frac{2M_{\rm BH}r}{\Sigma} - \right)dt^2 - \frac{4M_{\rm BH} ar sin^2 \theta}  {\Sigma}dt d\phi + \\ \frac{\Sigma}{\Delta}dr^2 + \Sigma d\theta^2 + \frac{(r^2+a^2)^2 - a^2 \Delta sin^2\theta}{\Sigma}sin^2\theta d\phi^2 \,,
\label{eq:boyer}
\end{equation}
\end{widetext}
which is invariant under time translations, possessing a Killing vector. The metric functions are defined as,

\begin{figure*}
  \centering
  \subfloat[xz / $a_*=0.7$.]{\includegraphics[width=0.32\textwidth]{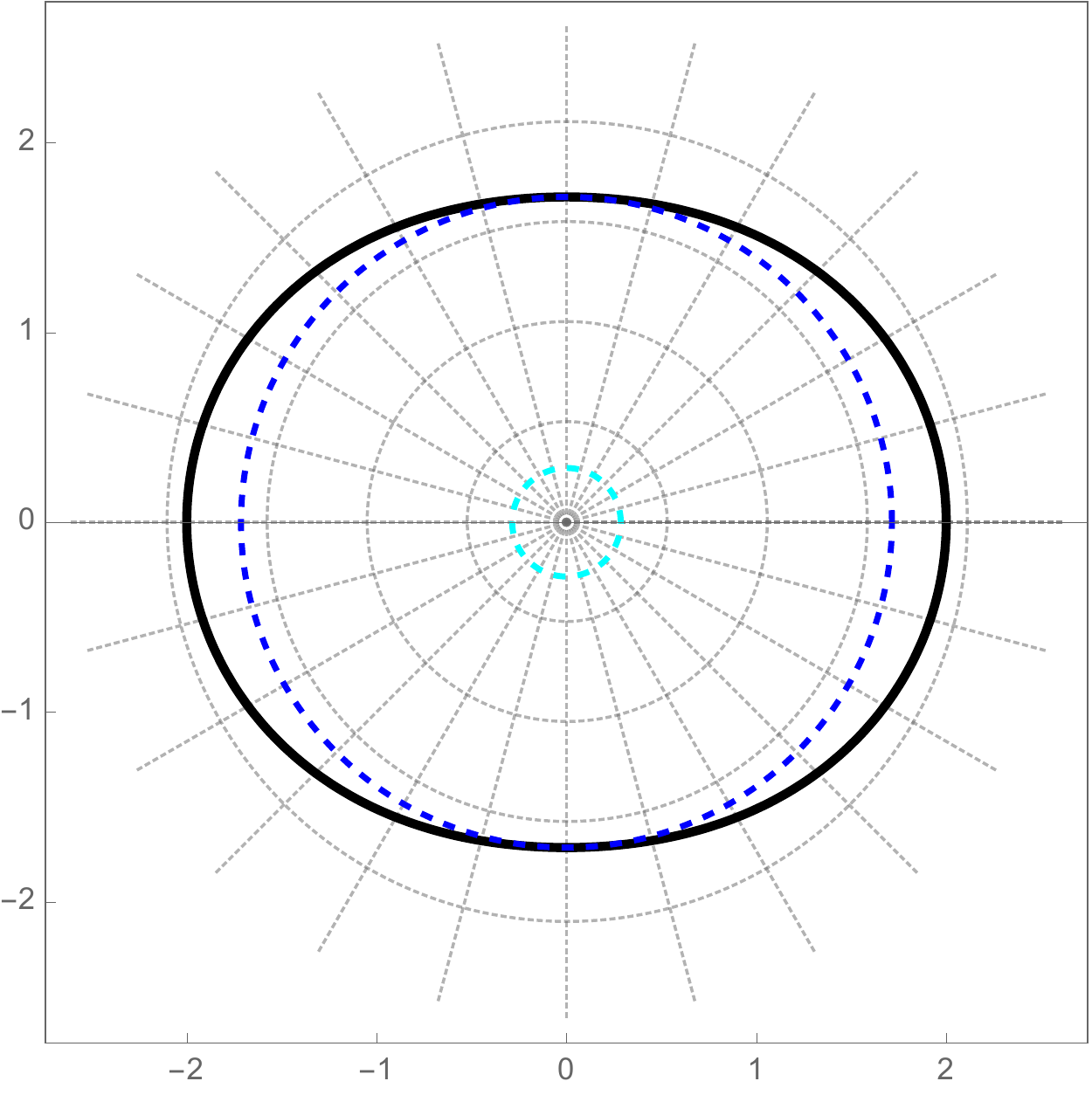}\label{fig:polar3}}
  \hfill
  \subfloat[xz / $a_*=0.9$.]{\includegraphics[width=0.32\textwidth]{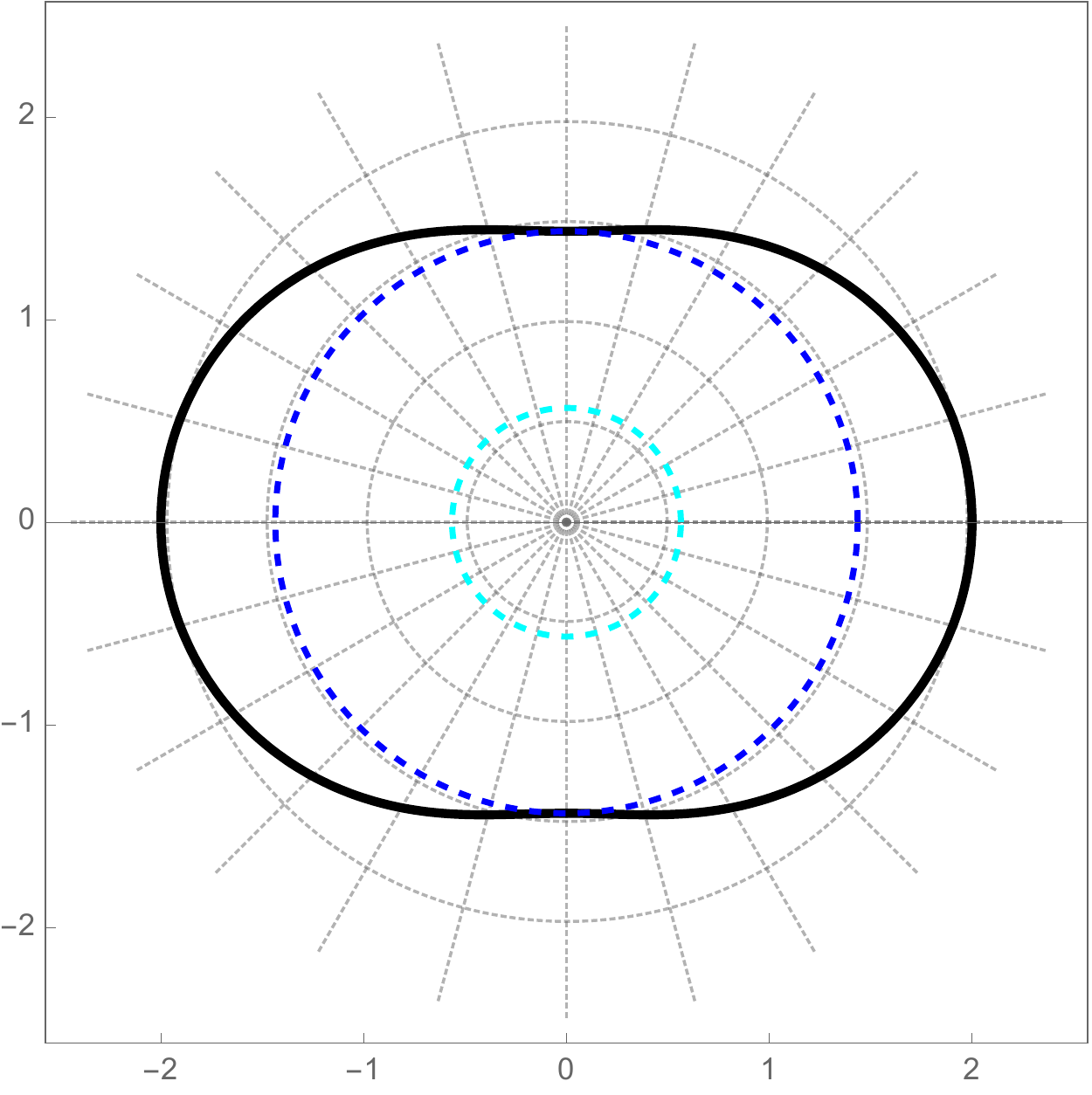}\label{fig:polar2}}
    \hfill
  \subfloat[xz / $a_*=0.999$.]{\includegraphics[width=0.32\textwidth]{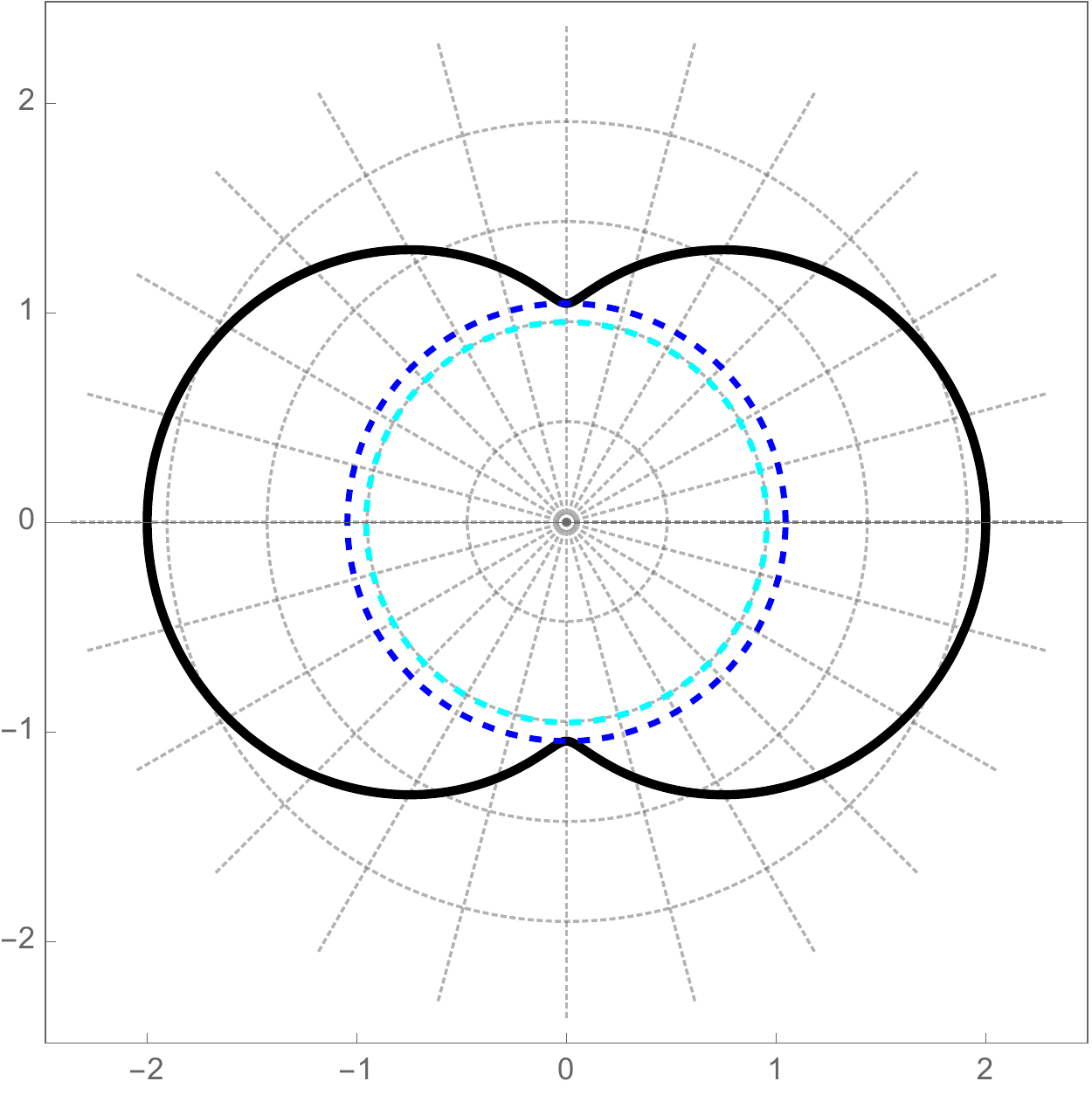}\label{fig:polar1}} \\
  \centering
  \subfloat[xyz / $a_*=0.7$.]{\includegraphics[width=0.32\textwidth]{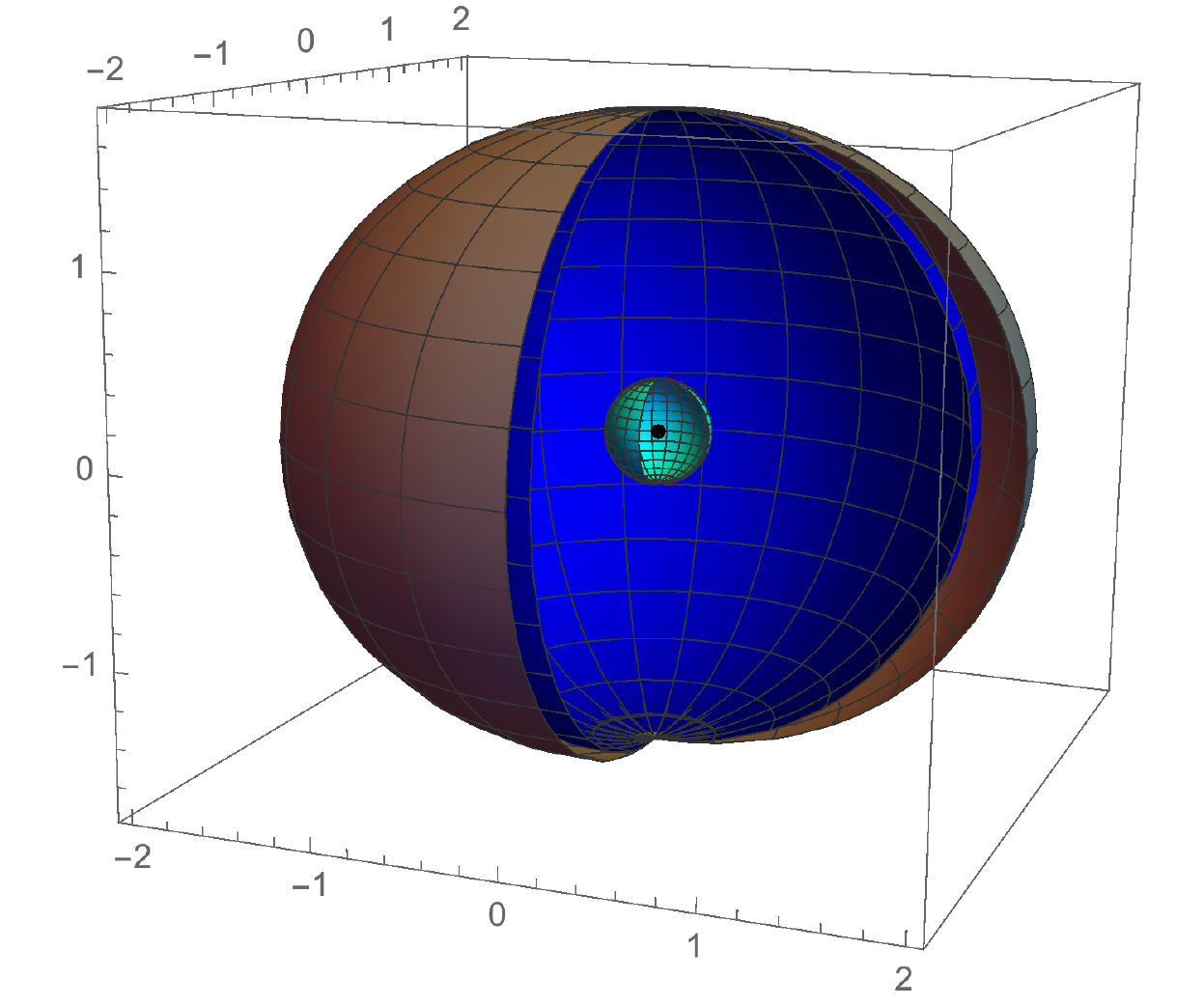}\label{fig:polar3d3}}
  \hfill
  \subfloat[xyz / $a_*=0.9$.]{\includegraphics[width=0.32\textwidth]{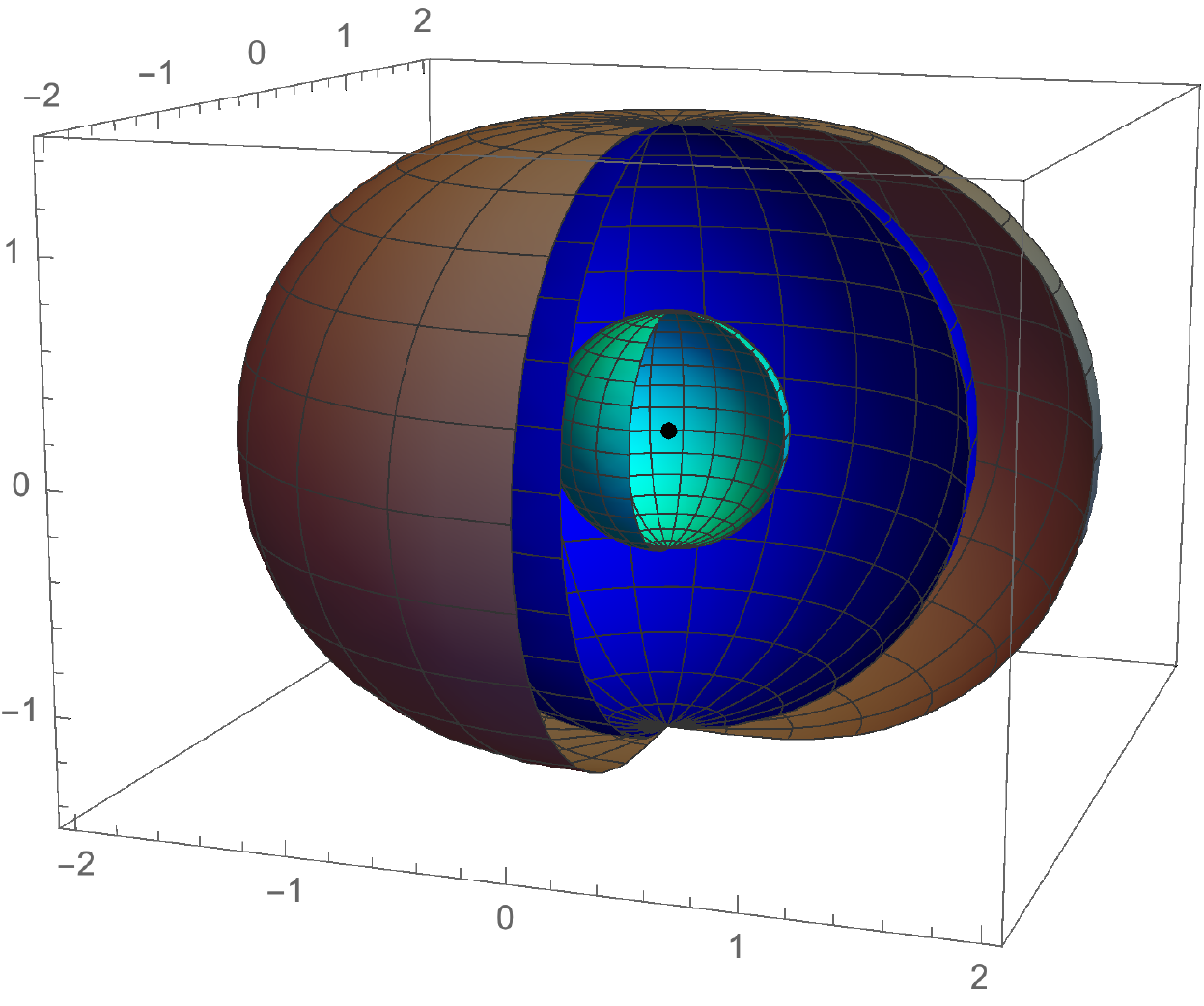}\label{fig:polar3d2}}
    \hfill
  \subfloat[xyz / $a_*=0.999$.]{\includegraphics[width=0.32\textwidth]{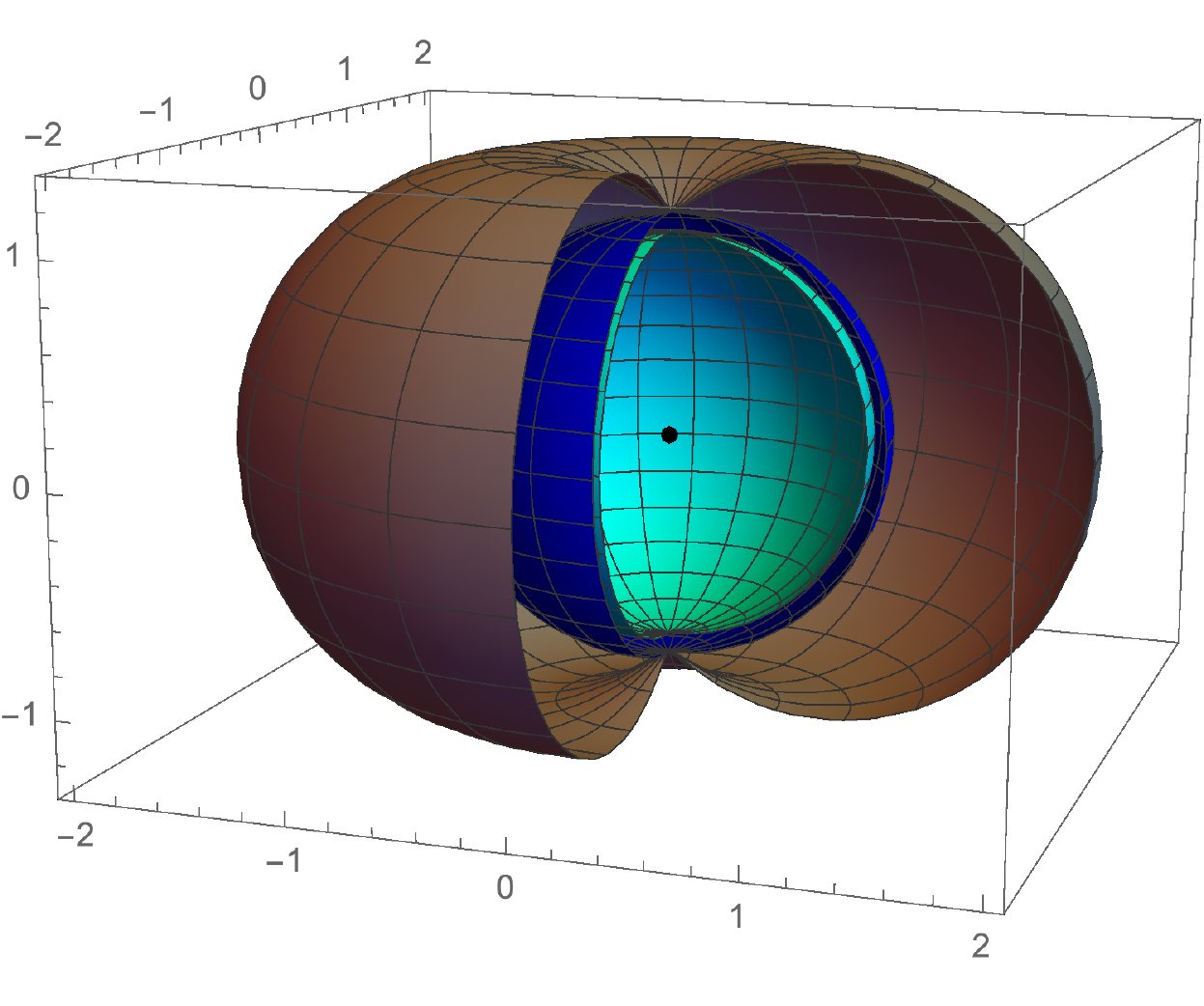}\label{fig:polar3d1}} 
  \caption{Kerr BH horizons in the xz-plane and xyz-volume for fixed values of the dimensionless spin parameter, $a_*$ approaching the limit for an extremal BH. The x-axis represents the radial distance from the BH in polar coordinates. The solid black line/surface defines the ergoregion, the dashed/solid blue and cyan lines/surfaces represent the outer and inner horizons respectively in the xz plane/xyz volume. The two hypersurfaces of the event horizon and the ergosphere meet at the co-latitude pole of $0$ degrees.}
   \label{fig:polar2d}
\end{figure*}

\begin{align}
\Sigma &= r^2 + a^2cos^2 \theta\,,	\\
\Delta &= r^2 + a^2 - 2Mr\,, \label{eq:delta} \\
r_{\pm} &= M_{\rm BH} \pm \sqrt{M_{\rm BH}^2 - a^2}\,. \label{eq:rpm}
\end{align}
The zero solutions of Eq.~(\ref{eq:delta}) define two horizons, an inner Cauchy horizon at $r_{-}$ with the larger root at $r_{+}$ defining the outer physical event horizon. The characteristic limits of each BH horizon as a function of the dimensionless spin are displayed in the panels of Fig.~\ref{fig:polar2d}. As the spin of the BH approaches the extremal limit, $a_{*} = 1$ the inner and outer horizons coincide. A defining property of Kerr BHs is existence of an a surface external to the outer horizon known as the ergosurface. The ergosurface is defined by the static limit roots, $g_{tt}=0$ with the coordinates, 
\begin{equation}
r_{\rm ergo} = M_{\rm BH} + \sqrt{M^2_{\rm BH} - a^2cos^2\theta}	\,.
\end{equation}
As the BH spin approaches the static Schwarzschild solution, $a_{*} \rightarrow 0$ the ergosurface and outer horizon coincide. The region between the outer horizon and ergosurface defines the ergoregion. Inside the ergoregion the vector, $\xi^{\mu}$ in the time coordinate basis becomes spacelike, $\xi^{\mu}\xi^{\nu}g_{\mu\nu} = g_{tt}>0$. This property allows for a Killing energy in the presence of a BH to be negative inside the ergoregion, leading to the superradiant amplification of the infalling waves associated to the bosonic field. The event horizon angular velocity for observers at spacial infinity for the BH is, 
\begin{equation}
\Omega_{H} = \frac{a}{r^2_{+}+a^2}\,.	
\end{equation}
The dynamics of the linearised massive scalar in the Kerr spacetime are governed by the Klein-Gordon wave equation.

\subsection{The Klein-Gordon Wave Equation}
\label{sec:klein}

The classical massive scalar field obeys the Klein-Gordon wave equation, 
\begin{equation}
(\nabla^\mu \nabla_\mu - \mu_{\rm ax})\Psi = 0\,.	
\label{eq:wave}
\end{equation}
The massive Klein-Gordon equation on a Kerr spacetime background allows for a separation of variables
\begin{equation}
	\Psi = \sum_{l,m} e^{- i\omega t + i m \psi} S_{lm}(\theta)R_{lm}(r) + h.c.\,,
\end{equation}
with an infinite discrete set of complex eigenfrequencies $\omega_{lmn}$, of the form in Eq.~(\ref{eq:eigenfrequencies}). The Klein-Gordon wave equation following a separation of variables is expressed by two coupled ordinary differential equations. Using the Teukolsky formulism \cite{PhysRevLett.29.1114} the separated ODEs for the radial and angular parts, $R_{lm}(r)$ and $S_{lm}(\theta)$  respectively are,
\begin{widetext}
\begin{equation}
	\frac{1}{sin(\theta)}\frac{d}{d\theta}\left(sin(\theta)\frac{dS}{d\theta}\right) \left[a^2(\omega^2-\mu^2)cos^2(\theta) -\frac{m^2}{sin^2(\theta)}+\Lambda_{lm}\right]S_{lm}(\theta) = 0\,,
	\label{eq:firstode} 
\end{equation}

\begin{equation}
\Delta \partial_r (\partial_r R) + (\omega^2(r^2+a^2)^2 - \\ 4ar_g rm\omega + a^2m^2 - \Delta(\mu_{\rm ax}r^2 a^2\omega^2+l(l+1))R(r)=0\,.
\label{eq:radial}	
\end{equation}
\end{widetext}
The first ODE in Eq.~(\ref{eq:firstode}) determines the angular component, $S_{lm}(\theta)$ of the scalar eigenfunction. The angular solutions of Eq.~(\ref{eq:firstode}), $S_{lm}$ are the the spheroidal harmonics which are required to be regular at the pole boundaries, $\theta=0$ and $\theta=\pi$. These boundary conditions single out a discrete family $\{K_{lm}\}$ of angular eigenvalues also known as the \emph{coupling constant} which characterise the massive scalar. The angular eigenvalues can either be found using an expansion in the limit that $a\omega$ and $a\mu_{\rm ax} \rightarrow 0$ where the expansion of $K_{lm}$,
 \begin{equation}
 \Lambda_{lm} = l(l+1)+\sum_{k=1}^{\infty}c_k[a^2(\mu^2_{\rm ax} - \omega^2 )]^k\,,
 \label{eq:angulareigenbvalue}	
 \end{equation}
 gives $\Lambda_{lm} \rightarrow l(l+1) +\mathcal{O}(a^2\omega^2)$ in the non rotating limit where, when $k=0$, an analytical expression can be extracted. Higher orders of $k$ require numerical solutions. The function inside the sum defines the so called \emph{spheroidicity}. These angular eigenvalues can also be found via Leavers' continued fraction method (Appendix~\ref{app:numerics}) or Hughes' spectral decomposition method.   

A rescaling of the radial function introducing, $\psi_{lm} = \sqrt{r^2+a^2}R_{lm}$, along with a definition of the Regge-Wheeler tortoise coordinate 
\begin{equation}
dr^* = 	\frac{(r^2+a^2)}{\Delta}dr\,,
\end{equation}
where, 
\begin{equation}
r^* = r + \frac{2M}{r_+ - r_-}\left( r_+\ln\left|\frac{r-r+}{2M}\right| - r_- \ln \left|\frac{r-r-}{2M}\right| \right)	\,,
\end{equation}
allows for the radial Teukolsky equation (Eq.~(\ref{eq:radial})) to be expressed in the form of a Schr\"{o}dinger like wave equation,
\begin{equation}
\frac{d^2\psi_{lm}}{dr^{*2}} = \left[\omega^2 -V(r,\omega)\right]\psi_{lm}\,.	
\end{equation}
The effective potential is defined as: 
\begin{widetext}
\begin{equation}
V = \frac{4r_g r a m \omega -a^2 m^2}{(r^2 + a^2)^2}	 + \frac{\Delta}{(r^2 +a^2)}\left( \mu_{\rm ax} + \frac{l(l+1)+(\mu_{\rm ax} + \omega^2 )a^2}{r^2+a^2} + \frac{3r^2-4r_g r +a^2}{(r^2+a^2)^2} - \frac{3\Delta r^2}{(r^2+a^2)^3} \right)\,.
\label{eq:effectivepot}
\end{equation}
\end{widetext}
We require solutions to Eq.~(\ref{eq:radial}) with boundary conditions defining an outgoing solution tending to zero at spacial infinity and purely incoming waves at the event horizon. In the limit where $\omega m\ll 1$ and $\mu_{\rm ax }m \ll 1$, Eq.~(\ref{eq:radial}) is susceptible to analytic methods. These boundary conditions correspond to modifications of the radial solutions in the limits, 
\begin{empheq}[left=\empheqlbrace]{align}
\lim_{r^*\to -\infty} R_{lm} &\sim e^{-ik_+r^*}\,, \\
\lim_{r^*\to \infty} R_{lm} &\sim \frac{1}{r} e^{i\sqrt{(\omega^2-\mu^2)}r^*}\,,\\
\end{empheq}
where $k_+ \equiv \omega - m\Omega_H$. In the low energy limit, $\omega M_{\rm BH} \ll 1$, the radial equation is amenable to the method of matched asymptotics.

\subsection{Analytic Approximations for Non-Relativistic Bound States}

It has been shown that analytic solutions for small values of $\alpha$ can be found using approximate solutions at large and small radii in terms of hypergeometric functions, where matching techniques are used at an intermediate radius to obtain the superradiance rates to leading order in $\alpha$ \cite{PhysRevD.22.2323}. In this limit analytical methods utilise the fact that the radial mode functions, $R_{lm}(r)$ can be approximated in asymptotic regimes by known analytical functions. For each region the equations can be reduced to the form of a confluent hypergeometric function.

Regions far from the BH outer horizon adhering to $r\gg r_g$ whilst ensuring we are in the $\mu_{\rm ax} M_{\rm BH} \ll 1$ regime allow the ODE in Eq.~(\ref{eq:radial}) to be approximated as 
\begin{equation}
\frac{d^2}{dr^2}(rR)+\left[ \omega^2 - \mu^2_{\rm ax} + \frac{2M\mu^2_{\rm ax}}{r} - \frac{l(l+1)}{r^2}\right]rR = 0 \,,	
\label{eq:approxode}
\end{equation}
where the axion momentum is given by, 
\begin{equation}
k^2 \equiv \mu^2 - \omega^2_{\rm ax}	\,. 
\end{equation}
The solutions can be extracted by defining:
\begin{equation}
\nu \equiv \frac{\mu^2_{\rm ax}M}{k} = n+l+1+\delta \nu\,,
\end{equation}
where the value of $\delta \nu$ represents a small complex number which describes the deviation away from the pure hydrogenic spectrum. When the axion momentum satisfies $k^2>0$ we are presented with a series of quasi-bound state solutions. This equation is the same form of the Schr\"{o}dinger equation which governs the electron in the hydrogen atom. The solution to Eq.~(\ref{eq:approxode}) can be expressed as 
\begin{equation}
R(r) = (2kr)^l e^{-kr}U(l+1- \frac{\alpha}{r_g k },2(l+1),2kr)\,,
\label{eq:confluent}	
\end{equation}
where U is the confluent hypergeometric function of the second kind.
In the limit $r\ll r_g$, Eq.~(\ref{eq:radial}) is solved analytically where the approximate solution takes the form 
\begin{equation}
z(z+1)\frac{d}{dz}\left[z(z+1)\frac{dR}{dz}\right] + \left[P^2 - l(l+1)z(z+1)\right] R = 0\,, 
\label{eq:formhyper}	
\end{equation}
with the values, 
 \begin{align}
 z &= \frac{r-r_+}{r_+-r_-}\,,\\
 P &=  \frac{2r_+(\omega - m\omega_+)}{r_+ - r_-}\,.	
 \end{align}
The form of the equation in Eq.~(\ref{eq:formhyper}) presents a solution infalling at the horizon

\begin{equation}
R(r) = \left( \frac{r-r_+}{r-r_-} \right)^{-iP} {}_2F_1\left(-l,l+1,1+2iP,\frac{r-r_-}{r_+-r_-}\right)	\,,
\label{eq:Gauss}
\end{equation}
where ${}_2F_1$ is the Gauss hypergeometric function. Enforcing the condition that $\mu_{\rm ax}M \ll 1$ the two approximate solutions in Eq.~(\ref{eq:confluent}) and Eq.~(\ref{eq:Gauss}) have an overlap in their respective regions of validity. Matching the lowest terms for $r$ in Eq.~(\ref{eq:confluent}) with the asymptotic form of Eq.~(\ref{eq:Gauss}) yields the solutions for the imaginary component of the frequency encapsulating the superradiance rate given in Section \ref{sec:rates}.

\subsection{Numerical Solutions for Bound States}
\label{app:numerics}
Analytical based methods for approximating the superradiance rates suggest a maximal value for the approximate regime $\mu_{\rm ax}M_{\rm BH} \sim 1$. In order to probe this region of the parameter space it is required to solve the radial mode function ODEs eigenvalue problem using numerical techniques. See \cite{Furuhashi:2004jk,Cardoso:2005vk} for an initial study incorporating Leaver's continued fraction method \cite{10.2307/2397876} for numerical calculations and Dolan's work \cite{Dolan:2007mj} for an extensive study of the expanded parameter space, providing numerical solutions using a three-term recurrence relation and the continued fraction method. 

The radial function $R(r)$ is assumed to take the following form of the the infinite series 
\begin{figure}
\centering
\includegraphics[width=0.5\textwidth]{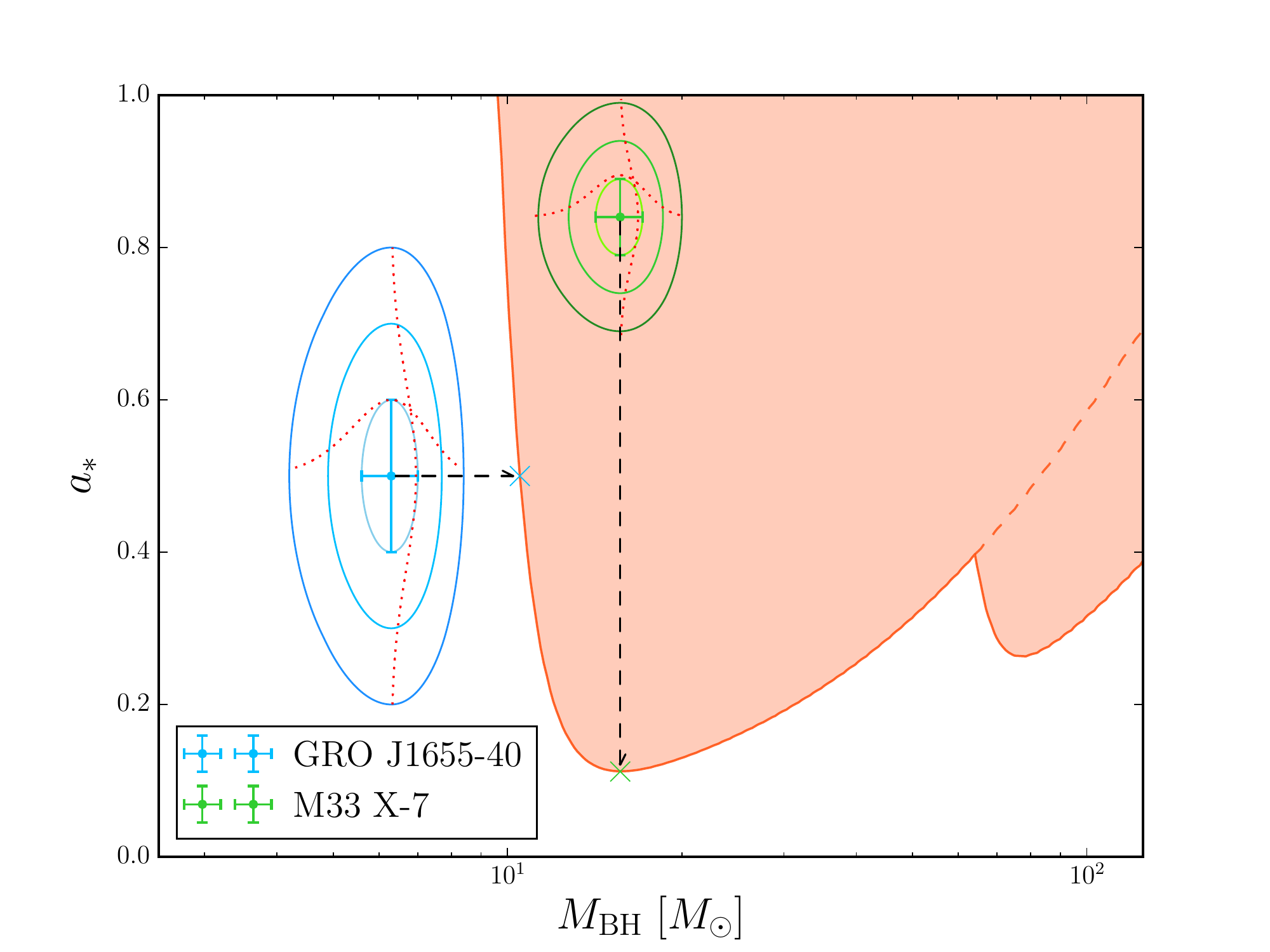}	
\caption{Visual representation of the statistical model methodology using two example stellar BHs, GRO J1655-40 and M33 X-7 with instability bounds for an axion mass, $\mu_{\rm ax} = 10^{-12.75}\ {\rm eV}$. Each data point is shown with 1$\sigma$, 2$\sigma$ and 3$\sigma$ contour levels. Effective errors are calculated by projection in either the x or y axis (crossed points) depending on weather the BH falls inside the instability bounds where $f(x)$ is defined.}
\label{fig:effective_error_example}
\end{figure}
\begin{equation}
R(r) = (r-r_+)^{-i\sigma}(r-r_-)^{i\sigma+\chi-1}e^{qr} \sum_{n=0}^{\infty} a_n \left( \frac{r-r_+}{r-r_{-}} \right)^n \,,
\label{eq:formradial}
\end{equation}
where, 
\begin{align}
\sigma &= \frac{2r_+(\omega - \omega_c)}{r_+ - r_-}\,, \\
q & = \pm \sqrt{\mu^2 - \omega^2}\,,	\\
\chi &= \frac{\mu-2\omega^2}{q}\,.
\end{align}
A substitution of Eq.~(\ref{eq:formradial}) into Eq.~(\ref{eq:radial}) obtains the three term relation for the expansion coefficients $a_n$ for $n>0, n\in \mathbb{N}$ 
\begin{align}
\alpha_0 a_1 &+ \beta_0 a_0 = 0\,, \label{eq:numcoef1} \\
\alpha_n a_{n+1} + &\beta_n a_n + \gamma_n a_{n-1} = 0\,, 	
\end{align}
where, 
\begin{align}
\alpha_n = n^2 + (c_0+1)n +c_0\,, \\
\beta_n = -2n^2 + (c1+2)n + c3\,, \\
\gamma_n = n^2 + (c_2-3)n + c_4	\,.
\end{align}
The values of the constants $c1,c2,c3$ and $c_4$ are expressed as functions dependant on the parameters, $\omega, \sigma, m$ and angular eigenvalues, $\Lambda_{lm}$ (Eq.~(\ref{eq:angulareigenbvalue})), where 
\begin{align}
c_0 &= 1 - 2i\omega - \frac{2 i }{b}\left( \omega - \frac{am}{2}\right) \,,  \\
c_1 &= -4 + 4i \left(\omega - iq(1+b)\right) + \frac{4i}{b}\left(\omega - \frac{am}{2}\right) - \frac{2(\omega^2+q^2)}{q}\,, \\
c_2 &= 3 - 2i\omega - \frac{2(q^2-\omega^2)}{q} - \frac{2i}{b}\left(\omega - \frac{am}{2} \right)\,,
\end{align}
\begin{widetext}
\begin{align}
c_3 &= \frac{2i(\omega - iq)^3}{q} + 2 (\omega -iq)^2 b +q^2 a^2 + 2iqam - \Lambda_{lm} - 1 - \frac{\left( \omega - iq\right)^2}{q} + 2qb + \frac{2i}{b}\left( \frac{\left( \omega - iq\right)^2}{q}+1 \right) \left(\omega - \frac{am}{2}\right)\,,
\end{align}	
\end{widetext}
\begin{align}
c_4 &= \frac{\left(\omega - iq\right)^4}{q^2} + \frac{2i\omega\left(\omega-iq\right)^2}{q}-\frac{2i\left(\omega-iq\right)^2}{bq}  \left(\omega - \frac{am}{2}\right)\,.   
\end{align}
with,
\begin{equation}
b = \sqrt{1-a^2}\,.	
\end{equation}

The three factor recurrence relation can be solved in terms of a continued fraction if we take the assumption that the factor $\sfrac{a_{n+1}}{a_n}\rightarrow 0$ as $n \rightarrow \infty$ obtaining, 
\begin{widetext}
\begin{equation}
\frac{\left( a_{n+1}\right)}{\left(a_n\right)} = - \frac{\left(\gamma_{n+1}\right)}{\left(\beta_{n+1}\right)+\left(\alpha_{n+1}\right)\left(\frac{\alpha_{n+2}}{\alpha_{n+1}}\right)} =  - \frac{\left( \gamma_{n+1}\right)}{\left(\beta_{n+1}-\right)} \frac{\left(\alpha_{n+1}\right)\left(\gamma_{n+2}\right)}{\left( \beta_{n+2}-\right)} \frac{\left(\alpha_{n+2}\right)\left( \gamma_{n+3}\right)}{\left(\beta_{n+3}-\right)}\ \ldots \ \ . 
\label{eq:continuedfrac}
\end{equation}	
\end{widetext}
Rearranging Eq.~(\ref{eq:numcoef1}) to give
\begin{equation}
\frac{a_1}{a_0} = \frac{-\beta_0}{\alpha_0}\,,
\end{equation}
and substituting in $n=0$ into Eq.~(\ref{eq:continuedfrac}) gives the condition for the eigenvalue equation for bound state eigenfrequencies of the form in Eq.~(\ref{eq:eigenfrequencies}), 
\begin{equation}
\beta_0 - \frac{\alpha_0\gamma_1}{\beta_1-}   \frac{\alpha_1\gamma_2}{\beta_{2}-} \frac{\alpha_2 \gamma_3}{\beta_3-} \ldots = 0\,, 
\end{equation}
which can be solved using numerical method techniques.  
\section{Statistical Model}
\label{appendix:stats}


We model the BH data in Table~\ref{tab:cosmopar} with two dimensional multivariate gaussian distributions for both $x=M_{\rm BH}$ and $y=a_{*}$. There are $N_d$ data points $d_i$ comprising the dataset $\{d_i\}$. For each point in the data set the values of $M_{\rm BH}$ and $a_*$ and their associated errors become centred data values $(\bar{x},\bar{y})$ with errors $(\sigma_x,\sigma_y)$. We are interested in the probability that a given model, $\mathcal{M}$, is excluded given the data, $\{d_i\}$: $P_{\rm ex}(\mathcal{M}|\{d_i\})$. Since a single data point in the disallowed region would exclude the model, $P_{\rm ex}(\mathcal{M}|\{d_i\})$ is given by the probability that any single data point is above the BH superradiance isocontour boundaries for each value of $l$. For a large number of data points, this is a relatively tricky combinatorial problem. However, the probability is normalised such that:
\be
\label{eq:probex}
P_{\rm ex}(\mathcal{M}|\{d_i\}) =1-P_{\rm allowed}(\mathcal{M}|\{d_i\}) \, .
\ee
Now we can use the binomial theorem (or a simple probability tree) to note that $P_{\rm allowed}(\mathcal{M}|\{d_i\})$ is simply the cumulative probability that all data points simultaneously fluctuate below the isoctontour:
\be
P_{\rm allowed}(\mathcal{M}|\{d_i\})=\prod_iP_{\rm allowed}(\mathcal{M}|d_i)\, ,
\ee
and $P_{\rm allowed}(\mathcal{M}|d_i)$ is simply the volume of the bivariate Gaussian contained outside the isocontour boundary given by the function $y=f(x)$. 

To evaluate $P_{\rm allowed}(\mathcal{M}|d_i)$ in a numerically efficient manner, we make two simplifying assumptions. Firstly, we assume zero covariance between $x$ and $y$. Secondly, the error on the two-dimensional data can be evaluated using an effective one dimensional error \cite{Marsh:2015wka,Ma:2013kun}. These two simplifications allow us to use the standard error function to evaluate $P_{\rm allowed}(\mathcal{M}|d_i)$, rather than the more numerically expensive integral under the curve. 

The shape of the BH superradiance contours $y=f(x)$, which only have support over finite $x$, requires this procedure to be evaluated in two separate regimes. Where the contour is defined, we use the contour as $y=f(x)$ and evaluate the effective one dimensional error in $y$, $\Sigma_y$, as:
\begin{equation}
\Sigma^2_y = \sigma_y^2 + f'(\bar{x})^2\sigma_x^2\, .
\label{eqn:effective_error_y}
\end{equation}
When the contour is not defined for a given $x$, we instead use the inverse function $x=g(y)$ and evaluate the effective error in $x$, $\Sigma_x$, as:
\begin{equation}
\Sigma^2_x = \sigma_x^2 + g'(\bar{y})^2\sigma_y^2\, .
\label{eqn:effective_error_x}
\end{equation}
The effective errors are represented visually in Fig.~\ref{fig:effective_error_example}. Since our functions are all given numerically, the inverse function and its derivative are trivial to evaluate given the original function.

A complication arises since $g(y)$ is multivalued, taking two values $g_1$ and $g_2$ for a single $y$. We choose to evaluate the derivative $g'(\bar{y})$ at the nearest part of the contour (i.e. the value $g_i$ which minimises $\bar{x}-g(\bar{y})$), and evaluate the error function between the two values $g_1$ and $g_2$. This approximation only affects $P_{\rm allowed}(\mathcal{M}|d_i)$ for values close to unity, while $P_{\rm ex}(\mathcal{M}|\{d_i\})$ is dominated by the smallest values of $P_{\rm allowed}(\mathcal{M}|d_i)$ contained well within the contours where $f(x)$ has support and is single valued.

The use of the effective errors, Eqs.~(\ref{eqn:effective_error_y}, \ref{eqn:effective_error_x}), assumes that, for a given data point, the functions $f(x)$ and $g(y)$ are smooth at the mean value over the range of the errors. When a BH data point with large errors sits close to a cusp in the contours the exclusion probability computed from the effective error is smaller than the true answer. Cusps in the total contour are caused by the meeting of individual contours with different $l$ values, each of which are smooth. A more exact procedure would thus be to compute the probability individually for each $l$ contour, and then compute the cumulative probability from a product over $l$. This would increase the number of likelihood evaluations by $l_{\rm max}\times N_d$, and for speed of computation we do not perform this more accurate calculation. The more accurate calculation would give larger exclusion probabilities (reducing the overall effective size of BH errors), and so the approximate computation is more conservative in the sense that it does not give overly strong exclusions.

\section{The Axiverse Mass Spectrum}
\label{appendix:axiverse}

\subsection{Diagonalising the Lagrangian}
\label{sec:multi}

The most general form for the multi-axion action for fields below any compactification, moduli stabilisation or PQ symmetry scales is of the form given in Eq.~(\ref{eq:multiaxion}). We set the axion field alignment used to determine the diameter of the fundamental domain to only include $P=N$, where we always possess sufficient instanton contributions $N$ for each axion field, $P$. We restrict our considerations to non-perturbative terms with trivial charges, $\mathcal{Q}_{j,i} = \mathds{1}_{N}$. We therefore only need consider enhancements to the $N_{\rm ax}$ field space diameter defined as the longest distance between vertices in the polytopes defining the field ranges via the pythagorean sum from the N-flation model and kinetic alignment in our models considered in Section~\ref{sec:rmt_models}. Lattice alignment as well as alignment theories possessing $P\geq N$ are beyond the scope of this work. See Refs.~\cite{2017JHEP...02..014L,Bachlechner:2017hsj,Bachlechner:2015qja,Bachlechner:2014gfa} for extensive details of axion field alignment.

We begin in the lattice basis, with an \emph{axion} defined by a single cosine potential possessing a shift symmetry obeying, $\theta_i \rightarrow \theta_i + 2\pi $. We diagonalise and canonically normalise the axion field space metric $\mathcal{K}_{ij}$ moving to the kinetic basis with the unitary rotation $U_{ij}$ where, 
\begin{equation}
\mathcal{K}_{ij} = U_{ik}^T{\rm diag}(\mathcal{K}_{kl}) U_{lj} = \frac{1}{2}U^T{\rm diag}(f_a){\rm diag}(f_a)U \, .
 \end{equation}
We define the axion decay constants, $f_a$, from the eigenvalues of $\kahlerij$ in the lattice basis in Planck units,
\begin{equation}
\vec{f_a} = \sqrt{2 {\rm eig}(\mathcal{K}_{ij})}\, .
\label{eq:fbasis}
 \end{equation}
We can now define the canonically normalised field as,
\begin{equation}
\tilde{\phi}_{i}=M_{pl}{\rm diag}(f_a) U_{ij} \theta_{j} \, .
 \end{equation}
In the kinetic basis the effective Lagrangian takes the form,
\begin{equation}
\label{eq:nflatbasis}
\mathcal{L} = -\frac{1}{2}\partial_\mu\tilde{\phi}_i\partial^\mu\tilde{\phi}_j - \frac{1}{2}\tilde{\phi}_i\tilde{\mathcal{M}}_{ij}\tilde{\phi}_j \, ,
 \end{equation}
where the new mass matrix is defined as,
\begin{equation}
\tilde{\mathcal{M}}=2{\rm diag}(1/f_a)U\mathcal{M}U^T{\rm diag}(1/f_a) \, .
 \end{equation}
 Moving to the mass eigenstate basis with a further unitary rotation, $V_{ij}$ gives,
\begin{equation}
\label{eq:massm}
\tilde{\mathcal{M}}=V^T{\rm diag}(m^2_a)V \, .
 \end{equation}
In this basis the mass eigenstate fields are defined as,
\begin{equation}
\phi = V\tilde{\phi} = M_{pl}V{\rm diag}(f_a)U\theta \, ,
\label{eqn:mass_eigenstates}
 \end{equation}
with the effective Lagrangian,
\beq
\label{eq:nflatbasis2}
\mathcal{L} = -\frac{1}{2}\partial_\mu {\phi}_i\partial^\mu {\phi}_j - \frac{1}{2} {\phi}_i {\mathcal{M}}_{ij} {\phi}_j \,. 
\eeq

\subsection{The Random Matrix Theory Mass Spectrum}
\label{sec:rmt_models}

A systematic construction of the axion decay constant and mass spectrum in explicit realisations of the string axiverse is an extremely complex and numerically comprehensive task to undertake. In general, considerations need to be made for leading instanton corrections to the superpotential (Eq.~(\ref{eq:superpot})), a calculation of the full scalar potential  along with a minimisation of polynomial expressions with potentially many variables when considering realistic numbers of apparent axions or moduli. See Ref.~\cite{Halverson:2018xge} for a detailed discussion of the complexities of the string landscape. The effective field theory approach in Section~\ref{sec:multi} can benefit from the simplistic nature of RMT inspired models on the grounds of universality. In these models universality dictates that the distribution of the physical dimensional parameters are characterised by some mean scale and variance. For generic field space metric considerations the kinetic matrix in Eq.~(\ref{eq:multiaxion}) can be well described by a matrix belonging a class of matrices of the Wishart form, 
\begin{figure*}
  \centering
  \subfloat[White Wishart class mass eigenstate basis spectrum.]{\includegraphics[width=0.32\textwidth]{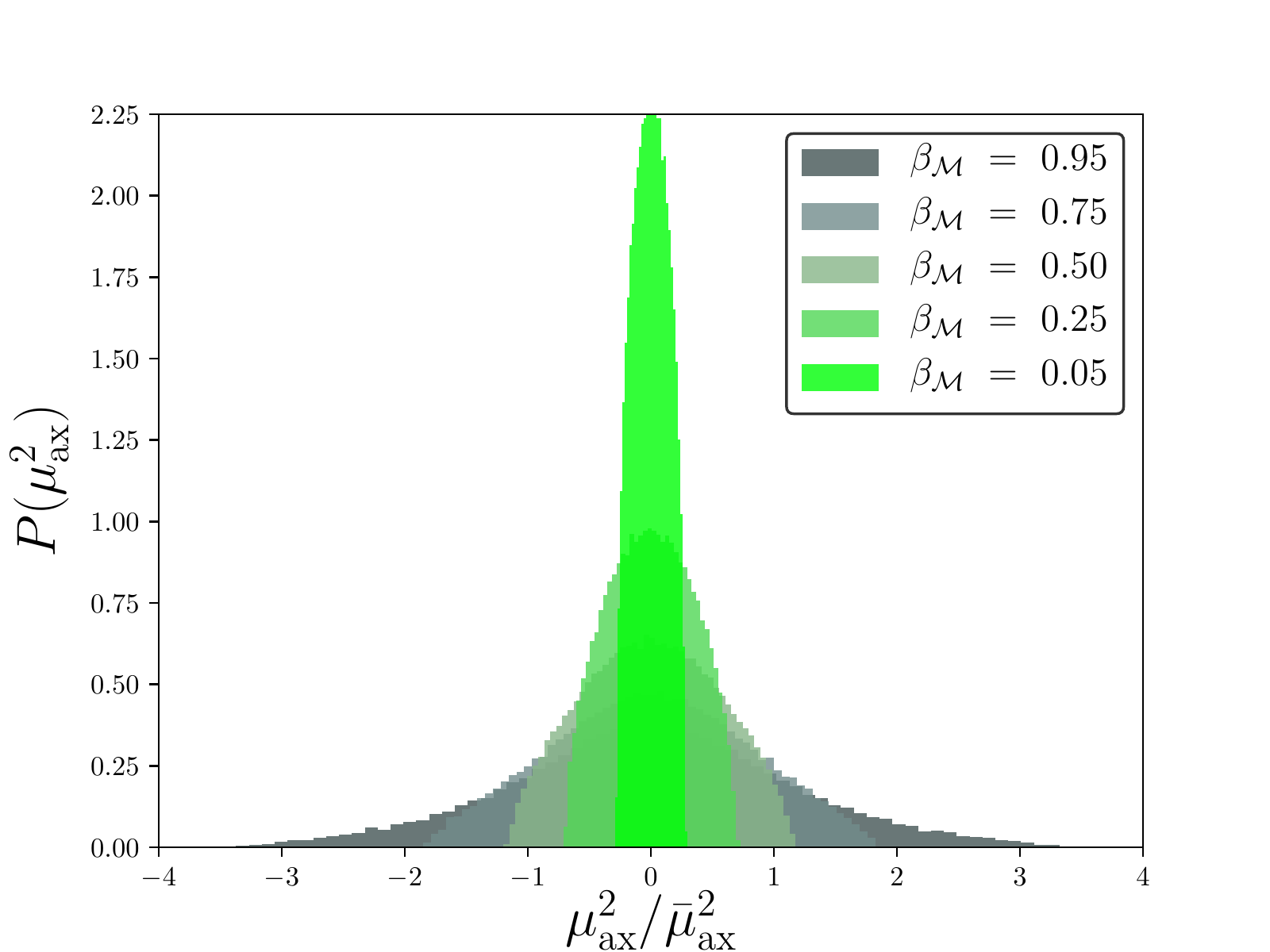}\label{fig:kinetic_spectra}}
  \hfill
  \subfloat[Spiked Wishart class kinetic basis spectrum.]{\includegraphics[width=0.32\textwidth]{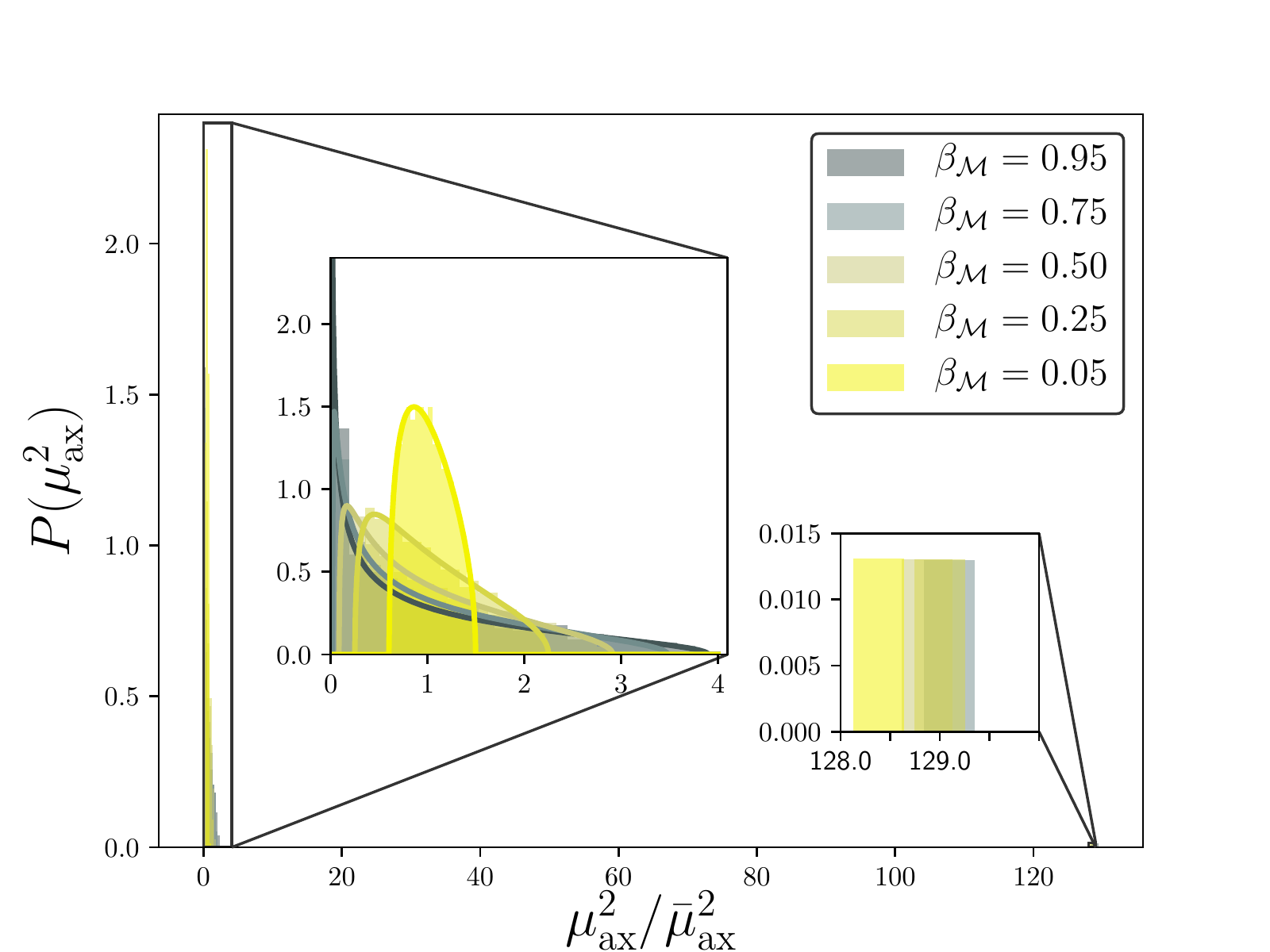}\label{fig:spiked_eigenvalue}}
    \hfill
  \subfloat[Spiked Wishart class mass eigenstate basis spectrum.]{\includegraphics[width=0.32\textwidth]{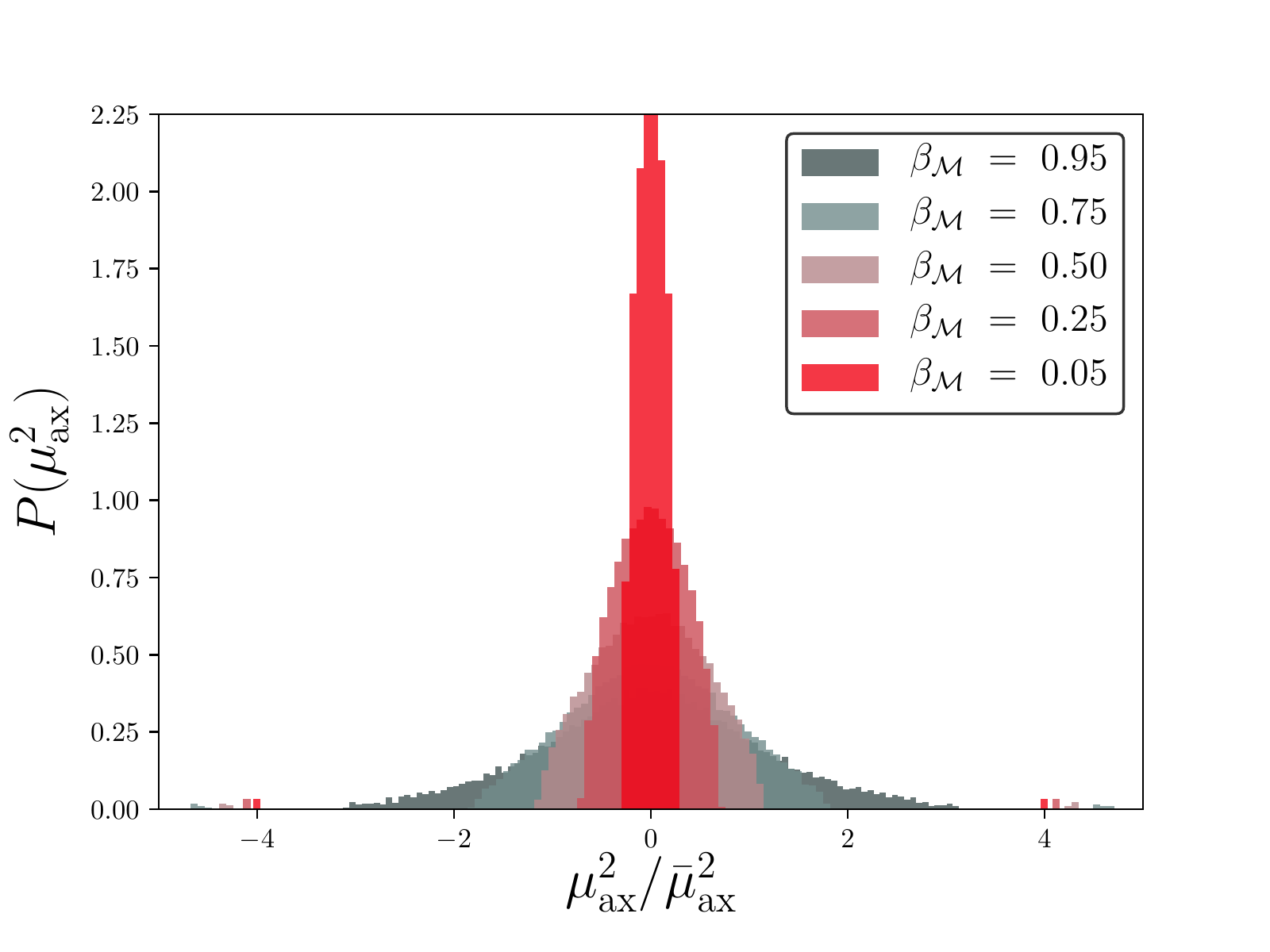}\label{fig:spiked_spectra}}
  \caption{Normalised eigenvalue spectra for axion masses $\mu_{\rm ax}^2$ for different values of the spectrum shaping parameter $\beta_{\mathcal{M}}$ in each Wishart class model in Table~\ref{tab:models}. }
\end{figure*}
\begin{align}
X_{ij} = \frac{1}{N}Y_{ik}^T Y_{kj}\,.
\label{eq:generalmatrix}
\end{align}
This formalism can also be extended in the small field approximation to govern the properties of the axion mass matrix. The mean scale of the axion population defines the phenomenological properties of the fields. The spread of the spectrum is controlled by a shaping index $\beta_{\mathcal{K},\mathcal{M}} \in (0,1]$ which has been shown to have theoretical foundations relating to the total dimension of the moduli space \cite{Easther:2005zr,2017PhRvD..96h3510S}. Below we present a series of RMT inspired models based on charge quantisation and field space alignment considerations.   

{
\renewcommand{\arraystretch}{1.0}
\begin{table*}[t]
\begin{center}
\caption{RMT models considered in this work and extensively covered in Ref.~\cite{2017PhRvD..96h3510S}. Detailed are the initial basis considerations for each effective model along with the relevant sampling procedures and field space diameters. The values of $\sigma$ represent the approximate spread a population or subpopulation of axions would have in each model.}   
\label{tab:models}
\begin{tabular*}{\textwidth}{K{0.145\textwidth}K{0.085\textwidth}K{0.10\textwidth}K{0.25\textwidth}K{0.17\textwidth}K{0.09\textwidth}K{0.09\textwidth}}
\toprule
    \text{$Model$}   & \text{$Alignment$} &\text{$Initial\ Basis$}       & \text{$Sampling$}& \text{$Diameter$}&\text{$Spectra$}&\text{$\sigma$}  \\ 
   \hline
   Mar\v{c}henko-Pastur &N-flation &Kinetic $Eq.~(\ref{eq:nflatbasis2})$ & $B_{ij} \in \mathcal{N}(0,1)$&$D = 2\pi \sqrt{N_{\rm ax}\bar{f_a}^2}.	$& Fig.~\ref{fig:MP_spectra}&$\sim \mathcal{O}(1)$
\\ \hline
White Wishart &Kinetic&Lattice $Eq.~(\ref{eq:multiaxion})$  & $A_{ij}, \ B_{ij} \in \mathcal{N}(0,1)$&$D = 2\pi \sqrt{N_{\rm ax}} f_{a,\rm max}.$&Fig.~\ref{fig:kinetic_spectra} &$\sim (0.1-5)$
\\ \hline
Spiked Wishart& Kinetic &Lattice $Eq.~(\ref{eq:multiaxion})$ & $A_{ij}, \ B_{ij} \in \log_{10}\mathcal{U}(min,max)$&$D = 2\pi \sqrt{N_{\rm ax}} f_{a,\rm max}.$&Fig.~\ref{fig:spiked_spectra} &$\sim (0.1-5)$
\\

\botrule  
\end{tabular*}
\end{center}
\end{table*}
}

The defining features of each of our models we consider are presented in Table~\ref{tab:models}. Each model presents a modest hierarchy in complexity regarding the initial basis and determined mass eigenstate spectrum. Our first model is based on \emph{N-flation} type field alignment \cite{Easther:2005zr} acting as our simplest \emph{strawman} model. The fields in canonical coordinates are defined by the effective field space metric, $\mathcal{K}_{ij} = diag(f^2_a)$. The canonical field ranges in this basis are defined as $\phi_i = \bar{f_a}\theta_i$ where $\bar{f_a}$ is a scaling factor introduced to represent the degenerate decay constant scales arising from the diagonal kinetic matrix. The field space diameter is given by the pythagorean sum over the N-Dimensional hyper-rectangle (see Table~\ref{tab:models}). The resulting mass spectra is displayed in Fig.~\ref{fig:MP_spectra}.

When beginning in the lattice basis a spectrum of decay constants now scale the initial field sampling. It has been shown the kinetic matrix $\mathcal{K}_{ij}$, could belong to the Gaussian orthogonal Wishart ensemble with i.i.d gaussian entries \cite{Bachlechner:2014hsa,Bachlechner:2017hsj}. The entries for the sub-matrices $Y_{ij}$ in Eq.~(\ref{eq:generalmatrix}) composing each of the kinetic and mass matrices are drawn from normal distributions with zero mean and unit variance. The axion decay constant spectrum is given by the limiting Mar\v{c}henko-Pastur law (Fig.~\ref{fig:MP_spectra}). In the mass eigenstate basis the mass spectrum is now rotated by the non trivial rotations between the lattice and kinetic basis. Universality dictates a convergent mass spectrum well modelled by a log-normal distribution with its limited variance, $\sigma$ defined by the bounded nature of the initial Wishart structure. The mass spectrum in this model is presented in Fig.~\ref{fig:kinetic_spectra}. 

If the entries of the sub-matrices are not selected as i.i.d gaussian entries and instead selected from a log-uniform distribution, the resulting matrix of the Wishart form will now reside in a class of rank one spiked Wishart matrices. In the original models \cite{2004math......3022B,2010arXiv1011.1877B,2011arXiv1101.5144M} the matrices are defined by the class, $W_R (\Sigma, M)$ where a single element of the covariance matrix deviates from unity inducing a phase transition in the distribution for the largest eigenvalues. In the limit $N_{\rm ax}\rightarrow \infty$ a bulk region forms supported by the Mar\v{c}henko-Pastur limiting law and one singular eigenvalue is repulsed from the bulk as shown in Fig.~\ref{fig:spiked_eigenvalue}. The approximate order of the singular eigenvalue is $\lambda_{\rm ax}\approx \mathcal{O}(N_{\rm ax})$, which defines the enhancement of the diameter of field space by a factor of $\sqrt{N_{\rm ax}}$ due to the large hierarchy between $f_{a,\rm max}$ and the second largest eigenvalue. This behaviour governs the axion decay constant spectrum in the model (up to canonical normalisation factors) as shown in Fig.~\ref{fig:spiked_spectra}. In the limit $\beta_{\mathcal{M}} = 1$ the mass spectra converges to the white Wishart case. For values of $\beta_{\mathcal{M}} < 1$ the convergent mass spectrum is well modelled by a log-normal distribution plus two positively and negatively logarithmically repulsed regions enhancing the total spectral width.

\subsection{The M-Theory Mass Spectrum}
\label{app:m_theory}
It has been shown in M-theory compactified on $G_2$ manifolds with an absence of fluxes it is possible to stabilise both the moduli and axions in order to realise a spectrum of ultra-light axions in the low energy spectrum of its four-dimensional effective supergravity theory. We follow the explicit realisation of the string axiverse in \cite{Acharya:2006ia,Acharya:2007rc,Acharya:2010zx}. In such models the moduli are stabilised in a non-supersymmetric minima, with all axions pairing up with geometric moduli where all moduli superfields possess PQ symmetries. The superpotential in this model takes the form 
\begin{equation}
W = A_1\phi_1^a e^{ib_1 F_1} + A_2e^{ib_2 F1} + \sum^\infty_{k=3}A_ke^{ib_k F_k}\,,	
\label{eq:superpot}
\end{equation}
with order $\mathcal{O}(1)$ constants, $A_{k}$. The first two terms in Eq.~(\ref{eq:superpot}) come from strong gauge dynamics in the hidden sector using up one combination of axions where $\phi_1$ is a holomorphic composite field made of hidden sector matter fields. In general the rest of the fields present in the summation come from non-perturbative physics such as membrane instantons and serve as a fundamental feature of such compactification models. We only need to consider the higher order correctional terms assumed to be generated from membrane instantons where $b_k = 2\pi I$ and $I\in \mathbb{Z}$ with the gauge kinetic functions, $F_k  = \sum^N_{i=1} N_{K}^iz_i$. It is always possible to find realistic arguments determining the number of non-perturbative effects as larger than the number of axions, $N_{\rm Inst}>N_{\rm ax}$ giving rise to sufficient independent terms in the superpotential. To consider a spectrum of axions we integrate out the moduli and heavy axion combinations such that the relevant effective superpotential becomes 
\begin{equation}
W_{\rm Inst} = \sum^{N}_{i=1} \tilde{\Lambda}_i^3 e^{ib_iF_i}\,,	
\label{eq:general_super}
\end{equation}
where $\tilde{\Lambda}_i$ are the associated mass scales for each non-perturbative effect. The potential now takes the following form, 
\begin{align}
V \approx & F\left( \sum_{i=1}^{N_{\rm ax}} \frac{\partial}{\partial z_i} \sum_{j=1}^{N} \widetilde{\Lambda}_j^3 e^{i b_j F_j}\right) + \text{c.c.} \nonumber\,, \\
\approx & \sum_{i=1}^{N_{\rm ax}} \sum_{j=1}^N \frac{2 F \widetilde{\Lambda}_j^3 b_j N_j^i}{M_{S}} e^{- b_j \sum_k^{N_{\rm ax}} N_j^k s_k} \cos{\left(\sum_{k=1}^{N_{\rm ax}} b_j N_j^k t_k \right)}\,.
\label{eq:mtheorysuper}
\end{align}
In Ref.~\cite{2017PhRvD..96h3510S} it was shown an expansion of the periodic potentials to quadratic order reveals the axion mass matrix, 
\begin{align}
\mathcal{M}_{ij} =& \sum_{k=1}^{N_{\rm ax}} \sum_{r=1}^N \frac{4F \widetilde{\Lambda}_r^3 b_r N_r^k}{M_{S}^3} e^{- b_r \sum_m^{N_{\rm ax}} N_r^m s_m} b_r N_r^i b_r N_r^j\,, \\
=& \sum_{r=1}^N \frac{4 F \widetilde{\Lambda}_r^3 C_r}{M_{S}^3} e^{- S_r} \widetilde{N}_r^i \widetilde{N}_r^j \,,
\label{eq:mtheory_massmatrix}
\end{align}
where $\widetilde{N}_i^j = b_i N_i^j$ is a rectangular matrix of size $(N_{\rm ax}, N)$, $C_r = \sum_k ^{N_{\rm ax}}\widetilde{N}^k_r$ and $S_r = \sum_m^{N_{\rm ax}} \widetilde{N}_r^m s_m$. The dimensions of $\widetilde{N}_i^j$ are controlled by the axion population size, $N_{\rm ax}$ and the number of non-perturbative instantons, N. Using this form the mass matrix can be parameterised as 
\beq
\mathcal{M}_{ij} = \sum_{r=1}^N 4 F \widetilde{\Lambda}_r^3 C_r e^{- S_r} \tilde{N}_r^i \tilde{N}_r^j\,, \label{eq:mmassmatrix}
\eeq
which in terms of sub-matrix structure following the philosophy of Eq.~(\ref{eq:generalmatrix}) gives, 
\beq
\mathcal{M}_{ij} = {1\over N} A_{ir}A_{jr} \label{eq:AxiverseM1}\,.
\eeq
This defines the following form for the sample sub-matrix, 
\beq
A_{ir} = \left(2 \sqrt{F \widetilde{\Lambda}_r^3 C_r} \right) e^{- S_r/2} \widetilde{N}_r^i \, , \label{eq:AxiverseM2}
\eeq                      
where $i,j = 1,\ldots,N_{\rm ax}$ and $r = 1,\ldots,N$, $A_{ir}$ is a rectangular matrix of size $(N_{\rm ax}, N)$ with a normalisation factor ${\sfrac{1}{N}}$.

In order to define our mass scales of interest in the M-theory axiverse consider the general form for the superpotential in Eq.~(\ref{eq:general_super}). $F_i$ represents the gauge kinetic functions which are linear combinations of the moduli superfields, 
\beq
F_i = \sum_k^{n_{\rm ax}} N_i^k z_k = \sum_k^{n_{\rm ax}} N_i^k (t_k + is_k)\,.
\eeq
The generalised volume of the corresponding 3-cycles is calculated from,
\beq
V_{X}^i = \text{Im}(F_i) = \sum_{k=1}^{n_{\rm ax}} N_{i}^k s_k = \frac{1}{2\pi} \sum_{k=1}^{n_{\rm ax}} \widetilde{N}_{i}^k s_k \,.  
\eeq
The geometric moduli are stabilised in terms of a single parameter $\langle V_X\rangle$ which represents the stabilised volume of the three-cycle supporting the hidden sector. In order to realise a GUT in the low energy limit of the theory, at least one of the gauge kinetic functions must give rise to the expected value of the GUT coupling constant, 
\begin{equation}
\alpha_{\rm GUT} = \frac{1}{V_X} \approx \frac{1}{25}\,.
\end{equation} 
The average value of $\langle V_X\rangle$ therefore fixes the mass scales of the spectrum of axions appearing in the visible sector (Fig.~\ref{fig:m-theory_mass}). We parameterise the axion mass distribution in terms of the average value of the three-cycle volume distribution, $\langle V_X \rangle$ via the relationship
 \begin{equation}
 \langle V_X \rangle = \frac{N_{\rm ax} \tilde{N}_{\rm max} \langle s \rangle}{4\pi}\,,
 \label{eqn:cycle_volume}	
 \end{equation}
which contains the parameters we statistically sample to determine the nature of $\sigma$ used throughout the basis of this work (see Section III D of Ref.~\cite{2017PhRvD..96h3510S} for details of the parameters used). The number of parameters and hierarchy of scales involved in the statistical sampling of the three-cycle volume dictate a large spread in the mass eigenstates covering many decades. The nature of universality ensures a convergence to a normal distribution over these scales \cite{2017PhRvD..96h3510S}.

\pagebreak
\bibliography{supperradiance_limitsNotes}

\begin{thebibliography}{127}%
\makeatletter
\providecommand \@ifxundefined [1]{%
 \@ifx{#1\undefined}
}%
\providecommand \@ifnum [1]{%
 \ifnum #1\expandafter \@firstoftwo
 \else \expandafter \@secondoftwo
 \fi
}%
\providecommand \@ifx [1]{%
 \ifx #1\expandafter \@firstoftwo
 \else \expandafter \@secondoftwo
 \fi
}%
\providecommand \natexlab [1]{#1}%
\providecommand \enquote  [1]{``#1''}%
\providecommand \bibnamefont  [1]{#1}%
\providecommand \bibfnamefont [1]{#1}%
\providecommand \citenamefont [1]{#1}%
\providecommand \href@noop [0]{\@secondoftwo}%
\providecommand \href [0]{\begingroup \@sanitize@url \@href}%
\providecommand \@href[1]{\@@startlink{#1}\@@href}%
\providecommand \@@href[1]{\endgroup#1\@@endlink}%
\providecommand \@sanitize@url [0]{\catcode `\\12\catcode `\$12\catcode
  `\&12\catcode `\#12\catcode `\^12\catcode `\_12\catcode `\%12\relax}%
\providecommand \@@startlink[1]{}%
\providecommand \@@endlink[0]{}%
\providecommand \url  [0]{\begingroup\@sanitize@url \@url }%
\providecommand \@url [1]{\endgroup\@href {#1}{\urlprefix }}%
\providecommand \urlprefix  [0]{URL }%
\providecommand \Eprint [0]{\href }%
\providecommand \doibase [0]{http://dx.doi.org/}%
\providecommand \selectlanguage [0]{\@gobble}%
\providecommand \bibinfo  [0]{\@secondoftwo}%
\providecommand \bibfield  [0]{\@secondoftwo}%
\providecommand \translation [1]{[#1]}%
\providecommand \BibitemOpen [0]{}%
\providecommand \bibitemStop [0]{}%
\providecommand \bibitemNoStop [0]{.\EOS\space}%
\providecommand \EOS [0]{\spacefactor3000\relax}%
\providecommand \BibitemShut  [1]{\csname bibitem#1\endcsname}%
\let\auto@bib@innerbib\@empty
\bibitem [{\citenamefont {{Penrose}}(1969)}]{1969NCimR...1..252P}%
  \BibitemOpen
  \bibfield  {author} {\bibinfo {author} {\bibfnamefont {R.}~\bibnamefont
  {{Penrose}}},\ }\bibfield  {title} {\enquote {\bibinfo {title}
  {{Gravitational Collapse: the Role of General Relativity}},}\ }\href@noop {}
  {\bibfield  {journal} {\bibinfo  {journal} {Nuovo Cimento Rivista Serie}\
  }\textbf {\bibinfo {volume} {1}},\ \bibinfo {pages} {252} (\bibinfo {year}
  {1969})}\BibitemShut {NoStop}%
\bibitem [{\citenamefont {{Press}}\ and\ \citenamefont
  {{Teukolsky}}(1972)}]{1972Natur.238..211P}%
  \BibitemOpen
  \bibfield  {author} {\bibinfo {author} {\bibfnamefont {W.~H.}\ \bibnamefont
  {{Press}}}\ and\ \bibinfo {author} {\bibfnamefont {S.~A.}\ \bibnamefont
  {{Teukolsky}}},\ }\bibfield  {title} {\enquote {\bibinfo {title} {{Floating
  Orbits, Superradiant Scattering and the Black-hole Bomb}},}\ }\href {\doibase
  10.1038/238211a0} {\bibfield  {journal} {\bibinfo  {journal} {\nat}\ }\textbf
  {\bibinfo {volume} {238}},\ \bibinfo {pages} {211--212} (\bibinfo {year}
  {1972})}\BibitemShut {NoStop}%
\bibitem [{\citenamefont {{Press}}\ and\ \citenamefont
  {{Teukolsky}}(1973)}]{1973ApJ...185..649P}%
  \BibitemOpen
  \bibfield  {author} {\bibinfo {author} {\bibfnamefont {W.~H.}\ \bibnamefont
  {{Press}}}\ and\ \bibinfo {author} {\bibfnamefont {S.~A.}\ \bibnamefont
  {{Teukolsky}}},\ }\bibfield  {title} {\enquote {\bibinfo {title}
  {{Perturbations of a Rotating Black Hole. II. Dynamical Stability of the Kerr
  Metric}},}\ }\href {\doibase 10.1086/152445} {\bibfield  {journal} {\bibinfo
  {journal} {\apj}\ }\textbf {\bibinfo {volume} {185}},\ \bibinfo {pages}
  {649--674} (\bibinfo {year} {1973})}\BibitemShut {NoStop}%
\bibitem [{\citenamefont {Brito}\ \emph {et~al.}(2015)\citenamefont {Brito},
  \citenamefont {Cardoso},\ and\ \citenamefont {Pani}}]{Brito:2015oca}%
  \BibitemOpen
  \bibfield  {author} {\bibinfo {author} {\bibfnamefont {Richard}\ \bibnamefont
  {Brito}}, \bibinfo {author} {\bibfnamefont {Vitor}\ \bibnamefont {Cardoso}},
  \ and\ \bibinfo {author} {\bibfnamefont {Paolo}\ \bibnamefont {Pani}},\
  }\bibfield  {title} {\enquote {\bibinfo {title} {{Superradiance}},}\ }\href
  {\doibase 10.1007/978-3-319-19000-6} {\bibfield  {journal} {\bibinfo
  {journal} {Lect. Notes Phys.}\ }\textbf {\bibinfo {volume} {906}},\ \bibinfo
  {pages} {pp.1--237} (\bibinfo {year} {2015})},\ \Eprint
  {http://arxiv.org/abs/1501.06570} {arXiv:1501.06570 [gr-qc]} \BibitemShut
  {NoStop}%
\bibitem [{\citenamefont {{Abbott}}\ \emph {et~al.}(2016)\citenamefont
  {{Abbott}}, \citenamefont {{Abbott}}, \citenamefont {{Abbott}}, \citenamefont
  {{Abernathy}}, \citenamefont {{Acernese}}, \citenamefont {{Ackley}},
  \citenamefont {{Adams}}, \citenamefont {{Adams}}, \citenamefont {{Addesso}},
  \citenamefont {{Adhikari}},\ and\ \citenamefont
  {et~al.}}]{2016PhRvL.116f1102A}%
  \BibitemOpen
  \bibfield  {author} {\bibinfo {author} {\bibfnamefont {B.~P.}\ \bibnamefont
  {{Abbott}}}, \bibinfo {author} {\bibfnamefont {R.}~\bibnamefont {{Abbott}}},
  \bibinfo {author} {\bibfnamefont {T.~D.}\ \bibnamefont {{Abbott}}}, \bibinfo
  {author} {\bibfnamefont {M.~R.}\ \bibnamefont {{Abernathy}}}, \bibinfo
  {author} {\bibfnamefont {F.}~\bibnamefont {{Acernese}}}, \bibinfo {author}
  {\bibfnamefont {K.}~\bibnamefont {{Ackley}}}, \bibinfo {author}
  {\bibfnamefont {C.}~\bibnamefont {{Adams}}}, \bibinfo {author} {\bibfnamefont
  {T.}~\bibnamefont {{Adams}}}, \bibinfo {author} {\bibfnamefont
  {P.}~\bibnamefont {{Addesso}}}, \bibinfo {author} {\bibfnamefont {R.~X.}\
  \bibnamefont {{Adhikari}}}, \ and\ \bibinfo {author} {\bibnamefont
  {et~al.}},\ }\bibfield  {title} {\enquote {\bibinfo {title} {{Observation of
  Gravitational Waves from a Binary Black Hole Merger}},}\ }\href {\doibase
  10.1103/PhysRevLett.116.061102} {\bibfield  {journal} {\bibinfo  {journal}
  {Physical Review Letters}\ }\textbf {\bibinfo {volume} {116}},\ \bibinfo
  {eid} {061102} (\bibinfo {year} {2016})},\ \Eprint
  {http://arxiv.org/abs/1602.03837} {arXiv:1602.03837 [gr-qc]} \BibitemShut
  {NoStop}%
\bibitem [{\citenamefont {{Arvanitaki}}\ \emph {et~al.}(2017)\citenamefont
  {{Arvanitaki}}, \citenamefont {{Baryakhtar}}, \citenamefont {{Dimopoulos}},
  \citenamefont {{Dubovsky}},\ and\ \citenamefont
  {{Lasenby}}}]{2017PhRvD..95d3001A}%
  \BibitemOpen
  \bibfield  {author} {\bibinfo {author} {\bibfnamefont {A.}~\bibnamefont
  {{Arvanitaki}}}, \bibinfo {author} {\bibfnamefont {M.}~\bibnamefont
  {{Baryakhtar}}}, \bibinfo {author} {\bibfnamefont {S.}~\bibnamefont
  {{Dimopoulos}}}, \bibinfo {author} {\bibfnamefont {S.}~\bibnamefont
  {{Dubovsky}}}, \ and\ \bibinfo {author} {\bibfnamefont {R.}~\bibnamefont
  {{Lasenby}}},\ }\bibfield  {title} {\enquote {\bibinfo {title} {{Black hole
  mergers and the QCD axion at Advanced LIGO}},}\ }\href {\doibase
  10.1103/PhysRevD.95.043001} {\bibfield  {journal} {\bibinfo  {journal}
  {\prd}\ }\textbf {\bibinfo {volume} {95}},\ \bibinfo {eid} {043001} (\bibinfo
  {year} {2017})},\ \Eprint {http://arxiv.org/abs/1604.03958} {arXiv:1604.03958
  [hep-ph]} \BibitemShut {NoStop}%
\bibitem [{\citenamefont {{Baryakhtar}}\ \emph {et~al.}(2017)\citenamefont
  {{Baryakhtar}}, \citenamefont {{Lasenby}},\ and\ \citenamefont
  {{Teo}}}]{2017PhRvD..96c5019B}%
  \BibitemOpen
  \bibfield  {author} {\bibinfo {author} {\bibfnamefont {M.}~\bibnamefont
  {{Baryakhtar}}}, \bibinfo {author} {\bibfnamefont {R.}~\bibnamefont
  {{Lasenby}}}, \ and\ \bibinfo {author} {\bibfnamefont {M.}~\bibnamefont
  {{Teo}}},\ }\bibfield  {title} {\enquote {\bibinfo {title} {{Black hole
  superradiance signatures of ultralight vectors}},}\ }\href {\doibase
  10.1103/PhysRevD.96.035019} {\bibfield  {journal} {\bibinfo  {journal}
  {\prd}\ }\textbf {\bibinfo {volume} {96}},\ \bibinfo {eid} {035019} (\bibinfo
  {year} {2017})},\ \Eprint {http://arxiv.org/abs/1704.05081} {arXiv:1704.05081
  [hep-ph]} \BibitemShut {NoStop}%
\bibitem [{\citenamefont {{Brito}}\ \emph {et~al.}(2017)\citenamefont
  {{Brito}}, \citenamefont {{Ghosh}}, \citenamefont {{Barausse}}, \citenamefont
  {{Berti}}, \citenamefont {{Cardoso}}, \citenamefont {{Dvorkin}},
  \citenamefont {{Klein}},\ and\ \citenamefont {{Pani}}}]{2017PhRvD..96f4050B}%
  \BibitemOpen
  \bibfield  {author} {\bibinfo {author} {\bibfnamefont {R.}~\bibnamefont
  {{Brito}}}, \bibinfo {author} {\bibfnamefont {S.}~\bibnamefont {{Ghosh}}},
  \bibinfo {author} {\bibfnamefont {E.}~\bibnamefont {{Barausse}}}, \bibinfo
  {author} {\bibfnamefont {E.}~\bibnamefont {{Berti}}}, \bibinfo {author}
  {\bibfnamefont {V.}~\bibnamefont {{Cardoso}}}, \bibinfo {author}
  {\bibfnamefont {I.}~\bibnamefont {{Dvorkin}}}, \bibinfo {author}
  {\bibfnamefont {A.}~\bibnamefont {{Klein}}}, \ and\ \bibinfo {author}
  {\bibfnamefont {P.}~\bibnamefont {{Pani}}},\ }\bibfield  {title} {\enquote
  {\bibinfo {title} {{Gravitational wave searches for ultralight bosons with
  LIGO and LISA}},}\ }\href {\doibase 10.1103/PhysRevD.96.064050} {\bibfield
  {journal} {\bibinfo  {journal} {\prd}\ }\textbf {\bibinfo {volume} {96}},\
  \bibinfo {eid} {064050} (\bibinfo {year} {2017})},\ \Eprint
  {http://arxiv.org/abs/1706.06311} {arXiv:1706.06311 [gr-qc]} \BibitemShut
  {NoStop}%
\bibitem [{\citenamefont {{Cardoso}}\ \emph {et~al.}(2018)\citenamefont
  {{Cardoso}}, \citenamefont {{Dias}}, \citenamefont {{Hartnett}},
  \citenamefont {{Middleton}}, \citenamefont {{Pani}},\ and\ \citenamefont
  {{Santos}}}]{2018arXiv180101420C}%
  \BibitemOpen
  \bibfield  {author} {\bibinfo {author} {\bibfnamefont {V.}~\bibnamefont
  {{Cardoso}}}, \bibinfo {author} {\bibfnamefont {{\'O}.~J.~C.}\ \bibnamefont
  {{Dias}}}, \bibinfo {author} {\bibfnamefont {G.~S.}\ \bibnamefont
  {{Hartnett}}}, \bibinfo {author} {\bibfnamefont {M.}~\bibnamefont
  {{Middleton}}}, \bibinfo {author} {\bibfnamefont {P.}~\bibnamefont {{Pani}}},
  \ and\ \bibinfo {author} {\bibfnamefont {J.~E.}\ \bibnamefont {{Santos}}},\
  }\bibfield  {title} {\enquote {\bibinfo {title} {{Constraining the mass of
  dark photons and axion-like particles through black-hole superradiance}},}\
  }\href@noop {} {\bibfield  {journal} {\bibinfo  {journal} {ArXiv e-prints}\ }
  (\bibinfo {year} {2018})},\ \Eprint {http://arxiv.org/abs/1801.01420}
  {arXiv:1801.01420 [gr-qc]} \BibitemShut {NoStop}%
\bibitem [{\citenamefont {Hannuksela}\ \emph {et~al.}(2018)\citenamefont
  {Hannuksela}, \citenamefont {Brito}, \citenamefont {Berti},\ and\
  \citenamefont {Li}}]{Hannuksela:2018izj}%
  \BibitemOpen
  \bibfield  {author} {\bibinfo {author} {\bibfnamefont {Otto~A.}\ \bibnamefont
  {Hannuksela}}, \bibinfo {author} {\bibfnamefont {Richard}\ \bibnamefont
  {Brito}}, \bibinfo {author} {\bibfnamefont {Emanuele}\ \bibnamefont {Berti}},
  \ and\ \bibinfo {author} {\bibfnamefont {Tjonnie G.~F.}\ \bibnamefont {Li}},\
  }\bibfield  {title} {\enquote {\bibinfo {title} {{Probing the existence of
  ultralight bosons with a single gravitational-wave measurement}},}\
  }\href@noop {} {\  (\bibinfo {year} {2018})},\ \Eprint
  {http://arxiv.org/abs/1804.09659} {arXiv:1804.09659 [astro-ph.HE]}
  \BibitemShut {NoStop}%
\bibitem [{\citenamefont {Arvanitaki}\ and\ \citenamefont
  {Dubovsky}(2011)}]{Arvanitaki:2010sy}%
  \BibitemOpen
  \bibfield  {author} {\bibinfo {author} {\bibfnamefont {Asimina}\ \bibnamefont
  {Arvanitaki}}\ and\ \bibinfo {author} {\bibfnamefont {Sergei}\ \bibnamefont
  {Dubovsky}},\ }\bibfield  {title} {\enquote {\bibinfo {title} {{Exploring the
  String Axiverse with Precision Black Hole Physics}},}\ }\href {\doibase
  10.1103/PhysRevD.83.044026} {\bibfield  {journal} {\bibinfo  {journal} {Phys.
  Rev.}\ }\textbf {\bibinfo {volume} {D83}},\ \bibinfo {pages} {044026}
  (\bibinfo {year} {2011})},\ \Eprint {http://arxiv.org/abs/1004.3558}
  {arXiv:1004.3558 [hep-th]} \BibitemShut {NoStop}%
\bibitem [{\citenamefont {Baryakhtar}\ \emph {et~al.}(2017)\citenamefont
  {Baryakhtar}, \citenamefont {Lasenby},\ and\ \citenamefont
  {Teo}}]{Baryakhtar:2017ngi}%
  \BibitemOpen
  \bibfield  {author} {\bibinfo {author} {\bibfnamefont {Masha}\ \bibnamefont
  {Baryakhtar}}, \bibinfo {author} {\bibfnamefont {Robert}\ \bibnamefont
  {Lasenby}}, \ and\ \bibinfo {author} {\bibfnamefont {Mae}\ \bibnamefont
  {Teo}},\ }\bibfield  {title} {\enquote {\bibinfo {title} {{Black Hole
  Superradiance Signatures of Ultralight Vectors}},}\ }\href {\doibase
  10.1103/PhysRevD.96.035019} {\bibfield  {journal} {\bibinfo  {journal} {Phys.
  Rev.}\ }\textbf {\bibinfo {volume} {D96}},\ \bibinfo {pages} {035019}
  (\bibinfo {year} {2017})},\ \Eprint {http://arxiv.org/abs/1704.05081}
  {arXiv:1704.05081 [hep-ph]} \BibitemShut {NoStop}%
\bibitem [{\citenamefont {Brito}\ \emph {et~al.}(2017)\citenamefont {Brito},
  \citenamefont {Ghosh}, \citenamefont {Barausse}, \citenamefont {Berti},
  \citenamefont {Cardoso}, \citenamefont {Dvorkin}, \citenamefont {Klein},\
  and\ \citenamefont {Pani}}]{Brito:2017zvb}%
  \BibitemOpen
  \bibfield  {author} {\bibinfo {author} {\bibfnamefont {Richard}\ \bibnamefont
  {Brito}}, \bibinfo {author} {\bibfnamefont {Shrobana}\ \bibnamefont {Ghosh}},
  \bibinfo {author} {\bibfnamefont {Enrico}\ \bibnamefont {Barausse}}, \bibinfo
  {author} {\bibfnamefont {Emanuele}\ \bibnamefont {Berti}}, \bibinfo {author}
  {\bibfnamefont {Vitor}\ \bibnamefont {Cardoso}}, \bibinfo {author}
  {\bibfnamefont {Irina}\ \bibnamefont {Dvorkin}}, \bibinfo {author}
  {\bibfnamefont {Antoine}\ \bibnamefont {Klein}}, \ and\ \bibinfo {author}
  {\bibfnamefont {Paolo}\ \bibnamefont {Pani}},\ }\bibfield  {title} {\enquote
  {\bibinfo {title} {{Gravitational wave searches for ultralight bosons with
  LIGO and LISA}},}\ }\href {\doibase 10.1103/PhysRevD.96.064050} {\bibfield
  {journal} {\bibinfo  {journal} {Phys. Rev.}\ }\textbf {\bibinfo {volume}
  {D96}},\ \bibinfo {pages} {064050} (\bibinfo {year} {2017})},\ \Eprint
  {http://arxiv.org/abs/1706.06311} {arXiv:1706.06311 [gr-qc]} \BibitemShut
  {NoStop}%
\bibitem [{\citenamefont {Baumann}\ \emph {et~al.}(2018)\citenamefont
  {Baumann}, \citenamefont {Chia},\ and\ \citenamefont
  {Porto}}]{Baumann:2018vus}%
  \BibitemOpen
  \bibfield  {author} {\bibinfo {author} {\bibfnamefont {Daniel}\ \bibnamefont
  {Baumann}}, \bibinfo {author} {\bibfnamefont {Horng~Sheng}\ \bibnamefont
  {Chia}}, \ and\ \bibinfo {author} {\bibfnamefont {Rafael~A.}\ \bibnamefont
  {Porto}},\ }\bibfield  {title} {\enquote {\bibinfo {title} {{Probing
  Ultralight Bosons with Binary Black Holes}},}\ }\href@noop {} {\  (\bibinfo
  {year} {2018})},\ \Eprint {http://arxiv.org/abs/1804.03208} {arXiv:1804.03208
  [gr-qc]} \BibitemShut {NoStop}%
\bibitem [{\citenamefont {{Brito}}\ \emph {et~al.}(2015)\citenamefont
  {{Brito}}, \citenamefont {{Cardoso}},\ and\ \citenamefont
  {{Pani}}}]{2015CQGra..32m4001B}%
  \BibitemOpen
  \bibfield  {author} {\bibinfo {author} {\bibfnamefont {R.}~\bibnamefont
  {{Brito}}}, \bibinfo {author} {\bibfnamefont {V.}~\bibnamefont {{Cardoso}}},
  \ and\ \bibinfo {author} {\bibfnamefont {P.}~\bibnamefont {{Pani}}},\
  }\bibfield  {title} {\enquote {\bibinfo {title} {{Black holes as particle
  detectors: evolution of superradiant instabilities}},}\ }\href {\doibase
  10.1088/0264-9381/32/13/134001} {\bibfield  {journal} {\bibinfo  {journal}
  {Classical and Quantum Gravity}\ }\textbf {\bibinfo {volume} {32}},\ \bibinfo
  {eid} {134001} (\bibinfo {year} {2015})},\ \Eprint
  {http://arxiv.org/abs/1411.0686} {arXiv:1411.0686 [gr-qc]} \BibitemShut
  {NoStop}%
\bibitem [{\citenamefont {{Yoshino}}\ and\ \citenamefont
  {{Kodama}}(2014)}]{2014PTEP.2014d3E02Y}%
  \BibitemOpen
  \bibfield  {author} {\bibinfo {author} {\bibfnamefont {H.}~\bibnamefont
  {{Yoshino}}}\ and\ \bibinfo {author} {\bibfnamefont {H.}~\bibnamefont
  {{Kodama}}},\ }\bibfield  {title} {\enquote {\bibinfo {title} {{Gravitational
  radiation from an axion cloud around a black hole: Superradiant phase}},}\
  }\href {\doibase 10.1093/ptep/ptu029} {\bibfield  {journal} {\bibinfo
  {journal} {Progress of Theoretical and Experimental Physics}\ }\textbf
  {\bibinfo {volume} {2014}},\ \bibinfo {eid} {043E02} (\bibinfo {year}
  {2014})},\ \Eprint {http://arxiv.org/abs/1312.2326} {arXiv:1312.2326 [gr-qc]}
  \BibitemShut {NoStop}%
\bibitem [{\citenamefont {{Pani}}\ \emph {et~al.}(2012)\citenamefont {{Pani}},
  \citenamefont {{Cardoso}}, \citenamefont {{Gualtieri}}, \citenamefont
  {{Berti}},\ and\ \citenamefont {{Ishibashi}}}]{2012PhRvL.109m1102P}%
  \BibitemOpen
  \bibfield  {author} {\bibinfo {author} {\bibfnamefont {P.}~\bibnamefont
  {{Pani}}}, \bibinfo {author} {\bibfnamefont {V.}~\bibnamefont {{Cardoso}}},
  \bibinfo {author} {\bibfnamefont {L.}~\bibnamefont {{Gualtieri}}}, \bibinfo
  {author} {\bibfnamefont {E.}~\bibnamefont {{Berti}}}, \ and\ \bibinfo
  {author} {\bibfnamefont {A.}~\bibnamefont {{Ishibashi}}},\ }\bibfield
  {title} {\enquote {\bibinfo {title} {{Black-Hole Bombs and Photon-Mass
  Bounds}},}\ }\href {\doibase 10.1103/PhysRevLett.109.131102} {\bibfield
  {journal} {\bibinfo  {journal} {\prl}\ }\textbf {\bibinfo {volume} {109}},\
  \bibinfo {eid} {131102} (\bibinfo {year} {2012})},\ \Eprint
  {http://arxiv.org/abs/1209.0465} {arXiv:1209.0465 [gr-qc]} \BibitemShut
  {NoStop}%
\bibitem [{\citenamefont {{Kodama}}(2008)}]{2008PThPS.172...11K}%
  \BibitemOpen
  \bibfield  {author} {\bibinfo {author} {\bibfnamefont {H.}~\bibnamefont
  {{Kodama}}},\ }\bibfield  {title} {\enquote {\bibinfo {title} {{Superradiance
  and Instability of Black Holes}},}\ }\href {\doibase 10.1143/PTPS.172.11}
  {\bibfield  {journal} {\bibinfo  {journal} {Progress of Theoretical Physics
  Supplement}\ }\textbf {\bibinfo {volume} {172}},\ \bibinfo {pages} {11--20}
  (\bibinfo {year} {2008})},\ \Eprint {http://arxiv.org/abs/0711.4184}
  {arXiv:0711.4184 [hep-th]} \BibitemShut {NoStop}%
\bibitem [{\citenamefont {Arvanitaki}\ \emph {et~al.}(2010)\citenamefont
  {Arvanitaki}, \citenamefont {Dimopoulos}, \citenamefont {Dubovsky},
  \citenamefont {Kaloper},\ and\ \citenamefont {March-Russell}}]{axiverse}%
  \BibitemOpen
  \bibfield  {author} {\bibinfo {author} {\bibfnamefont {Asimina}\ \bibnamefont
  {Arvanitaki}}, \bibinfo {author} {\bibfnamefont {Savas}\ \bibnamefont
  {Dimopoulos}}, \bibinfo {author} {\bibfnamefont {Sergei}\ \bibnamefont
  {Dubovsky}}, \bibinfo {author} {\bibfnamefont {Nemanja}\ \bibnamefont
  {Kaloper}}, \ and\ \bibinfo {author} {\bibfnamefont {John}\ \bibnamefont
  {March-Russell}},\ }\bibfield  {title} {\enquote {\bibinfo {title} {{String
  Axiverse}},}\ }\href {\doibase 10.1103/PhysRevD.81.123530} {\bibfield
  {journal} {\bibinfo  {journal} {Phys. Rev.}\ }\textbf {\bibinfo {volume}
  {D81}},\ \bibinfo {pages} {123530} (\bibinfo {year} {2010})},\ \Eprint
  {http://arxiv.org/abs/0905.4720} {arXiv:0905.4720 [hep-th]} \BibitemShut
  {NoStop}%
\bibitem [{\citenamefont {{Arvanitaki}}\ \emph {et~al.}(2015)\citenamefont
  {{Arvanitaki}}, \citenamefont {{Baryakhtar}},\ and\ \citenamefont
  {{Huang}}}]{2015PhRvD..91h4011A}%
  \BibitemOpen
  \bibfield  {author} {\bibinfo {author} {\bibfnamefont {A.}~\bibnamefont
  {{Arvanitaki}}}, \bibinfo {author} {\bibfnamefont {M.}~\bibnamefont
  {{Baryakhtar}}}, \ and\ \bibinfo {author} {\bibfnamefont {X.}~\bibnamefont
  {{Huang}}},\ }\bibfield  {title} {\enquote {\bibinfo {title} {{Discovering
  the QCD axion with black holes and gravitational waves}},}\ }\href {\doibase
  10.1103/PhysRevD.91.084011} {\bibfield  {journal} {\bibinfo  {journal}
  {\prd}\ }\textbf {\bibinfo {volume} {91}},\ \bibinfo {eid} {084011} (\bibinfo
  {year} {2015})},\ \Eprint {http://arxiv.org/abs/1411.2263} {arXiv:1411.2263
  [hep-ph]} \BibitemShut {NoStop}%
\bibitem [{\citenamefont {Peccei}\ and\ \citenamefont
  {Quinn}(1977)}]{pecceiquinn1977}%
  \BibitemOpen
  \bibfield  {author} {\bibinfo {author} {\bibfnamefont {R.D.}\ \bibnamefont
  {Peccei}}\ and\ \bibinfo {author} {\bibfnamefont {Helen~R.}\ \bibnamefont
  {Quinn}},\ }\bibfield  {title} {\enquote {\bibinfo {title} {{CP Conservation
  in the Presence of Instantons}},}\ }\href {\doibase
  10.1103/PhysRevLett.38.1440} {\bibfield  {journal} {\bibinfo  {journal}
  {\prl}\ }\textbf {\bibinfo {volume} {38}},\ \bibinfo {pages} {1440--1443}
  (\bibinfo {year} {1977})}\BibitemShut {NoStop}%
\bibitem [{\citenamefont {Weinberg}(1978)}]{weinberg1978}%
  \BibitemOpen
  \bibfield  {author} {\bibinfo {author} {\bibfnamefont {Steven}\ \bibnamefont
  {Weinberg}},\ }\bibfield  {title} {\enquote {\bibinfo {title} {{A New Light
  Boson?}}}\ }\href {\doibase 10.1103/PhysRevLett.40.223} {\bibfield  {journal}
  {\bibinfo  {journal} {\prl}\ }\textbf {\bibinfo {volume} {40}},\ \bibinfo
  {pages} {223--226} (\bibinfo {year} {1978})}\BibitemShut {NoStop}%
\bibitem [{\citenamefont {Wilczek}(1978)}]{wilczek1978}%
  \BibitemOpen
  \bibfield  {author} {\bibinfo {author} {\bibfnamefont {Frank}\ \bibnamefont
  {Wilczek}},\ }\bibfield  {title} {\enquote {\bibinfo {title} {{Problem of
  Strong p and t Invariance in the Presence of Instantons}},}\ }\href {\doibase
  10.1103/PhysRevLett.40.279} {\bibfield  {journal} {\bibinfo  {journal}
  {\prl}\ }\textbf {\bibinfo {volume} {40}},\ \bibinfo {pages} {279--282}
  (\bibinfo {year} {1978})}\BibitemShut {NoStop}%
\bibitem [{\citenamefont {{Press}}\ \emph {et~al.}(1990)\citenamefont
  {{Press}}, \citenamefont {{Ryden}},\ and\ \citenamefont
  {{Spergel}}}]{1990PhRvL..64.1084P}%
  \BibitemOpen
  \bibfield  {author} {\bibinfo {author} {\bibfnamefont {W.~H.}\ \bibnamefont
  {{Press}}}, \bibinfo {author} {\bibfnamefont {B.~S.}\ \bibnamefont
  {{Ryden}}}, \ and\ \bibinfo {author} {\bibfnamefont {D.~N.}\ \bibnamefont
  {{Spergel}}},\ }\bibfield  {title} {\enquote {\bibinfo {title} {{Single
  mechanism for generating large-scale structure and providing dark missing
  matter}},}\ }\href {\doibase 10.1103/PhysRevLett.64.1084} {\bibfield
  {journal} {\bibinfo  {journal} {\prl}\ }\textbf {\bibinfo {volume} {64}},\
  \bibinfo {pages} {1084--1087} (\bibinfo {year} {1990})}\BibitemShut {NoStop}%
\bibitem [{\citenamefont {Hu}\ \emph {et~al.}(2000)\citenamefont {Hu},
  \citenamefont {Barkana},\ and\ \citenamefont {Gruzinov}}]{hu2000}%
  \BibitemOpen
  \bibfield  {author} {\bibinfo {author} {\bibfnamefont {Wayne}\ \bibnamefont
  {Hu}}, \bibinfo {author} {\bibfnamefont {Rennan}\ \bibnamefont {Barkana}}, \
  and\ \bibinfo {author} {\bibfnamefont {Andrei}\ \bibnamefont {Gruzinov}},\
  }\bibfield  {title} {\enquote {\bibinfo {title} {{Cold and fuzzy dark
  matter}},}\ }\href {\doibase 10.1103/PhysRevLett.85.1158} {\bibfield
  {journal} {\bibinfo  {journal} {\prl}\ }\textbf {\bibinfo {volume} {85}},\
  \bibinfo {pages} {1158--1161} (\bibinfo {year} {2000})},\ \Eprint
  {http://arxiv.org/abs/astro-ph/0003365} {astro-ph/0003365} \BibitemShut
  {NoStop}%
\bibitem [{\citenamefont {{Marsh}}\ and\ \citenamefont
  {{Silk}}(2014)}]{Marsh:2013ywa}%
  \BibitemOpen
  \bibfield  {author} {\bibinfo {author} {\bibfnamefont {D.~J.~E.}\
  \bibnamefont {{Marsh}}}\ and\ \bibinfo {author} {\bibfnamefont
  {J.}~\bibnamefont {{Silk}}},\ }\bibfield  {title} {\enquote {\bibinfo {title}
  {{A model for halo formation with axion mixed dark matter}},}\ }\href
  {\doibase 10.1093/mnras/stt2079} {\bibfield  {journal} {\bibinfo  {journal}
  {\mnras}\ }\textbf {\bibinfo {volume} {437}},\ \bibinfo {pages} {2652--2663}
  (\bibinfo {year} {2014})},\ \Eprint {http://arxiv.org/abs/1307.1705}
  {arXiv:1307.1705 [astro-ph.CO]} \BibitemShut {NoStop}%
\bibitem [{\citenamefont {{Schive}}\ \emph {et~al.}(2014)\citenamefont
  {{Schive}}, \citenamefont {{Chiueh}},\ and\ \citenamefont
  {{Broadhurst}}}]{2014NatPh..10..496S}%
  \BibitemOpen
  \bibfield  {author} {\bibinfo {author} {\bibfnamefont {H.-Y.}\ \bibnamefont
  {{Schive}}}, \bibinfo {author} {\bibfnamefont {T.}~\bibnamefont {{Chiueh}}},
  \ and\ \bibinfo {author} {\bibfnamefont {T.}~\bibnamefont {{Broadhurst}}},\
  }\bibfield  {title} {\enquote {\bibinfo {title} {{Cosmic structure as the
  quantum interference of a coherent dark wave}},}\ }\href {\doibase
  10.1038/nphys2996} {\bibfield  {journal} {\bibinfo  {journal} {Nature
  Physics}\ }\textbf {\bibinfo {volume} {10}},\ \bibinfo {pages} {496--499}
  (\bibinfo {year} {2014})},\ \Eprint {http://arxiv.org/abs/1406.6586}
  {arXiv:1406.6586} \BibitemShut {NoStop}%
\bibitem [{\citenamefont {{Hui}}\ \emph {et~al.}(2017)\citenamefont {{Hui}},
  \citenamefont {{Ostriker}}, \citenamefont {{Tremaine}},\ and\ \citenamefont
  {{Witten}}}]{2017PhRvD..95d3541H}%
  \BibitemOpen
  \bibfield  {author} {\bibinfo {author} {\bibfnamefont {L.}~\bibnamefont
  {{Hui}}}, \bibinfo {author} {\bibfnamefont {J.~P.}\ \bibnamefont
  {{Ostriker}}}, \bibinfo {author} {\bibfnamefont {S.}~\bibnamefont
  {{Tremaine}}}, \ and\ \bibinfo {author} {\bibfnamefont {E.}~\bibnamefont
  {{Witten}}},\ }\bibfield  {title} {\enquote {\bibinfo {title} {{Ultralight
  scalars as cosmological dark matter}},}\ }\href {\doibase
  10.1103/PhysRevD.95.043541} {\bibfield  {journal} {\bibinfo  {journal}
  {\prd}\ }\textbf {\bibinfo {volume} {95}},\ \bibinfo {eid} {043541} (\bibinfo
  {year} {2017})},\ \Eprint {http://arxiv.org/abs/1610.08297}
  {arXiv:1610.08297} \BibitemShut {NoStop}%
\bibitem [{\citenamefont {Acharya}\ \emph {et~al.}(2010)\citenamefont
  {Acharya}, \citenamefont {Bobkov},\ and\ \citenamefont
  {Kumar}}]{Acharya:2010zx}%
  \BibitemOpen
  \bibfield  {author} {\bibinfo {author} {\bibfnamefont {Bobby~Samir}\
  \bibnamefont {Acharya}}, \bibinfo {author} {\bibfnamefont {Konstantin}\
  \bibnamefont {Bobkov}}, \ and\ \bibinfo {author} {\bibfnamefont {Piyush}\
  \bibnamefont {Kumar}},\ }\bibfield  {title} {\enquote {\bibinfo {title} {{An
  M Theory Solution to the Strong CP Problem and Constraints on the
  Axiverse}},}\ }\href {\doibase 10.1007/JHEP11(2010)105} {\bibfield  {journal}
  {\bibinfo  {journal} {JHEP}\ }\textbf {\bibinfo {volume} {11}},\ \bibinfo
  {pages} {105} (\bibinfo {year} {2010})},\ \Eprint
  {http://arxiv.org/abs/1004.5138} {arXiv:1004.5138 [hep-th]} \BibitemShut
  {NoStop}%
\bibitem [{\citenamefont {{Stott}}\ \emph {et~al.}(2017)\citenamefont
  {{Stott}}, \citenamefont {{Marsh}}, \citenamefont {{Pongkitivanichkul}},
  \citenamefont {{Price}},\ and\ \citenamefont
  {{Acharya}}}]{2017PhRvD..96h3510S}%
  \BibitemOpen
  \bibfield  {author} {\bibinfo {author} {\bibfnamefont {M.~J.}\ \bibnamefont
  {{Stott}}}, \bibinfo {author} {\bibfnamefont {D.~J.~E.}\ \bibnamefont
  {{Marsh}}}, \bibinfo {author} {\bibfnamefont {C.}~\bibnamefont
  {{Pongkitivanichkul}}}, \bibinfo {author} {\bibfnamefont {L.~C.}\
  \bibnamefont {{Price}}}, \ and\ \bibinfo {author} {\bibfnamefont {B.~S.}\
  \bibnamefont {{Acharya}}},\ }\bibfield  {title} {\enquote {\bibinfo {title}
  {{Spectrum of the axion dark sector}},}\ }\href {\doibase
  10.1103/PhysRevD.96.083510} {\bibfield  {journal} {\bibinfo  {journal}
  {\prd}\ }\textbf {\bibinfo {volume} {96}},\ \bibinfo {eid} {083510} (\bibinfo
  {year} {2017})},\ \Eprint {http://arxiv.org/abs/1706.03236}
  {arXiv:1706.03236} \BibitemShut {NoStop}%
\bibitem [{\citenamefont {{Witten}}(1984)}]{1984PhLB..149..351W}%
  \BibitemOpen
  \bibfield  {author} {\bibinfo {author} {\bibfnamefont {E.}~\bibnamefont
  {{Witten}}},\ }\bibfield  {title} {\enquote {\bibinfo {title} {{Some
  properties of O(32) superstrings}},}\ }\href {\doibase
  10.1016/0370-2693(84)90422-2} {\bibfield  {journal} {\bibinfo  {journal}
  {Physics Letters B}\ }\textbf {\bibinfo {volume} {149}},\ \bibinfo {pages}
  {351--356} (\bibinfo {year} {1984})}\BibitemShut {NoStop}%
\bibitem [{\citenamefont {Svrcek}\ and\ \citenamefont
  {Witten}(2006)}]{Svrcek:2006yi}%
  \BibitemOpen
  \bibfield  {author} {\bibinfo {author} {\bibfnamefont {Peter}\ \bibnamefont
  {Svrcek}}\ and\ \bibinfo {author} {\bibfnamefont {Edward}\ \bibnamefont
  {Witten}},\ }\bibfield  {title} {\enquote {\bibinfo {title} {{Axions In
  String Theory}},}\ }\href {\doibase 10.1088/1126-6708/2006/06/051} {\bibfield
   {journal} {\bibinfo  {journal} {JHEP}\ }\textbf {\bibinfo {volume} {06}},\
  \bibinfo {pages} {051} (\bibinfo {year} {2006})},\ \Eprint
  {http://arxiv.org/abs/hep-th/0605206} {arXiv:hep-th/0605206 [hep-th]}
  \BibitemShut {NoStop}%
\bibitem [{\citenamefont {{Conlon}}(2006)}]{2006JHEP...05..078C}%
  \BibitemOpen
  \bibfield  {author} {\bibinfo {author} {\bibfnamefont {J.~P.}\ \bibnamefont
  {{Conlon}}},\ }\bibfield  {title} {\enquote {\bibinfo {title} {{The QCD axion
  and moduli stabilisation}},}\ }\href {\doibase 10.1088/1126-6708/2006/05/078}
  {\bibfield  {journal} {\bibinfo  {journal} {Journal of High Energy Physics}\
  }\textbf {\bibinfo {volume} {5}},\ \bibinfo {eid} {078} (\bibinfo {year}
  {2006})},\ \Eprint {http://arxiv.org/abs/hep-th/0602233} {hep-th/0602233}
  \BibitemShut {NoStop}%
\bibitem [{\citenamefont {{Cicoli}}\ \emph {et~al.}(2012)\citenamefont
  {{Cicoli}}, \citenamefont {{Goodsell}},\ and\ \citenamefont
  {{Ringwald}}}]{2012JHEP...10..146C}%
  \BibitemOpen
  \bibfield  {author} {\bibinfo {author} {\bibfnamefont {M.}~\bibnamefont
  {{Cicoli}}}, \bibinfo {author} {\bibfnamefont {M.~D.}\ \bibnamefont
  {{Goodsell}}}, \ and\ \bibinfo {author} {\bibfnamefont {A.}~\bibnamefont
  {{Ringwald}}},\ }\bibfield  {title} {\enquote {\bibinfo {title} {{The type
  IIB string axiverse and its low-energy phenomenology}},}\ }\href {\doibase
  10.1007/JHEP10(2012)146} {\bibfield  {journal} {\bibinfo  {journal} {Journal
  of High Energy Physics}\ }\textbf {\bibinfo {volume} {10}},\ \bibinfo {eid}
  {146} (\bibinfo {year} {2012})},\ \Eprint {http://arxiv.org/abs/1206.0819}
  {arXiv:1206.0819 [hep-th]} \BibitemShut {NoStop}%
\bibitem [{\citenamefont {Kreuzer}\ and\ \citenamefont
  {Skarke}(2002)}]{Kreuzer:2000xy}%
  \BibitemOpen
  \bibfield  {author} {\bibinfo {author} {\bibfnamefont {Maximilian}\
  \bibnamefont {Kreuzer}}\ and\ \bibinfo {author} {\bibfnamefont {Harald}\
  \bibnamefont {Skarke}},\ }\bibfield  {title} {\enquote {\bibinfo {title}
  {{Complete classification of reflexive polyhedra in four-dimensions}},}\
  }\href {\doibase 10.4310/ATMP.2000.v4.n6.a2} {\bibfield  {journal} {\bibinfo
  {journal} {Adv. Theor. Math. Phys.}\ }\textbf {\bibinfo {volume} {4}},\
  \bibinfo {pages} {1209--1230} (\bibinfo {year} {2002})},\ \Eprint
  {http://arxiv.org/abs/hep-th/0002240} {arXiv:hep-th/0002240 [hep-th]}
  \BibitemShut {NoStop}%
\bibitem [{\citenamefont {Altman}\ \emph {et~al.}(2015)\citenamefont {Altman},
  \citenamefont {Gray}, \citenamefont {He}, \citenamefont {Jejjala},\ and\
  \citenamefont {Nelson}}]{Altman:2014bfa}%
  \BibitemOpen
  \bibfield  {author} {\bibinfo {author} {\bibfnamefont {Ross}\ \bibnamefont
  {Altman}}, \bibinfo {author} {\bibfnamefont {James}\ \bibnamefont {Gray}},
  \bibinfo {author} {\bibfnamefont {Yang-Hui}\ \bibnamefont {He}}, \bibinfo
  {author} {\bibfnamefont {Vishnu}\ \bibnamefont {Jejjala}}, \ and\ \bibinfo
  {author} {\bibfnamefont {Brent~D.}\ \bibnamefont {Nelson}},\ }\bibfield
  {title} {\enquote {\bibinfo {title} {{A Calabi-Yau Database: Threefolds
  Constructed from the Kreuzer-Skarke List}},}\ }\href {\doibase
  10.1007/JHEP02(2015)158} {\bibfield  {journal} {\bibinfo  {journal} {JHEP}\
  }\textbf {\bibinfo {volume} {02}},\ \bibinfo {pages} {158} (\bibinfo {year}
  {2015})},\ \Eprint {http://arxiv.org/abs/1411.1418} {arXiv:1411.1418
  [hep-th]} \BibitemShut {NoStop}%
\bibitem [{\citenamefont {Corti}\ \emph {et~al.}(2015)\citenamefont {Corti},
  \citenamefont {Haskins}, \citenamefont {Nordström},\ and\ \citenamefont
  {Pacini}}]{Corti:2012kd}%
  \BibitemOpen
  \bibfield  {author} {\bibinfo {author} {\bibfnamefont {Alessio}\ \bibnamefont
  {Corti}}, \bibinfo {author} {\bibfnamefont {Mark}\ \bibnamefont {Haskins}},
  \bibinfo {author} {\bibfnamefont {Johannes}\ \bibnamefont {Nordström}}, \
  and\ \bibinfo {author} {\bibfnamefont {Tommaso}\ \bibnamefont {Pacini}},\
  }\bibfield  {title} {\enquote {\bibinfo {title} {{$\mathrm{G}_{2}$-manifolds
  and associative submanifolds via semi-Fano $3$-folds}},}\ }\href {\doibase
  10.1215/00127094-3120743} {\bibfield  {journal} {\bibinfo  {journal} {Duke
  Math. J.}\ }\textbf {\bibinfo {volume} {164}},\ \bibinfo {pages} {1971--2092}
  (\bibinfo {year} {2015})},\ \Eprint {http://arxiv.org/abs/1207.4470}
  {arXiv:1207.4470 [math.DG]} \BibitemShut {NoStop}%
\bibitem [{\citenamefont {Halverson}\ and\ \citenamefont
  {Morrison}(2015)}]{Halverson:2014tya}%
  \BibitemOpen
  \bibfield  {author} {\bibinfo {author} {\bibfnamefont {James}\ \bibnamefont
  {Halverson}}\ and\ \bibinfo {author} {\bibfnamefont {David~R.}\ \bibnamefont
  {Morrison}},\ }\bibfield  {title} {\enquote {\bibinfo {title} {{The landscape
  of M-theory compactifications on seven-manifolds with G$_{2}$ holonomy}},}\
  }\href {\doibase 10.1007/JHEP04(2015)047} {\bibfield  {journal} {\bibinfo
  {journal} {JHEP}\ }\textbf {\bibinfo {volume} {04}},\ \bibinfo {pages} {047}
  (\bibinfo {year} {2015})},\ \Eprint {http://arxiv.org/abs/1412.4123}
  {arXiv:1412.4123 [hep-th]} \BibitemShut {NoStop}%
\bibitem [{\citenamefont {Halverson}\ and\ \citenamefont
  {Morrison}(2016)}]{Halverson:2015vta}%
  \BibitemOpen
  \bibfield  {author} {\bibinfo {author} {\bibfnamefont {James}\ \bibnamefont
  {Halverson}}\ and\ \bibinfo {author} {\bibfnamefont {David~R.}\ \bibnamefont
  {Morrison}},\ }\bibfield  {title} {\enquote {\bibinfo {title} {{On gauge
  enhancement and singular limits in G$_{2}$ compactifications of M-theory}},}\
  }\href {\doibase 10.1007/JHEP04(2016)100} {\bibfield  {journal} {\bibinfo
  {journal} {JHEP}\ }\textbf {\bibinfo {volume} {04}},\ \bibinfo {pages} {100}
  (\bibinfo {year} {2016})},\ \Eprint {http://arxiv.org/abs/1507.05965}
  {arXiv:1507.05965 [hep-th]} \BibitemShut {NoStop}%
\bibitem [{\citenamefont {Braun}(2017)}]{Braun:2016igl}%
  \BibitemOpen
  \bibfield  {author} {\bibinfo {author} {\bibfnamefont {Andreas~P.}\
  \bibnamefont {Braun}},\ }\bibfield  {title} {\enquote {\bibinfo {title}
  {{Tops as building blocks for G$_{2}$ manifolds}},}\ }\href {\doibase
  10.1007/JHEP10(2017)083} {\bibfield  {journal} {\bibinfo  {journal} {JHEP}\
  }\textbf {\bibinfo {volume} {10}},\ \bibinfo {pages} {083} (\bibinfo {year}
  {2017})},\ \Eprint {http://arxiv.org/abs/1602.03521} {arXiv:1602.03521
  [hep-th]} \BibitemShut {NoStop}%
\bibitem [{\citenamefont {Braun}\ and\ \citenamefont
  {Schfer-Nameki}(2017)}]{Braun:2017uku}%
  \BibitemOpen
  \bibfield  {author} {\bibinfo {author} {\bibfnamefont {Andreas~P.}\
  \bibnamefont {Braun}}\ and\ \bibinfo {author} {\bibfnamefont {Sakura}\
  \bibnamefont {Schfer-Nameki}},\ }\bibfield  {title} {\enquote {\bibinfo
  {title} {{Compact, Singular G2-Holonomy Manifolds and M/Heterotic/F-Theory
  Duality}},}\ }\href@noop {} {\  (\bibinfo {year} {2017})},\ \Eprint
  {http://arxiv.org/abs/1708.07215} {arXiv:1708.07215 [hep-th]} \BibitemShut
  {NoStop}%
\bibitem [{\citenamefont {Braun}\ and\ \citenamefont
  {Del~Zotto}(2017)}]{Braun:2017ryx}%
  \BibitemOpen
  \bibfield  {author} {\bibinfo {author} {\bibfnamefont {Andreas~P.}\
  \bibnamefont {Braun}}\ and\ \bibinfo {author} {\bibfnamefont {Michele}\
  \bibnamefont {Del~Zotto}},\ }\bibfield  {title} {\enquote {\bibinfo {title}
  {{Mirror Symmetry for $G_2$-Manifolds: Twisted Connected Sums and Dual
  Tops}},}\ }\href {\doibase 10.1007/JHEP05(2017)080} {\bibfield  {journal}
  {\bibinfo  {journal} {JHEP}\ }\textbf {\bibinfo {volume} {05}},\ \bibinfo
  {pages} {080} (\bibinfo {year} {2017})},\ \Eprint
  {http://arxiv.org/abs/1701.05202} {arXiv:1701.05202 [hep-th]} \BibitemShut
  {NoStop}%
\bibitem [{\citenamefont {Braun}\ and\ \citenamefont
  {Del~Zotto}(2018)}]{Braun:2017csz}%
  \BibitemOpen
  \bibfield  {author} {\bibinfo {author} {\bibfnamefont {Andreas~P.}\
  \bibnamefont {Braun}}\ and\ \bibinfo {author} {\bibfnamefont {Michele}\
  \bibnamefont {Del~Zotto}},\ }\bibfield  {title} {\enquote {\bibinfo {title}
  {{Towards Generalized Mirror Symmetry for Twisted Connected Sum $G_2$
  Manifolds}},}\ }\href {\doibase 10.1007/JHEP03(2018)082} {\bibfield
  {journal} {\bibinfo  {journal} {JHEP}\ }\textbf {\bibinfo {volume} {03}},\
  \bibinfo {pages} {082} (\bibinfo {year} {2018})},\ \Eprint
  {http://arxiv.org/abs/1712.06571} {arXiv:1712.06571 [hep-th]} \BibitemShut
  {NoStop}%
\bibitem [{\citenamefont {{Douglas}}\ and\ \citenamefont
  {{Kachru}}(2007)}]{2007RvMP...79..733D}%
  \BibitemOpen
  \bibfield  {author} {\bibinfo {author} {\bibfnamefont {M.~R.}\ \bibnamefont
  {{Douglas}}}\ and\ \bibinfo {author} {\bibfnamefont {S.}~\bibnamefont
  {{Kachru}}},\ }\bibfield  {title} {\enquote {\bibinfo {title} {{Flux
  compactification}},}\ }\href {\doibase 10.1103/RevModPhys.79.733} {\bibfield
  {journal} {\bibinfo  {journal} {Reviews of Modern Physics}\ }\textbf
  {\bibinfo {volume} {79}},\ \bibinfo {pages} {733--796} (\bibinfo {year}
  {2007})},\ \Eprint {http://arxiv.org/abs/hep-th/0610102} {hep-th/0610102}
  \BibitemShut {NoStop}%
\bibitem [{\citenamefont {{Mehta}}(1991)}]{mehta}%
  \BibitemOpen
  \bibfield  {author} {\bibinfo {author} {\bibfnamefont {M.~L.}\ \bibnamefont
  {{Mehta}}},\ }\href@noop {} {\emph {\bibinfo {title} {{Random Matrices}}}}\
  (\bibinfo  {publisher} {Academic Press},\ \bibinfo {year} {1991})\BibitemShut
  {NoStop}%
\bibitem [{\citenamefont {Easther}\ and\ \citenamefont
  {McAllister}(2006)}]{Easther:2005zr}%
  \BibitemOpen
  \bibfield  {author} {\bibinfo {author} {\bibfnamefont {Richard}\ \bibnamefont
  {Easther}}\ and\ \bibinfo {author} {\bibfnamefont {Liam}\ \bibnamefont
  {McAllister}},\ }\bibfield  {title} {\enquote {\bibinfo {title} {{Random
  matrices and the spectrum of N-flation}},}\ }\href {\doibase
  10.1088/1475-7516/2006/05/018} {\bibfield  {journal} {\bibinfo  {journal}
  {JCAP}\ }\textbf {\bibinfo {volume} {0605}},\ \bibinfo {pages} {018}
  (\bibinfo {year} {2006})},\ \Eprint {http://arxiv.org/abs/hep-th/0512102}
  {arXiv:hep-th/0512102 [hep-th]} \BibitemShut {NoStop}%
\bibitem [{\citenamefont {{Bachlechner}}\ \emph {et~al.}(2013)\citenamefont
  {{Bachlechner}}, \citenamefont {{Marsh}}, \citenamefont {{McAllister}},\ and\
  \citenamefont {{Wrase}}}]{2013JHEP...01..136B}%
  \BibitemOpen
  \bibfield  {author} {\bibinfo {author} {\bibfnamefont {T.~C.}\ \bibnamefont
  {{Bachlechner}}}, \bibinfo {author} {\bibfnamefont {D.}~\bibnamefont
  {{Marsh}}}, \bibinfo {author} {\bibfnamefont {L.}~\bibnamefont
  {{McAllister}}}, \ and\ \bibinfo {author} {\bibfnamefont {T.}~\bibnamefont
  {{Wrase}}},\ }\bibfield  {title} {\enquote {\bibinfo {title} {{Supersymmetric
  vacua in random supergravity}},}\ }\href {\doibase 10.1007/JHEP01(2013)136}
  {\bibfield  {journal} {\bibinfo  {journal} {Journal of High Energy Physics}\
  }\textbf {\bibinfo {volume} {1}},\ \bibinfo {eid} {136} (\bibinfo {year}
  {2013})},\ \Eprint {http://arxiv.org/abs/1207.2763} {arXiv:1207.2763
  [hep-th]} \BibitemShut {NoStop}%
\bibitem [{\citenamefont {Long}\ \emph {et~al.}(2014)\citenamefont {Long},
  \citenamefont {McAllister},\ and\ \citenamefont {McGuirk}}]{Long:2014fba}%
  \BibitemOpen
  \bibfield  {author} {\bibinfo {author} {\bibfnamefont {Cody}\ \bibnamefont
  {Long}}, \bibinfo {author} {\bibfnamefont {Liam}\ \bibnamefont {McAllister}},
  \ and\ \bibinfo {author} {\bibfnamefont {Paul}\ \bibnamefont {McGuirk}},\
  }\bibfield  {title} {\enquote {\bibinfo {title} {{Heavy Tails in Calabi-Yau
  Moduli Spaces}},}\ }\href {\doibase 10.1007/JHEP10(2014)187} {\bibfield
  {journal} {\bibinfo  {journal} {JHEP}\ }\textbf {\bibinfo {volume} {10}},\
  \bibinfo {pages} {187} (\bibinfo {year} {2014})},\ \Eprint
  {http://arxiv.org/abs/1407.0709} {arXiv:1407.0709 [hep-th]} \BibitemShut
  {NoStop}%
\bibitem [{\citenamefont {Brodie}\ and\ \citenamefont
  {Marsh}(2016)}]{Brodie:2015kza}%
  \BibitemOpen
  \bibfield  {author} {\bibinfo {author} {\bibfnamefont {Callum}\ \bibnamefont
  {Brodie}}\ and\ \bibinfo {author} {\bibfnamefont {M.~C.~David}\ \bibnamefont
  {Marsh}},\ }\bibfield  {title} {\enquote {\bibinfo {title} {{The Spectra of
  Type IIB Flux Compactifications at Large Complex Structure}},}\ }\href
  {\doibase 10.1007/JHEP01(2016)037} {\bibfield  {journal} {\bibinfo  {journal}
  {JHEP}\ }\textbf {\bibinfo {volume} {01}},\ \bibinfo {pages} {037} (\bibinfo
  {year} {2016})},\ \Eprint {http://arxiv.org/abs/1509.06761} {arXiv:1509.06761
  [hep-th]} \BibitemShut {NoStop}%
\bibitem [{\citenamefont {Bachlechner}\ \emph {et~al.}(2017)\citenamefont
  {Bachlechner}, \citenamefont {Eckerle}, \citenamefont {Janssen},\ and\
  \citenamefont {Kleban}}]{Bachlechner:2017hsj}%
  \BibitemOpen
  \bibfield  {author} {\bibinfo {author} {\bibfnamefont {Thomas~C.}\
  \bibnamefont {Bachlechner}}, \bibinfo {author} {\bibfnamefont {Kate}\
  \bibnamefont {Eckerle}}, \bibinfo {author} {\bibfnamefont {Oliver}\
  \bibnamefont {Janssen}}, \ and\ \bibinfo {author} {\bibfnamefont {Matthew}\
  \bibnamefont {Kleban}},\ }\bibfield  {title} {\enquote {\bibinfo {title}
  {{Systematics of Aligned Axions}},}\ }\href {\doibase
  10.1007/JHEP11(2017)036} {\bibfield  {journal} {\bibinfo  {journal} {JHEP}\
  }\textbf {\bibinfo {volume} {11}},\ \bibinfo {pages} {036} (\bibinfo {year}
  {2017})},\ \Eprint {http://arxiv.org/abs/1709.01080} {arXiv:1709.01080
  [hep-th]} \BibitemShut {NoStop}%
\bibitem [{\citenamefont {Aasi}\ \emph {et~al.}(2015)\citenamefont {Aasi} \emph
  {et~al.}}]{TheLIGOScientific:2014jea}%
  \BibitemOpen
  \bibfield  {author} {\bibinfo {author} {\bibfnamefont {J.}~\bibnamefont
  {Aasi}} \emph {et~al.} (\bibinfo {collaboration} {LIGO Scientific}),\
  }\bibfield  {title} {\enquote {\bibinfo {title} {{Advanced LIGO}},}\ }\href
  {\doibase 10.1088/0264-9381/32/7/074001} {\bibfield  {journal} {\bibinfo
  {journal} {Class. Quant. Grav.}\ }\textbf {\bibinfo {volume} {32}},\ \bibinfo
  {pages} {074001} (\bibinfo {year} {2015})},\ \Eprint
  {http://arxiv.org/abs/1411.4547} {arXiv:1411.4547 [gr-qc]} \BibitemShut
  {NoStop}%
\bibitem [{\citenamefont {Amaro-Seoane}\ \emph {et~al.}(2012)\citenamefont
  {Amaro-Seoane} \emph {et~al.}}]{AmaroSeoane:2012je}%
  \BibitemOpen
  \bibfield  {author} {\bibinfo {author} {\bibfnamefont {Pau}\ \bibnamefont
  {Amaro-Seoane}} \emph {et~al.},\ }\bibfield  {title} {\enquote {\bibinfo
  {title} {{Low-frequency gravitational-wave science with eLISA/NGO}},}\
  }\bibfield  {booktitle} {\emph {\bibinfo {booktitle} {{Gravitational waves.
  Numerical relativity - data analysis. Proceedings, 9th Edoardo Amaldi
  Conference, Amaldi 9, and meeting, NRDA 2011, Cardiff, UK, July 10-15,
  2011}}},\ }\href {\doibase 10.1088/0264-9381/29/12/124016} {\bibfield
  {journal} {\bibinfo  {journal} {Class. Quant. Grav.}\ }\textbf {\bibinfo
  {volume} {29}},\ \bibinfo {pages} {124016} (\bibinfo {year} {2012})},\
  \Eprint {http://arxiv.org/abs/1202.0839} {arXiv:1202.0839 [gr-qc]}
  \BibitemShut {NoStop}%
\bibitem [{\citenamefont {Amaro-Seoane}\ \emph {et~al.}(2013)\citenamefont
  {Amaro-Seoane} \emph {et~al.}}]{AmaroSeoane:2012km}%
  \BibitemOpen
  \bibfield  {author} {\bibinfo {author} {\bibfnamefont {Pau}\ \bibnamefont
  {Amaro-Seoane}} \emph {et~al.},\ }\bibfield  {title} {\enquote {\bibinfo
  {title} {{eLISA/NGO: Astrophysics and cosmology in the gravitational-wave
  millihertz regime}},}\ }\href@noop {} {\bibfield  {journal} {\bibinfo
  {journal} {GW Notes}\ }\textbf {\bibinfo {volume} {6}},\ \bibinfo {pages}
  {4--110} (\bibinfo {year} {2013})},\ \Eprint {http://arxiv.org/abs/1201.3621}
  {arXiv:1201.3621 [astro-ph.CO]} \BibitemShut {NoStop}%
\bibitem [{\citenamefont {Dimopoulos}\ \emph {et~al.}(2008)\citenamefont
  {Dimopoulos}, \citenamefont {Graham}, \citenamefont {Hogan}, \citenamefont
  {Kasevich},\ and\ \citenamefont {Rajendran}}]{Dimopoulos:2008sv}%
  \BibitemOpen
  \bibfield  {author} {\bibinfo {author} {\bibfnamefont {Savas}\ \bibnamefont
  {Dimopoulos}}, \bibinfo {author} {\bibfnamefont {Peter~W.}\ \bibnamefont
  {Graham}}, \bibinfo {author} {\bibfnamefont {Jason~M.}\ \bibnamefont
  {Hogan}}, \bibinfo {author} {\bibfnamefont {Mark~A.}\ \bibnamefont
  {Kasevich}}, \ and\ \bibinfo {author} {\bibfnamefont {Surjeet}\ \bibnamefont
  {Rajendran}},\ }\bibfield  {title} {\enquote {\bibinfo {title} {{An Atomic
  Gravitational Wave Interferometric Sensor (AGIS)}},}\ }\href {\doibase
  10.1103/PhysRevD.78.122002} {\bibfield  {journal} {\bibinfo  {journal} {Phys.
  Rev.}\ }\textbf {\bibinfo {volume} {D78}},\ \bibinfo {pages} {122002}
  (\bibinfo {year} {2008})},\ \Eprint {http://arxiv.org/abs/0806.2125}
  {arXiv:0806.2125 [gr-qc]} \BibitemShut {NoStop}%
\bibitem [{\citenamefont {Luo}\ \emph {et~al.}(2016)\citenamefont {Luo} \emph
  {et~al.}}]{Luo:2015ght}%
  \BibitemOpen
  \bibfield  {author} {\bibinfo {author} {\bibfnamefont {Jun}\ \bibnamefont
  {Luo}} \emph {et~al.} (\bibinfo {collaboration} {TianQin}),\ }\bibfield
  {title} {\enquote {\bibinfo {title} {{TianQin: a space-borne gravitational
  wave detector}},}\ }\href {\doibase 10.1088/0264-9381/33/3/035010} {\bibfield
   {journal} {\bibinfo  {journal} {Class. Quant. Grav.}\ }\textbf {\bibinfo
  {volume} {33}},\ \bibinfo {pages} {035010} (\bibinfo {year} {2016})},\
  \Eprint {http://arxiv.org/abs/1512.02076} {arXiv:1512.02076 [astro-ph.IM]}
  \BibitemShut {NoStop}%
\bibitem [{\citenamefont {Sathyaprakash}\ \emph {et~al.}(2012)\citenamefont
  {Sathyaprakash} \emph {et~al.}}]{Sathyaprakash:2012jk}%
  \BibitemOpen
  \bibfield  {author} {\bibinfo {author} {\bibfnamefont {B.}~\bibnamefont
  {Sathyaprakash}} \emph {et~al.},\ }\bibfield  {title} {\enquote {\bibinfo
  {title} {{Scientific Objectives of Einstein Telescope}},}\ }\bibfield
  {booktitle} {\emph {\bibinfo {booktitle} {{Gravitational waves. Numerical
  relativity - data analysis. Proceedings, 9th Edoardo Amaldi Conference,
  Amaldi 9, and meeting, NRDA 2011, Cardiff, UK, July 10-15, 2011}}},\ }\href
  {\doibase 10.1088/0264-9381/29/12/124013, 10.1088/0264-9381/30/7/079501}
  {\bibfield  {journal} {\bibinfo  {journal} {Class. Quant. Grav.}\ }\textbf
  {\bibinfo {volume} {29}},\ \bibinfo {pages} {124013} (\bibinfo {year}
  {2012})},\ \bibinfo {note} {[Erratum: Class. Quant. Grav.30,079501(2013)]},\
  \Eprint {http://arxiv.org/abs/1206.0331} {arXiv:1206.0331 [gr-qc]}
  \BibitemShut {NoStop}%
\bibitem [{\citenamefont {et~al}(2008)}]{1742-6596-120-3-032004}%
  \BibitemOpen
  \bibfield  {author} {\bibinfo {author} {\bibfnamefont {S~Kawamura}\
  \bibnamefont {et~al}},\ }\bibfield  {title} {\enquote {\bibinfo {title} {The
  japanese space gravitational wave antenna; decigo},}\ }\href
  {http://stacks.iop.org/1742-6596/120/i=3/a=032004} {\bibfield  {journal}
  {\bibinfo  {journal} {Journal of Physics: Conference Series}\ }\textbf
  {\bibinfo {volume} {120}},\ \bibinfo {pages} {032004} (\bibinfo {year}
  {2008})}\BibitemShut {NoStop}%
\bibitem [{\citenamefont {Brito}\ \emph {et~al.}(2015)\citenamefont {Brito},
  \citenamefont {Cardoso},\ and\ \citenamefont {Pani}}]{Brito:2014wla}%
  \BibitemOpen
  \bibfield  {author} {\bibinfo {author} {\bibfnamefont {Richard}\ \bibnamefont
  {Brito}}, \bibinfo {author} {\bibfnamefont {Vitor}\ \bibnamefont {Cardoso}},
  \ and\ \bibinfo {author} {\bibfnamefont {Paolo}\ \bibnamefont {Pani}},\
  }\bibfield  {title} {\enquote {\bibinfo {title} {{Black holes as particle
  detectors: evolution of superradiant instabilities}},}\ }\href {\doibase
  10.1088/0264-9381/32/13/134001} {\bibfield  {journal} {\bibinfo  {journal}
  {Class. Quant. Grav.}\ }\textbf {\bibinfo {volume} {32}},\ \bibinfo {pages}
  {134001} (\bibinfo {year} {2015})},\ \Eprint {http://arxiv.org/abs/1411.0686}
  {arXiv:1411.0686 [gr-qc]} \BibitemShut {NoStop}%
\bibitem [{\citenamefont {Detweiler}(1980)}]{PhysRevD.22.2323}%
  \BibitemOpen
  \bibfield  {author} {\bibinfo {author} {\bibfnamefont {Steven}\ \bibnamefont
  {Detweiler}},\ }\bibfield  {title} {\enquote {\bibinfo {title} {Klein-gordon
  equation and rotating black holes},}\ }\href {\doibase
  10.1103/PhysRevD.22.2323} {\bibfield  {journal} {\bibinfo  {journal} {Phys.
  Rev. D}\ }\textbf {\bibinfo {volume} {22}},\ \bibinfo {pages} {2323--2326}
  (\bibinfo {year} {1980})}\BibitemShut {NoStop}%
\bibitem [{\citenamefont {Zouros}\ and\ \citenamefont
  {Eardley}(1979)}]{ZOUROS1979139}%
  \BibitemOpen
  \bibfield  {author} {\bibinfo {author} {\bibfnamefont {Theodoros~J.M}\
  \bibnamefont {Zouros}}\ and\ \bibinfo {author} {\bibfnamefont {Douglas~M}\
  \bibnamefont {Eardley}},\ }\bibfield  {title} {\enquote {\bibinfo {title}
  {Instabilities of massive scalar perturbations of a rotating black hole},}\
  }\href {\doibase https://doi.org/10.1016/0003-4916(79)90237-9} {\bibfield
  {journal} {\bibinfo  {journal} {Annals of Physics}\ }\textbf {\bibinfo
  {volume} {118}},\ \bibinfo {pages} {139 -- 155} (\bibinfo {year}
  {1979})}\BibitemShut {NoStop}%
\bibitem [{\citenamefont {Dolan}(2007)}]{Dolan:2007mj}%
  \BibitemOpen
  \bibfield  {author} {\bibinfo {author} {\bibfnamefont {Sam~R.}\ \bibnamefont
  {Dolan}},\ }\bibfield  {title} {\enquote {\bibinfo {title} {{Instability of
  the massive Klein-Gordon field on the Kerr spacetime}},}\ }\href {\doibase
  10.1103/PhysRevD.76.084001} {\bibfield  {journal} {\bibinfo  {journal} {Phys.
  Rev.}\ }\textbf {\bibinfo {volume} {D76}},\ \bibinfo {pages} {084001}
  (\bibinfo {year} {2007})},\ \Eprint {http://arxiv.org/abs/0705.2880}
  {arXiv:0705.2880 [gr-qc]} \BibitemShut {NoStop}%
\bibitem [{\citenamefont {Dolan}(2013)}]{Dolan:2012yt}%
  \BibitemOpen
  \bibfield  {author} {\bibinfo {author} {\bibfnamefont {Sam~R.}\ \bibnamefont
  {Dolan}},\ }\bibfield  {title} {\enquote {\bibinfo {title} {{Superradiant
  instabilities of rotating black holes in the time domain}},}\ }\href
  {\doibase 10.1103/PhysRevD.87.124026} {\bibfield  {journal} {\bibinfo
  {journal} {Phys. Rev.}\ }\textbf {\bibinfo {volume} {D87}},\ \bibinfo {pages}
  {124026} (\bibinfo {year} {2013})},\ \Eprint {http://arxiv.org/abs/1212.1477}
  {arXiv:1212.1477 [gr-qc]} \BibitemShut {NoStop}%
\bibitem [{\citenamefont {Furuhashi}\ and\ \citenamefont
  {Nambu}(2004)}]{Furuhashi:2004jk}%
  \BibitemOpen
  \bibfield  {author} {\bibinfo {author} {\bibfnamefont {Hironobu}\
  \bibnamefont {Furuhashi}}\ and\ \bibinfo {author} {\bibfnamefont {Yasusada}\
  \bibnamefont {Nambu}},\ }\bibfield  {title} {\enquote {\bibinfo {title}
  {{Instability of massive scalar fields in Kerr-Newman space-time}},}\ }\href
  {\doibase 10.1143/PTP.112.983} {\bibfield  {journal} {\bibinfo  {journal}
  {Prog. Theor. Phys.}\ }\textbf {\bibinfo {volume} {112}},\ \bibinfo {pages}
  {983--995} (\bibinfo {year} {2004})},\ \Eprint
  {http://arxiv.org/abs/gr-qc/0402037} {arXiv:gr-qc/0402037 [gr-qc]}
  \BibitemShut {NoStop}%
\bibitem [{\citenamefont {Cardoso}\ and\ \citenamefont
  {Yoshida}(2005)}]{Cardoso:2005vk}%
  \BibitemOpen
  \bibfield  {author} {\bibinfo {author} {\bibfnamefont {Vitor}\ \bibnamefont
  {Cardoso}}\ and\ \bibinfo {author} {\bibfnamefont {Shijun}\ \bibnamefont
  {Yoshida}},\ }\bibfield  {title} {\enquote {\bibinfo {title} {{Superradiant
  instabilities of rotating black branes and strings}},}\ }\href {\doibase
  10.1088/1126-6708/2005/07/009} {\bibfield  {journal} {\bibinfo  {journal}
  {JHEP}\ }\textbf {\bibinfo {volume} {07}},\ \bibinfo {pages} {009} (\bibinfo
  {year} {2005})},\ \Eprint {http://arxiv.org/abs/hep-th/0502206}
  {arXiv:hep-th/0502206 [hep-th]} \BibitemShut {NoStop}%
\bibitem [{\citenamefont {Yoshino}\ and\ \citenamefont
  {Kodama}(2015)}]{Yoshino:2015nsa}%
  \BibitemOpen
  \bibfield  {author} {\bibinfo {author} {\bibfnamefont {Hirotaka}\
  \bibnamefont {Yoshino}}\ and\ \bibinfo {author} {\bibfnamefont {Hideo}\
  \bibnamefont {Kodama}},\ }\bibfield  {title} {\enquote {\bibinfo {title}
  {{The bosenova and axiverse}},}\ }\href {\doibase
  10.1088/0264-9381/32/21/214001} {\bibfield  {journal} {\bibinfo  {journal}
  {Class. Quant. Grav.}\ }\textbf {\bibinfo {volume} {32}},\ \bibinfo {pages}
  {214001} (\bibinfo {year} {2015})},\ \Eprint
  {http://arxiv.org/abs/1505.00714} {arXiv:1505.00714 [gr-qc]} \BibitemShut
  {NoStop}%
\bibitem [{\citenamefont {Shankar}\ \emph {et~al.}(2009)\citenamefont
  {Shankar}, \citenamefont {Weinberg},\ and\ \citenamefont
  {Miralda-Escude}}]{Shankar:2007zg}%
  \BibitemOpen
  \bibfield  {author} {\bibinfo {author} {\bibfnamefont {Francesco}\
  \bibnamefont {Shankar}}, \bibinfo {author} {\bibfnamefont {David~H.}\
  \bibnamefont {Weinberg}}, \ and\ \bibinfo {author} {\bibfnamefont {Jordi}\
  \bibnamefont {Miralda-Escude}},\ }\bibfield  {title} {\enquote {\bibinfo
  {title} {{Self-Consistent Models of the AGN and Black Hole Populations: Duty
  Cycles, Accretion Rates, and the Mean Radiative Efficiency}},}\ }\href
  {\doibase 10.1088/0004-637X/690/1/20} {\bibfield  {journal} {\bibinfo
  {journal} {Astrophys. J.}\ }\textbf {\bibinfo {volume} {690}},\ \bibinfo
  {pages} {20--41} (\bibinfo {year} {2009})},\ \Eprint
  {http://arxiv.org/abs/0710.4488} {arXiv:0710.4488 [astro-ph]} \BibitemShut
  {NoStop}%
\bibitem [{\citenamefont {Cardoso}\ \emph {et~al.}(2018)\citenamefont
  {Cardoso}, \citenamefont {Dias}, \citenamefont {Hartnett}, \citenamefont
  {Middleton}, \citenamefont {Pani},\ and\ \citenamefont
  {Santos}}]{Cardoso:2018tly}%
  \BibitemOpen
  \bibfield  {author} {\bibinfo {author} {\bibfnamefont {Vitor}\ \bibnamefont
  {Cardoso}}, \bibinfo {author} {\bibfnamefont {Óscar J.~C.}\ \bibnamefont
  {Dias}}, \bibinfo {author} {\bibfnamefont {Gavin~S.}\ \bibnamefont
  {Hartnett}}, \bibinfo {author} {\bibfnamefont {Matthew}\ \bibnamefont
  {Middleton}}, \bibinfo {author} {\bibfnamefont {Paolo}\ \bibnamefont {Pani}},
  \ and\ \bibinfo {author} {\bibfnamefont {Jorge~E.}\ \bibnamefont {Santos}},\
  }\bibfield  {title} {\enquote {\bibinfo {title} {{Constraining the mass of
  dark photons and axion-like particles through black-hole superradiance}},}\
  }\href {\doibase 10.1088/1475-7516/2018/03/043} {\bibfield  {journal}
  {\bibinfo  {journal} {JCAP}\ }\textbf {\bibinfo {volume} {1803}},\ \bibinfo
  {pages} {043} (\bibinfo {year} {2018})},\ \Eprint
  {http://arxiv.org/abs/1801.01420} {arXiv:1801.01420 [gr-qc]} \BibitemShut
  {NoStop}%
\bibitem [{\citenamefont {Abbott}\ \emph {et~al.}(2016)\citenamefont {Abbott}
  \emph {et~al.}}]{TheLIGOScientific:2016pea}%
  \BibitemOpen
  \bibfield  {author} {\bibinfo {author} {\bibfnamefont {B.~P.}\ \bibnamefont
  {Abbott}} \emph {et~al.} (\bibinfo {collaboration} {Virgo, LIGO
  Scientific}),\ }\bibfield  {title} {\enquote {\bibinfo {title} {{Binary Black
  Hole Mergers in the first Advanced LIGO Observing Run}},}\ }\href {\doibase
  10.1103/PhysRevX.6.041015} {\bibfield  {journal} {\bibinfo  {journal} {Phys.
  Rev.}\ }\textbf {\bibinfo {volume} {X6}},\ \bibinfo {pages} {041015}
  (\bibinfo {year} {2016})},\ \Eprint {http://arxiv.org/abs/1606.04856}
  {arXiv:1606.04856 [gr-qc]} \BibitemShut {NoStop}%
\bibitem [{\citenamefont {et~al}\ \emph {et~al.}(2017)\citenamefont {et~al},
  \citenamefont {Collaboration},\ and\ \citenamefont
  {Collaboration)}}]{2041-8205-851-2-L35}%
  \BibitemOpen
  \bibfield  {author} {\bibinfo {author} {\bibfnamefont {B.~P.~Abbott}\
  \bibnamefont {et~al}}, \bibinfo {author} {\bibfnamefont {(LIGO~Scientific}\
  \bibnamefont {Collaboration}}, \ and\ \bibinfo {author} {\bibfnamefont
  {Virgo}\ \bibnamefont {Collaboration)}},\ }\bibfield  {title} {\enquote
  {\bibinfo {title} {Gw170608: Observation of a 19 solar-mass binary black hole
  coalescence},}\ }\href {http://stacks.iop.org/2041-8205/851/i=2/a=L35}
  {\bibfield  {journal} {\bibinfo  {journal} {The Astrophysical Journal
  Letters}\ }\textbf {\bibinfo {volume} {851}},\ \bibinfo {pages} {L35}
  (\bibinfo {year} {2017})}\BibitemShut {NoStop}%
\bibitem [{\citenamefont {Abbott}(2017)}]{PhysRevLett.119.141101}%
  \BibitemOpen
  \bibfield  {author} {\bibinfo {author} {\bibfnamefont {B.~P. et~al}\
  \bibnamefont {Abbott}} (\bibinfo {collaboration} {LIGO Scientific
  Collaboration and Virgo Collaboration}),\ }\bibfield  {title} {\enquote
  {\bibinfo {title} {Gw170814: A three-detector observation of gravitational
  waves from a binary black hole coalescence},}\ }\href {\doibase
  10.1103/PhysRevLett.119.141101} {\bibfield  {journal} {\bibinfo  {journal}
  {Phys. Rev. Lett.}\ }\textbf {\bibinfo {volume} {119}},\ \bibinfo {pages}
  {141101} (\bibinfo {year} {2017})}\BibitemShut {NoStop}%
\bibitem [{\citenamefont {Klein}\ \emph {et~al.}(2016)\citenamefont {Klein}
  \emph {et~al.}}]{Klein:2015hvg}%
  \BibitemOpen
  \bibfield  {author} {\bibinfo {author} {\bibfnamefont {Antoine}\ \bibnamefont
  {Klein}} \emph {et~al.},\ }\bibfield  {title} {\enquote {\bibinfo {title}
  {{Science with the space-based interferometer eLISA: Supermassive black hole
  binaries}},}\ }\href {\doibase 10.1103/PhysRevD.93.024003} {\bibfield
  {journal} {\bibinfo  {journal} {Phys. Rev.}\ }\textbf {\bibinfo {volume}
  {D93}},\ \bibinfo {pages} {024003} (\bibinfo {year} {2016})},\ \Eprint
  {http://arxiv.org/abs/1511.05581} {arXiv:1511.05581 [gr-qc]} \BibitemShut
  {NoStop}%
\bibitem [{\citenamefont {McClintock}\ \emph {et~al.}(2014)\citenamefont
  {McClintock}, \citenamefont {Narayan},\ and\ \citenamefont
  {Steiner}}]{McClintock:2013vwa}%
  \BibitemOpen
  \bibfield  {author} {\bibinfo {author} {\bibfnamefont {Jeffrey~E.}\
  \bibnamefont {McClintock}}, \bibinfo {author} {\bibfnamefont {Ramesh}\
  \bibnamefont {Narayan}}, \ and\ \bibinfo {author} {\bibfnamefont {James~F.}\
  \bibnamefont {Steiner}},\ }\bibfield  {title} {\enquote {\bibinfo {title}
  {{Black Hole Spin via Continuum Fitting and the Role of Spin in Powering
  Transient Jets}},}\ }\href {\doibase 10.1007/s11214-013-0003-9} {\bibfield
  {journal} {\bibinfo  {journal} {Space Sci. Rev.}\ }\textbf {\bibinfo {volume}
  {183}},\ \bibinfo {pages} {295--322} (\bibinfo {year} {2014})},\ \Eprint
  {http://arxiv.org/abs/1303.1583} {arXiv:1303.1583 [astro-ph.HE]} \BibitemShut
  {NoStop}%
\bibitem [{\citenamefont {Middleton}(2016)}]{Middleton:2015osa}%
  \BibitemOpen
  \bibfield  {author} {\bibinfo {author} {\bibfnamefont {Matthew}\ \bibnamefont
  {Middleton}},\ }\bibfield  {title} {\enquote {\bibinfo {title} {{Black hole
  spin: theory and observation}},}\ }\href {\doibase
  10.1007/978-3-662-52859-4_3} {\ ,\ \bibinfo {pages} {99--151} (\bibinfo
  {year} {2016})},\ \Eprint {http://arxiv.org/abs/1507.06153} {arXiv:1507.06153
  [astro-ph.HE]} \BibitemShut {NoStop}%
\bibitem [{\citenamefont {{Brenneman}}\ \emph
  {et~al.}(2011{\natexlab{a}})\citenamefont {{Brenneman}}, \citenamefont
  {{Reynolds}}, \citenamefont {{Nowak}}, \citenamefont {{Reis}}, \citenamefont
  {{Trippe}}, \citenamefont {{Fabian}}, \citenamefont {{Iwasawa}},
  \citenamefont {{Lee}}, \citenamefont {{Miller}}, \citenamefont {{Mushotzky}},
  \citenamefont {{Nandra}},\ and\ \citenamefont
  {{Volonteri}}}]{2011ApJ...736..103B}%
  \BibitemOpen
  \bibfield  {author} {\bibinfo {author} {\bibfnamefont {L.~W.}\ \bibnamefont
  {{Brenneman}}}, \bibinfo {author} {\bibfnamefont {C.~S.}\ \bibnamefont
  {{Reynolds}}}, \bibinfo {author} {\bibfnamefont {M.~A.}\ \bibnamefont
  {{Nowak}}}, \bibinfo {author} {\bibfnamefont {R.~C.}\ \bibnamefont {{Reis}}},
  \bibinfo {author} {\bibfnamefont {M.}~\bibnamefont {{Trippe}}}, \bibinfo
  {author} {\bibfnamefont {A.~C.}\ \bibnamefont {{Fabian}}}, \bibinfo {author}
  {\bibfnamefont {K.}~\bibnamefont {{Iwasawa}}}, \bibinfo {author}
  {\bibfnamefont {J.~C.}\ \bibnamefont {{Lee}}}, \bibinfo {author}
  {\bibfnamefont {J.~M.}\ \bibnamefont {{Miller}}}, \bibinfo {author}
  {\bibfnamefont {R.~F.}\ \bibnamefont {{Mushotzky}}}, \bibinfo {author}
  {\bibfnamefont {K.}~\bibnamefont {{Nandra}}}, \ and\ \bibinfo {author}
  {\bibfnamefont {M.}~\bibnamefont {{Volonteri}}},\ }\bibfield  {title}
  {\enquote {\bibinfo {title} {{The Spin of the Supermassive Black Hole in NGC
  3783}},}\ }\href {\doibase 10.1088/0004-637X/736/2/103} {\bibfield  {journal}
  {\bibinfo  {journal} {\apj}\ }\textbf {\bibinfo {volume} {736}},\ \bibinfo
  {eid} {103} (\bibinfo {year} {2011}{\natexlab{a}})},\ \Eprint
  {http://arxiv.org/abs/1104.1172} {arXiv:1104.1172 [astro-ph.HE]} \BibitemShut
  {NoStop}%
\bibitem [{\citenamefont {Reynolds}(2014)}]{Reynolds:2013qqa}%
  \BibitemOpen
  \bibfield  {author} {\bibinfo {author} {\bibfnamefont {Christopher~S.}\
  \bibnamefont {Reynolds}},\ }\bibfield  {title} {\enquote {\bibinfo {title}
  {{Measuring Black Hole Spin using X-ray Reflection Spectroscopy}},}\ }\href
  {\doibase 10.1007/s11214-013-0006-6} {\bibfield  {journal} {\bibinfo
  {journal} {Space Sci. Rev.}\ }\textbf {\bibinfo {volume} {183}},\ \bibinfo
  {pages} {277--294} (\bibinfo {year} {2014})},\ \Eprint
  {http://arxiv.org/abs/1302.3260} {arXiv:1302.3260 [astro-ph.HE]} \BibitemShut
  {NoStop}%
\bibitem [{\citenamefont {Miller}\ and\ \citenamefont
  {Miller}(2014)}]{Miller:2014aaa}%
  \BibitemOpen
  \bibfield  {author} {\bibinfo {author} {\bibfnamefont {M.~Coleman}\
  \bibnamefont {Miller}}\ and\ \bibinfo {author} {\bibfnamefont {Jon~M.}\
  \bibnamefont {Miller}},\ }\bibfield  {title} {\enquote {\bibinfo {title}
  {{The Masses and Spins of Neutron Stars and Stellar-Mass Black Holes}},}\
  }\href {\doibase 10.1016/j.physrep.2014.09.003} {\bibfield  {journal}
  {\bibinfo  {journal} {Phys. Rept.}\ }\textbf {\bibinfo {volume} {548}},\
  \bibinfo {pages} {1--34} (\bibinfo {year} {2014})},\ \Eprint
  {http://arxiv.org/abs/1408.4145} {arXiv:1408.4145 [astro-ph.HE]} \BibitemShut
  {NoStop}%
\bibitem [{\citenamefont {Reynolds}(2013)}]{Reynolds:2013rva}%
  \BibitemOpen
  \bibfield  {author} {\bibinfo {author} {\bibfnamefont {Christopher~S.}\
  \bibnamefont {Reynolds}},\ }\bibfield  {title} {\enquote {\bibinfo {title}
  {{The Spin of Supermassive Black Holes}},}\ }\href {\doibase
  10.1088/0264-9381/30/24/244004} {\bibfield  {journal} {\bibinfo  {journal}
  {Class. Quant. Grav.}\ }\textbf {\bibinfo {volume} {30}},\ \bibinfo {pages}
  {244004} (\bibinfo {year} {2013})},\ \Eprint {http://arxiv.org/abs/1307.3246}
  {arXiv:1307.3246 [astro-ph.HE]} \BibitemShut {NoStop}%
\bibitem [{\citenamefont {Abbott}\ \emph {et~al.}(2017)\citenamefont {Abbott}
  \emph {et~al.}}]{Abbott:2017vtc}%
  \BibitemOpen
  \bibfield  {author} {\bibinfo {author} {\bibfnamefont {Benjamin~P.}\
  \bibnamefont {Abbott}} \emph {et~al.} (\bibinfo {collaboration} {VIRGO, LIGO
  Scientific}),\ }\bibfield  {title} {\enquote {\bibinfo {title} {{GW170104:
  Observation of a 50-Solar-Mass Binary Black Hole Coalescence at Redshift
  0.2}},}\ }\href {\doibase 10.1103/PhysRevLett.118.221101} {\bibfield
  {journal} {\bibinfo  {journal} {Phys. Rev. Lett.}\ }\textbf {\bibinfo
  {volume} {118}},\ \bibinfo {pages} {221101} (\bibinfo {year} {2017})},\
  \Eprint {http://arxiv.org/abs/1706.01812} {arXiv:1706.01812 [gr-qc]}
  \BibitemShut {NoStop}%
\bibitem [{\citenamefont {{Orosz}}\ \emph
  {et~al.}(2011{\natexlab{a}})\citenamefont {{Orosz}}, \citenamefont
  {{McClintock}}, \citenamefont {{Aufdenberg}}, \citenamefont {{Remillard}},
  \citenamefont {{Reid}}, \citenamefont {{Narayan}},\ and\ \citenamefont
  {{Gou}}}]{2011ApJ...742...84O}%
  \BibitemOpen
  \bibfield  {author} {\bibinfo {author} {\bibfnamefont {J.~A.}\ \bibnamefont
  {{Orosz}}}, \bibinfo {author} {\bibfnamefont {J.~E.}\ \bibnamefont
  {{McClintock}}}, \bibinfo {author} {\bibfnamefont {J.~P.}\ \bibnamefont
  {{Aufdenberg}}}, \bibinfo {author} {\bibfnamefont {R.~A.}\ \bibnamefont
  {{Remillard}}}, \bibinfo {author} {\bibfnamefont {M.~J.}\ \bibnamefont
  {{Reid}}}, \bibinfo {author} {\bibfnamefont {R.}~\bibnamefont {{Narayan}}}, \
  and\ \bibinfo {author} {\bibfnamefont {L.}~\bibnamefont {{Gou}}},\ }\bibfield
   {title} {\enquote {\bibinfo {title} {{The Mass of the Black Hole in Cygnus
  X-1}},}\ }\href {\doibase 10.1088/0004-637X/742/2/84} {\bibfield  {journal}
  {\bibinfo  {journal} {\apj}\ }\textbf {\bibinfo {volume} {742}},\ \bibinfo
  {eid} {84} (\bibinfo {year} {2011}{\natexlab{a}})},\ \Eprint
  {http://arxiv.org/abs/1106.3689} {arXiv:1106.3689 [astro-ph.HE]} \BibitemShut
  {NoStop}%
\bibitem [{\citenamefont {Gou}\ \emph {et~al.}(2014)\citenamefont {Gou},
  \citenamefont {McClintock}, \citenamefont {Remillard}, \citenamefont
  {Steiner}, \citenamefont {Reid}, \citenamefont {Orosz}, \citenamefont
  {Narayan}, \citenamefont {Hanke},\ and\ \citenamefont
  {García}}]{Gou:2013dna}%
  \BibitemOpen
  \bibfield  {author} {\bibinfo {author} {\bibfnamefont {Lijun}\ \bibnamefont
  {Gou}}, \bibinfo {author} {\bibfnamefont {Jeffrey~E.}\ \bibnamefont
  {McClintock}}, \bibinfo {author} {\bibfnamefont {Ronald~A.}\ \bibnamefont
  {Remillard}}, \bibinfo {author} {\bibfnamefont {James~F.}\ \bibnamefont
  {Steiner}}, \bibinfo {author} {\bibfnamefont {Mark~J.}\ \bibnamefont {Reid}},
  \bibinfo {author} {\bibfnamefont {Jerome~A.}\ \bibnamefont {Orosz}}, \bibinfo
  {author} {\bibfnamefont {Ramesh}\ \bibnamefont {Narayan}}, \bibinfo {author}
  {\bibfnamefont {Manfred}\ \bibnamefont {Hanke}}, \ and\ \bibinfo {author}
  {\bibfnamefont {Javier}\ \bibnamefont {García}},\ }\bibfield  {title}
  {\enquote {\bibinfo {title} {{Confirmation Via the Continuum-Fitting Method
  that the Spin of the Black Hole in Cygnus X-1 is Extreme}},}\ }\href
  {\doibase 10.1088/0004-637X/790/1/29} {\bibfield  {journal} {\bibinfo
  {journal} {Astrophys. J.}\ }\textbf {\bibinfo {volume} {790}},\ \bibinfo
  {pages} {29} (\bibinfo {year} {2014})},\ \Eprint
  {http://arxiv.org/abs/1308.4760} {arXiv:1308.4760 [astro-ph.HE]} \BibitemShut
  {NoStop}%
\bibitem [{\citenamefont {{Orosz}}\ \emph
  {et~al.}(2011{\natexlab{b}})\citenamefont {{Orosz}}, \citenamefont
  {{Steiner}}, \citenamefont {{McClintock}}, \citenamefont {{Torres}},
  \citenamefont {{Remillard}}, \citenamefont {{Bailyn}},\ and\ \citenamefont
  {{Miller}}}]{2011ApJ...730...75O}%
  \BibitemOpen
  \bibfield  {author} {\bibinfo {author} {\bibfnamefont {J.~A.}\ \bibnamefont
  {{Orosz}}}, \bibinfo {author} {\bibfnamefont {J.~F.}\ \bibnamefont
  {{Steiner}}}, \bibinfo {author} {\bibfnamefont {J.~E.}\ \bibnamefont
  {{McClintock}}}, \bibinfo {author} {\bibfnamefont {M.~A.~P.}\ \bibnamefont
  {{Torres}}}, \bibinfo {author} {\bibfnamefont {R.~A.}\ \bibnamefont
  {{Remillard}}}, \bibinfo {author} {\bibfnamefont {C.~D.}\ \bibnamefont
  {{Bailyn}}}, \ and\ \bibinfo {author} {\bibfnamefont {J.~M.}\ \bibnamefont
  {{Miller}}},\ }\bibfield  {title} {\enquote {\bibinfo {title} {{An Improved
  Dynamical Model for the Microquasar XTE J1550-564}},}\ }\href {\doibase
  10.1088/0004-637X/730/2/75} {\bibfield  {journal} {\bibinfo  {journal}
  {\apj}\ }\textbf {\bibinfo {volume} {730}},\ \bibinfo {eid} {75} (\bibinfo
  {year} {2011}{\natexlab{b}})},\ \Eprint {http://arxiv.org/abs/1101.2499}
  {arXiv:1101.2499 [astro-ph.SR]} \BibitemShut {NoStop}%
\bibitem [{\citenamefont {{Steiner}}\ \emph {et~al.}(2011)\citenamefont
  {{Steiner}}, \citenamefont {{Reis}}, \citenamefont {{McClintock}},
  \citenamefont {{Narayan}}, \citenamefont {{Remillard}}, \citenamefont
  {{Orosz}}, \citenamefont {{Gou}}, \citenamefont {{Fabian}},\ and\
  \citenamefont {{Torres}}}]{2011MNRAS.416..941S}%
  \BibitemOpen
  \bibfield  {author} {\bibinfo {author} {\bibfnamefont {J.~F.}\ \bibnamefont
  {{Steiner}}}, \bibinfo {author} {\bibfnamefont {R.~C.}\ \bibnamefont
  {{Reis}}}, \bibinfo {author} {\bibfnamefont {J.~E.}\ \bibnamefont
  {{McClintock}}}, \bibinfo {author} {\bibfnamefont {R.}~\bibnamefont
  {{Narayan}}}, \bibinfo {author} {\bibfnamefont {R.~A.}\ \bibnamefont
  {{Remillard}}}, \bibinfo {author} {\bibfnamefont {J.~A.}\ \bibnamefont
  {{Orosz}}}, \bibinfo {author} {\bibfnamefont {L.}~\bibnamefont {{Gou}}},
  \bibinfo {author} {\bibfnamefont {A.~C.}\ \bibnamefont {{Fabian}}}, \ and\
  \bibinfo {author} {\bibfnamefont {M.~A.~P.}\ \bibnamefont {{Torres}}},\
  }\bibfield  {title} {\enquote {\bibinfo {title} {{The spin of the black hole
  microquasar XTE J1550-564 via the continuum-fitting and Fe-line methods}},}\
  }\href {\doibase 10.1111/j.1365-2966.2011.19089.x} {\bibfield  {journal}
  {\bibinfo  {journal} {\mnras}\ }\textbf {\bibinfo {volume} {416}},\ \bibinfo
  {pages} {941--958} (\bibinfo {year} {2011})},\ \Eprint
  {http://arxiv.org/abs/1010.1013} {arXiv:1010.1013 [astro-ph.HE]} \BibitemShut
  {NoStop}%
\bibitem [{\citenamefont {{Cantrell}}\ \emph {et~al.}(2010)\citenamefont
  {{Cantrell}}, \citenamefont {{Bailyn}}, \citenamefont {{Orosz}},
  \citenamefont {{McClintock}}, \citenamefont {{Remillard}}, \citenamefont
  {{Froning}}, \citenamefont {{Neilsen}}, \citenamefont {{Gelino}},\ and\
  \citenamefont {{Gou}}}]{2010ApJ...710.1127C}%
  \BibitemOpen
  \bibfield  {author} {\bibinfo {author} {\bibfnamefont {A.~G.}\ \bibnamefont
  {{Cantrell}}}, \bibinfo {author} {\bibfnamefont {C.~D.}\ \bibnamefont
  {{Bailyn}}}, \bibinfo {author} {\bibfnamefont {J.~A.}\ \bibnamefont
  {{Orosz}}}, \bibinfo {author} {\bibfnamefont {J.~E.}\ \bibnamefont
  {{McClintock}}}, \bibinfo {author} {\bibfnamefont {R.~A.}\ \bibnamefont
  {{Remillard}}}, \bibinfo {author} {\bibfnamefont {C.~S.}\ \bibnamefont
  {{Froning}}}, \bibinfo {author} {\bibfnamefont {J.}~\bibnamefont
  {{Neilsen}}}, \bibinfo {author} {\bibfnamefont {D.~M.}\ \bibnamefont
  {{Gelino}}}, \ and\ \bibinfo {author} {\bibfnamefont {L.}~\bibnamefont
  {{Gou}}},\ }\bibfield  {title} {\enquote {\bibinfo {title} {{The Inclination
  of the Soft X-Ray Transient A0620-00 and the Mass of its Black Hole}},}\
  }\href {\doibase 10.1088/0004-637X/710/2/1127} {\bibfield  {journal}
  {\bibinfo  {journal} {\apj}\ }\textbf {\bibinfo {volume} {710}},\ \bibinfo
  {pages} {1127--1141} (\bibinfo {year} {2010})},\ \Eprint
  {http://arxiv.org/abs/1001.0261} {arXiv:1001.0261 [astro-ph.HE]} \BibitemShut
  {NoStop}%
\bibitem [{\citenamefont {{Gou}}\ \emph {et~al.}(2010)\citenamefont {{Gou}},
  \citenamefont {{McClintock}}, \citenamefont {{Steiner}}, \citenamefont
  {{Narayan}}, \citenamefont {{Cantrell}}, \citenamefont {{Bailyn}},\ and\
  \citenamefont {{Orosz}}}]{2010ApJ...718L.122G}%
  \BibitemOpen
  \bibfield  {author} {\bibinfo {author} {\bibfnamefont {L.}~\bibnamefont
  {{Gou}}}, \bibinfo {author} {\bibfnamefont {J.~E.}\ \bibnamefont
  {{McClintock}}}, \bibinfo {author} {\bibfnamefont {J.~F.}\ \bibnamefont
  {{Steiner}}}, \bibinfo {author} {\bibfnamefont {R.}~\bibnamefont
  {{Narayan}}}, \bibinfo {author} {\bibfnamefont {A.~G.}\ \bibnamefont
  {{Cantrell}}}, \bibinfo {author} {\bibfnamefont {C.~D.}\ \bibnamefont
  {{Bailyn}}}, \ and\ \bibinfo {author} {\bibfnamefont {J.~A.}\ \bibnamefont
  {{Orosz}}},\ }\bibfield  {title} {\enquote {\bibinfo {title} {{The Spin of
  the Black Hole in the Soft X-ray Transient A0620-00}},}\ }\href {\doibase
  10.1088/2041-8205/718/2/L122} {\bibfield  {journal} {\bibinfo  {journal}
  {\apjl}\ }\textbf {\bibinfo {volume} {718}},\ \bibinfo {pages} {L122--L126}
  (\bibinfo {year} {2010})},\ \Eprint {http://arxiv.org/abs/1002.2211}
  {arXiv:1002.2211 [astro-ph.HE]} \BibitemShut {NoStop}%
\bibitem [{\citenamefont {J.~A.~Orosz}\ and\ \citenamefont
  {(Eds.)}(2003)}]{Orosz67}%
  \BibitemOpen
  \bibfield  {author} {\bibinfo {author} {\bibfnamefont {A.~Herrero}\
  \bibnamefont {J.~A.~Orosz}, \bibfnamefont {K.~van der~Hucht}}\ and\ \bibinfo
  {author} {\bibfnamefont {C.~Esteban}\ \bibnamefont {(Eds.)}},\ }\bibfield
  {title} {\enquote {\bibinfo {title} {{ A Massive Star Odyssey: From Main
  Sequence to Supernova}},}\ }\href@noop {} {\bibfield  {journal} {\bibinfo
  {journal} {IAU Symposium}\ }\textbf {\bibinfo {volume} {Volume 212}},\
  \bibinfo {pages} {p. 365} (\bibinfo {year} {2003})}\BibitemShut {NoStop}%
\bibitem [{\citenamefont {Shafee}\ \emph
  {et~al.}(2006{\natexlab{a}})\citenamefont {Shafee}, \citenamefont
  {McClintock}, \citenamefont {Narayan}, \citenamefont {Davis}, \citenamefont
  {Li},\ and\ \citenamefont {Remillard}}]{Shafee:2005ef}%
  \BibitemOpen
  \bibfield  {author} {\bibinfo {author} {\bibfnamefont {Rebecca}\ \bibnamefont
  {Shafee}}, \bibinfo {author} {\bibfnamefont {Jeffrey~E.}\ \bibnamefont
  {McClintock}}, \bibinfo {author} {\bibfnamefont {Ramesh}\ \bibnamefont
  {Narayan}}, \bibinfo {author} {\bibfnamefont {Shane~W.}\ \bibnamefont
  {Davis}}, \bibinfo {author} {\bibfnamefont {Li-Xin}\ \bibnamefont {Li}}, \
  and\ \bibinfo {author} {\bibfnamefont {Ronald~A.}\ \bibnamefont
  {Remillard}},\ }\bibfield  {title} {\enquote {\bibinfo {title} {{Estimating
  the spin of stellar-mass black holes via spectral fitting of the x-ray
  continuum}},}\ }\href {\doibase 10.1086/498938} {\bibfield  {journal}
  {\bibinfo  {journal} {Astrophys. J.}\ }\textbf {\bibinfo {volume} {636}},\
  \bibinfo {pages} {L113--L116} (\bibinfo {year} {2006}{\natexlab{a}})},\
  \Eprint {http://arxiv.org/abs/astro-ph/0508302} {arXiv:astro-ph/0508302
  [astro-ph]} \BibitemShut {NoStop}%
\bibitem [{\citenamefont {Shafee}\ \emph
  {et~al.}(2006{\natexlab{b}})\citenamefont {Shafee}, \citenamefont
  {McClintock}, \citenamefont {Narayan}, \citenamefont {Davis}, \citenamefont
  {Li},\ and\ \citenamefont {Remillard}}]{1538-4357-636-2-L113}%
  \BibitemOpen
  \bibfield  {author} {\bibinfo {author} {\bibfnamefont {Rebecca}\ \bibnamefont
  {Shafee}}, \bibinfo {author} {\bibfnamefont {Jeffrey~E.}\ \bibnamefont
  {McClintock}}, \bibinfo {author} {\bibfnamefont {Ramesh}\ \bibnamefont
  {Narayan}}, \bibinfo {author} {\bibfnamefont {Shane~W.}\ \bibnamefont
  {Davis}}, \bibinfo {author} {\bibfnamefont {Li-Xin}\ \bibnamefont {Li}}, \
  and\ \bibinfo {author} {\bibfnamefont {Ronald~A.}\ \bibnamefont
  {Remillard}},\ }\bibfield  {title} {\enquote {\bibinfo {title} {Estimating
  the spin of stellar-mass black holes by spectral fitting of the x-ray
  continuum},}\ }\href {http://stacks.iop.org/1538-4357/636/i=2/a=L113}
  {\bibfield  {journal} {\bibinfo  {journal} {The Astrophysical Journal
  Letters}\ }\textbf {\bibinfo {volume} {636}},\ \bibinfo {pages} {L113}
  (\bibinfo {year} {2006}{\natexlab{b}})}\BibitemShut {NoStop}%
\bibitem [{\citenamefont {Greene}\ \emph {et~al.}(2001)\citenamefont {Greene},
  \citenamefont {Bailyn},\ and\ \citenamefont {Orosz}}]{Greene:2001wd}%
  \BibitemOpen
  \bibfield  {author} {\bibinfo {author} {\bibfnamefont {Jenny}\ \bibnamefont
  {Greene}}, \bibinfo {author} {\bibfnamefont {Charles~D.}\ \bibnamefont
  {Bailyn}}, \ and\ \bibinfo {author} {\bibfnamefont {Jerome~A.}\ \bibnamefont
  {Orosz}},\ }\bibfield  {title} {\enquote {\bibinfo {title} {{Optical and
  infrared photometry of the micro-quasar gro j1655-40 in quiescence}},}\
  }\href {\doibase 10.1086/321411} {\bibfield  {journal} {\bibinfo  {journal}
  {Astrophys. J.}\ }\textbf {\bibinfo {volume} {554}},\ \bibinfo {pages} {1290}
  (\bibinfo {year} {2001})},\ \Eprint {http://arxiv.org/abs/astro-ph/0101337}
  {arXiv:astro-ph/0101337 [astro-ph]} \BibitemShut {NoStop}%
\bibitem [{\citenamefont {Steeghs}\ \emph {et~al.}(2013)\citenamefont
  {Steeghs}, \citenamefont {McClintock}, \citenamefont {Parsons}, \citenamefont
  {Reid}, \citenamefont {Littlefair},\ and\ \citenamefont
  {Dhillon}}]{Steeghs:2013ksa}%
  \BibitemOpen
  \bibfield  {author} {\bibinfo {author} {\bibfnamefont {D.}~\bibnamefont
  {Steeghs}}, \bibinfo {author} {\bibfnamefont {J.~E.}\ \bibnamefont
  {McClintock}}, \bibinfo {author} {\bibfnamefont {S.~G.}\ \bibnamefont
  {Parsons}}, \bibinfo {author} {\bibfnamefont {M.~J.}\ \bibnamefont {Reid}},
  \bibinfo {author} {\bibfnamefont {S.}~\bibnamefont {Littlefair}}, \ and\
  \bibinfo {author} {\bibfnamefont {V.~S.}\ \bibnamefont {Dhillon}},\
  }\bibfield  {title} {\enquote {\bibinfo {title} {{The not-so-massive black
  hole in the microquasar GRS1915+105}},}\ }\href {\doibase
  10.1088/0004-637X/768/2/185} {\bibfield  {journal} {\bibinfo  {journal}
  {Astrophys. J.}\ }\textbf {\bibinfo {volume} {768}},\ \bibinfo {pages} {185}
  (\bibinfo {year} {2013})},\ \Eprint {http://arxiv.org/abs/1304.1808}
  {arXiv:1304.1808 [astro-ph.HE]} \BibitemShut {NoStop}%
\bibitem [{\citenamefont {McClintock}\ \emph {et~al.}(2006)\citenamefont
  {McClintock}, \citenamefont {Shafee}, \citenamefont {Narayan}, \citenamefont
  {Remillard}, \citenamefont {Davis},\ and\ \citenamefont
  {Li}}]{McClintock:2006xd}%
  \BibitemOpen
  \bibfield  {author} {\bibinfo {author} {\bibfnamefont {Jeffrey~E.}\
  \bibnamefont {McClintock}}, \bibinfo {author} {\bibfnamefont {Rebecca}\
  \bibnamefont {Shafee}}, \bibinfo {author} {\bibfnamefont {Ramesh}\
  \bibnamefont {Narayan}}, \bibinfo {author} {\bibfnamefont {Ronald~A.}\
  \bibnamefont {Remillard}}, \bibinfo {author} {\bibfnamefont {Shane~W.}\
  \bibnamefont {Davis}}, \ and\ \bibinfo {author} {\bibfnamefont {Li-Xin}\
  \bibnamefont {Li}},\ }\bibfield  {title} {\enquote {\bibinfo {title} {{The
  Spin of the Near-Extreme Kerr Black Hole GRS 1915+105}},}\ }\href {\doibase
  10.1086/508457} {\bibfield  {journal} {\bibinfo  {journal} {Astrophys. J.}\
  }\textbf {\bibinfo {volume} {652}},\ \bibinfo {pages} {518--539} (\bibinfo
  {year} {2006})},\ \Eprint {http://arxiv.org/abs/astro-ph/0606076}
  {arXiv:astro-ph/0606076 [astro-ph]} \BibitemShut {NoStop}%
\bibitem [{\citenamefont {Orosz}\ \emph {et~al.}(2009)\citenamefont {Orosz}
  \emph {et~al.}}]{Orosz:2008kk}%
  \BibitemOpen
  \bibfield  {author} {\bibinfo {author} {\bibfnamefont {Jerome~A.}\
  \bibnamefont {Orosz}} \emph {et~al.},\ }\bibfield  {title} {\enquote
  {\bibinfo {title} {{A New Dynamical Model for the Black Hole Binary LMC
  X-1}},}\ }\href {\doibase 10.1088/0004-637X/697/1/573} {\bibfield  {journal}
  {\bibinfo  {journal} {Astrophys. J.}\ }\textbf {\bibinfo {volume} {697}},\
  \bibinfo {pages} {573--591} (\bibinfo {year} {2009})},\ \Eprint
  {http://arxiv.org/abs/0810.3447} {arXiv:0810.3447 [astro-ph]} \BibitemShut
  {NoStop}%
\bibitem [{\citenamefont {{Gou}}\ \emph {et~al.}(2009)\citenamefont {{Gou}},
  \citenamefont {{McClintock}}, \citenamefont {{Liu}}, \citenamefont
  {{Narayan}}, \citenamefont {{Steiner}}, \citenamefont {{Remillard}},
  \citenamefont {{Orosz}}, \citenamefont {{Davis}}, \citenamefont {{Ebisawa}},\
  and\ \citenamefont {{Schlegel}}}]{2009ApJ...701.1076G}%
  \BibitemOpen
  \bibfield  {author} {\bibinfo {author} {\bibfnamefont {L.}~\bibnamefont
  {{Gou}}}, \bibinfo {author} {\bibfnamefont {J.~E.}\ \bibnamefont
  {{McClintock}}}, \bibinfo {author} {\bibfnamefont {J.}~\bibnamefont {{Liu}}},
  \bibinfo {author} {\bibfnamefont {R.}~\bibnamefont {{Narayan}}}, \bibinfo
  {author} {\bibfnamefont {J.~F.}\ \bibnamefont {{Steiner}}}, \bibinfo {author}
  {\bibfnamefont {R.~A.}\ \bibnamefont {{Remillard}}}, \bibinfo {author}
  {\bibfnamefont {J.~A.}\ \bibnamefont {{Orosz}}}, \bibinfo {author}
  {\bibfnamefont {S.~W.}\ \bibnamefont {{Davis}}}, \bibinfo {author}
  {\bibfnamefont {K.}~\bibnamefont {{Ebisawa}}}, \ and\ \bibinfo {author}
  {\bibfnamefont {E.~M.}\ \bibnamefont {{Schlegel}}},\ }\bibfield  {title}
  {\enquote {\bibinfo {title} {{A Determination of the Spin of the Black Hole
  Primary in LMC X-1}},}\ }\href {\doibase 10.1088/0004-637X/701/2/1076}
  {\bibfield  {journal} {\bibinfo  {journal} {\apj}\ }\textbf {\bibinfo
  {volume} {701}},\ \bibinfo {pages} {1076--1090} (\bibinfo {year} {2009})},\
  \Eprint {http://arxiv.org/abs/0901.0920} {arXiv:0901.0920 [astro-ph.HE]}
  \BibitemShut {NoStop}%
\bibitem [{\citenamefont {{Orosz}}\ \emph {et~al.}(2014)\citenamefont
  {{Orosz}}, \citenamefont {{Steiner}}, \citenamefont {{McClintock}},
  \citenamefont {{Buxton}}, \citenamefont {{Bailyn}}, \citenamefont
  {{Steeghs}}, \citenamefont {{Guberman}},\ and\ \citenamefont
  {{Torres}}}]{2014ApJ...794..154O}%
  \BibitemOpen
  \bibfield  {author} {\bibinfo {author} {\bibfnamefont {J.~A.}\ \bibnamefont
  {{Orosz}}}, \bibinfo {author} {\bibfnamefont {J.~F.}\ \bibnamefont
  {{Steiner}}}, \bibinfo {author} {\bibfnamefont {J.~E.}\ \bibnamefont
  {{McClintock}}}, \bibinfo {author} {\bibfnamefont {M.~M.}\ \bibnamefont
  {{Buxton}}}, \bibinfo {author} {\bibfnamefont {C.~D.}\ \bibnamefont
  {{Bailyn}}}, \bibinfo {author} {\bibfnamefont {D.}~\bibnamefont {{Steeghs}}},
  \bibinfo {author} {\bibfnamefont {A.}~\bibnamefont {{Guberman}}}, \ and\
  \bibinfo {author} {\bibfnamefont {M.~A.~P.}\ \bibnamefont {{Torres}}},\
  }\bibfield  {title} {\enquote {\bibinfo {title} {{The Mass of the Black Hole
  in LMC X-3}},}\ }\href {\doibase 10.1088/0004-637X/794/2/154} {\bibfield
  {journal} {\bibinfo  {journal} {\apj}\ }\textbf {\bibinfo {volume} {794}},\
  \bibinfo {eid} {154} (\bibinfo {year} {2014})},\ \Eprint
  {http://arxiv.org/abs/1402.0085} {arXiv:1402.0085 [astro-ph.SR]} \BibitemShut
  {NoStop}%
\bibitem [{\citenamefont {Steiner}\ \emph {et~al.}(2014)\citenamefont
  {Steiner}, \citenamefont {McClintock}, \citenamefont {Orosz}, \citenamefont
  {Remillard}, \citenamefont {Bailyn}, \citenamefont {Kolehmainen},\ and\
  \citenamefont {Straub}}]{Steiner:2014zha}%
  \BibitemOpen
  \bibfield  {author} {\bibinfo {author} {\bibfnamefont {James~F.}\
  \bibnamefont {Steiner}}, \bibinfo {author} {\bibfnamefont {Jeffrey~E.}\
  \bibnamefont {McClintock}}, \bibinfo {author} {\bibfnamefont {Jerome~A.}\
  \bibnamefont {Orosz}}, \bibinfo {author} {\bibfnamefont {Ronald~A.}\
  \bibnamefont {Remillard}}, \bibinfo {author} {\bibfnamefont {Charles~D.}\
  \bibnamefont {Bailyn}}, \bibinfo {author} {\bibfnamefont {Mari}\ \bibnamefont
  {Kolehmainen}}, \ and\ \bibinfo {author} {\bibfnamefont {Odele}\ \bibnamefont
  {Straub}},\ }\bibfield  {title} {\enquote {\bibinfo {title} {{The Low-Spin
  Black Hole in LMC X-3}},}\ }\href {\doibase 10.1088/2041-8205/793/2/L29}
  {\bibfield  {journal} {\bibinfo  {journal} {Astrophys. J.}\ }\textbf
  {\bibinfo {volume} {793}},\ \bibinfo {pages} {L29} (\bibinfo {year}
  {2014})},\ \Eprint {http://arxiv.org/abs/1402.0148} {arXiv:1402.0148
  [astro-ph.HE]} \BibitemShut {NoStop}%
\bibitem [{\citenamefont {Orosz}\ \emph {et~al.}(2007)\citenamefont {Orosz}
  \emph {et~al.}}]{Orosz:2007ng}%
  \BibitemOpen
  \bibfield  {author} {\bibinfo {author} {\bibfnamefont {Jerome~A.}\
  \bibnamefont {Orosz}} \emph {et~al.},\ }\bibfield  {title} {\enquote
  {\bibinfo {title} {{A 15.65 solar mass black hole in an eclipsing binary in
  the nearby spiral galaxy Messier 33}},}\ }\href {\doibase
  10.1038/nature06218} {\bibfield  {journal} {\bibinfo  {journal} {Nature}\
  }\textbf {\bibinfo {volume} {449}},\ \bibinfo {pages} {872} (\bibinfo {year}
  {2007})},\ \Eprint {http://arxiv.org/abs/0710.3165} {arXiv:0710.3165
  [astro-ph]} \BibitemShut {NoStop}%
\bibitem [{\citenamefont {Liu}\ \emph {et~al.}(2008)\citenamefont {Liu},
  \citenamefont {McClintock}, \citenamefont {Narayan}, \citenamefont {Davis},\
  and\ \citenamefont {Orosz}}]{1538-4357-679-1-L37}%
  \BibitemOpen
  \bibfield  {author} {\bibinfo {author} {\bibfnamefont {Jifeng}\ \bibnamefont
  {Liu}}, \bibinfo {author} {\bibfnamefont {Jeffrey~E.}\ \bibnamefont
  {McClintock}}, \bibinfo {author} {\bibfnamefont {Ramesh}\ \bibnamefont
  {Narayan}}, \bibinfo {author} {\bibfnamefont {Shane~W.}\ \bibnamefont
  {Davis}}, \ and\ \bibinfo {author} {\bibfnamefont {Jerome~A.}\ \bibnamefont
  {Orosz}},\ }\bibfield  {title} {\enquote {\bibinfo {title} {Precise
  measurement of the spin parameter of the stellar-mass black hole m33 x-7},}\
  }\href {http://stacks.iop.org/1538-4357/679/i=1/a=L37} {\bibfield  {journal}
  {\bibinfo  {journal} {The Astrophysical Journal Letters}\ }\textbf {\bibinfo
  {volume} {679}},\ \bibinfo {pages} {L37} (\bibinfo {year}
  {2008})}\BibitemShut {NoStop}%
\bibitem [{\citenamefont {Peterson}\ \emph {et~al.}(2004)\citenamefont
  {Peterson} \emph {et~al.}}]{Peterson:2004nu}%
  \BibitemOpen
  \bibfield  {author} {\bibinfo {author} {\bibfnamefont {Bradley~M.}\
  \bibnamefont {Peterson}} \emph {et~al.},\ }\bibfield  {title} {\enquote
  {\bibinfo {title} {{Central masses and broad-line region sizes of active
  galactic nuclei. II. A Homogeneous analysis of a large reverberation-mapping
  database}},}\ }\href {\doibase 10.1086/423269} {\bibfield  {journal}
  {\bibinfo  {journal} {Astrophys. J.}\ }\textbf {\bibinfo {volume} {613}},\
  \bibinfo {pages} {682--699} (\bibinfo {year} {2004})},\ \Eprint
  {http://arxiv.org/abs/astro-ph/0407299} {arXiv:astro-ph/0407299 [astro-ph]}
  \BibitemShut {NoStop}%
\bibitem [{\citenamefont {{Walton}}\ \emph {et~al.}(2013)\citenamefont
  {{Walton}}, \citenamefont {{Nardini}}, \citenamefont {{Fabian}},
  \citenamefont {{Gallo}},\ and\ \citenamefont {{Reis}}}]{2013MNRAS.428.2901W}%
  \BibitemOpen
  \bibfield  {author} {\bibinfo {author} {\bibfnamefont {D.~J.}\ \bibnamefont
  {{Walton}}}, \bibinfo {author} {\bibfnamefont {E.}~\bibnamefont {{Nardini}}},
  \bibinfo {author} {\bibfnamefont {A.~C.}\ \bibnamefont {{Fabian}}}, \bibinfo
  {author} {\bibfnamefont {L.~C.}\ \bibnamefont {{Gallo}}}, \ and\ \bibinfo
  {author} {\bibfnamefont {R.~C.}\ \bibnamefont {{Reis}}},\ }\bibfield  {title}
  {\enquote {\bibinfo {title} {{Suzaku observations of `bare' active galactic
  nuclei}},}\ }\href {\doibase 10.1093/mnras/sts227} {\bibfield  {journal}
  {\bibinfo  {journal} {\mnras}\ }\textbf {\bibinfo {volume} {428}},\ \bibinfo
  {pages} {2901--2920} (\bibinfo {year} {2013})},\ \Eprint
  {http://arxiv.org/abs/1210.4593} {arXiv:1210.4593 [astro-ph.HE]} \BibitemShut
  {NoStop}%
\bibitem [{\citenamefont {{Lohfink}}\ \emph {et~al.}(2012)\citenamefont
  {{Lohfink}}, \citenamefont {{Reynolds}}, \citenamefont {{Miller}},
  \citenamefont {{Brenneman}}, \citenamefont {{Mushotzky}}, \citenamefont
  {{Nowak}},\ and\ \citenamefont {{Fabian}}}]{2012ApJ...758...67L}%
  \BibitemOpen
  \bibfield  {author} {\bibinfo {author} {\bibfnamefont {A.~M.}\ \bibnamefont
  {{Lohfink}}}, \bibinfo {author} {\bibfnamefont {C.~S.}\ \bibnamefont
  {{Reynolds}}}, \bibinfo {author} {\bibfnamefont {J.~M.}\ \bibnamefont
  {{Miller}}}, \bibinfo {author} {\bibfnamefont {L.~W.}\ \bibnamefont
  {{Brenneman}}}, \bibinfo {author} {\bibfnamefont {R.~F.}\ \bibnamefont
  {{Mushotzky}}}, \bibinfo {author} {\bibfnamefont {M.~A.}\ \bibnamefont
  {{Nowak}}}, \ and\ \bibinfo {author} {\bibfnamefont {A.~C.}\ \bibnamefont
  {{Fabian}}},\ }\bibfield  {title} {\enquote {\bibinfo {title} {{The Black
  Hole Spin and Soft X-Ray Excess of the Luminous Seyfert Galaxy Fairall 9}},}\
  }\href {\doibase 10.1088/0004-637X/758/1/67} {\bibfield  {journal} {\bibinfo
  {journal} {\apj}\ }\textbf {\bibinfo {volume} {758}},\ \bibinfo {eid} {67}
  (\bibinfo {year} {2012})},\ \Eprint {http://arxiv.org/abs/1209.0468}
  {arXiv:1209.0468 [astro-ph.HE]} \BibitemShut {NoStop}%
\bibitem [{\citenamefont {{Gallo}}\ \emph {et~al.}(2011)\citenamefont
  {{Gallo}}, \citenamefont {{Miniutti}}, \citenamefont {{Miller}},
  \citenamefont {{Brenneman}}, \citenamefont {{Fabian}}, \citenamefont
  {{Guainazzi}},\ and\ \citenamefont {{Reynolds}}}]{2011MNRAS.411..607G}%
  \BibitemOpen
  \bibfield  {author} {\bibinfo {author} {\bibfnamefont {L.~C.}\ \bibnamefont
  {{Gallo}}}, \bibinfo {author} {\bibfnamefont {G.}~\bibnamefont {{Miniutti}}},
  \bibinfo {author} {\bibfnamefont {J.~M.}\ \bibnamefont {{Miller}}}, \bibinfo
  {author} {\bibfnamefont {L.~W.}\ \bibnamefont {{Brenneman}}}, \bibinfo
  {author} {\bibfnamefont {A.~C.}\ \bibnamefont {{Fabian}}}, \bibinfo {author}
  {\bibfnamefont {M.}~\bibnamefont {{Guainazzi}}}, \ and\ \bibinfo {author}
  {\bibfnamefont {C.~S.}\ \bibnamefont {{Reynolds}}},\ }\bibfield  {title}
  {\enquote {\bibinfo {title} {{Multi-epoch X-ray observations of the Seyfert
  1.2 galaxy Mrk 79: bulk motion of the illuminating X-ray source}},}\ }\href
  {\doibase 10.1111/j.1365-2966.2010.17705.x} {\bibfield  {journal} {\bibinfo
  {journal} {\mnras}\ }\textbf {\bibinfo {volume} {411}},\ \bibinfo {pages}
  {607--619} (\bibinfo {year} {2011})},\ \Eprint
  {http://arxiv.org/abs/1009.2987} {arXiv:1009.2987 [astro-ph.HE]} \BibitemShut
  {NoStop}%
\bibitem [{\citenamefont {{Brenneman}}\ \emph
  {et~al.}(2011{\natexlab{b}})\citenamefont {{Brenneman}}, \citenamefont
  {{Reynolds}}, \citenamefont {{Nowak}}, \citenamefont {{Reis}}, \citenamefont
  {{Trippe}}, \citenamefont {{Fabian}}, \citenamefont {{Iwasawa}},
  \citenamefont {{Lee}}, \citenamefont {{Miller}}, \citenamefont {{Mushotzky}},
  \citenamefont {{Nandra}},\ and\ \citenamefont {{Volonteri}}}]{2011Brenneman}%
  \BibitemOpen
  \bibfield  {author} {\bibinfo {author} {\bibfnamefont {L.~W.}\ \bibnamefont
  {{Brenneman}}}, \bibinfo {author} {\bibfnamefont {C.~S.}\ \bibnamefont
  {{Reynolds}}}, \bibinfo {author} {\bibfnamefont {M.~A.}\ \bibnamefont
  {{Nowak}}}, \bibinfo {author} {\bibfnamefont {R.~C.}\ \bibnamefont {{Reis}}},
  \bibinfo {author} {\bibfnamefont {M.}~\bibnamefont {{Trippe}}}, \bibinfo
  {author} {\bibfnamefont {A.~C.}\ \bibnamefont {{Fabian}}}, \bibinfo {author}
  {\bibfnamefont {K.}~\bibnamefont {{Iwasawa}}}, \bibinfo {author}
  {\bibfnamefont {J.~C.}\ \bibnamefont {{Lee}}}, \bibinfo {author}
  {\bibfnamefont {J.~M.}\ \bibnamefont {{Miller}}}, \bibinfo {author}
  {\bibfnamefont {R.~F.}\ \bibnamefont {{Mushotzky}}}, \bibinfo {author}
  {\bibfnamefont {K.}~\bibnamefont {{Nandra}}}, \ and\ \bibinfo {author}
  {\bibfnamefont {M.}~\bibnamefont {{Volonteri}}},\ }\bibfield  {title}
  {\enquote {\bibinfo {title} {{The Spin of the Supermassive Black Hole in NGC
  3783}},}\ }\href {\doibase 10.1088/0004-637X/736/2/103} {\bibfield  {journal}
  {\bibinfo  {journal} {\apj}\ }\textbf {\bibinfo {volume} {736}},\ \bibinfo
  {eid} {103} (\bibinfo {year} {2011}{\natexlab{b}})},\ \Eprint
  {http://arxiv.org/abs/1104.1172} {arXiv:1104.1172 [astro-ph.HE]} \BibitemShut
  {NoStop}%
\bibitem [{\citenamefont {McHardy}\ \emph {et~al.}(2005)\citenamefont
  {McHardy}, \citenamefont {Gunn}, \citenamefont {Uttley},\ and\ \citenamefont
  {Goad}}]{McHardy:2005ut}%
  \BibitemOpen
  \bibfield  {author} {\bibinfo {author} {\bibfnamefont {Ian~M.}\ \bibnamefont
  {McHardy}}, \bibinfo {author} {\bibfnamefont {K.~F.}\ \bibnamefont {Gunn}},
  \bibinfo {author} {\bibfnamefont {P.}~\bibnamefont {Uttley}}, \ and\ \bibinfo
  {author} {\bibfnamefont {M.~R.}\ \bibnamefont {Goad}},\ }\bibfield  {title}
  {\enquote {\bibinfo {title} {{MCG-6-30-15: Long timescale x-ray variability,
  black hole mass and AGN high states}},}\ }\href {\doibase
  10.1111/j.1365-2966.2005.08992.x} {\bibfield  {journal} {\bibinfo  {journal}
  {Mon. Not. Roy. Astron. Soc.}\ }\textbf {\bibinfo {volume} {359}},\ \bibinfo
  {pages} {1469--1480} (\bibinfo {year} {2005})},\ \Eprint
  {http://arxiv.org/abs/astro-ph/0503100} {arXiv:astro-ph/0503100 [astro-ph]}
  \BibitemShut {NoStop}%
\bibitem [{\citenamefont {Brenneman}\ and\ \citenamefont
  {Reynolds}(2006)}]{Brenneman:2006hw}%
  \BibitemOpen
  \bibfield  {author} {\bibinfo {author} {\bibfnamefont {Laura~W.}\
  \bibnamefont {Brenneman}}\ and\ \bibinfo {author} {\bibfnamefont
  {Christopher~S.}\ \bibnamefont {Reynolds}},\ }\bibfield  {title} {\enquote
  {\bibinfo {title} {{Constraining Black Hole Spin Via X-ray Spectroscopy}},}\
  }\href {\doibase 10.1086/508146} {\bibfield  {journal} {\bibinfo  {journal}
  {Astrophys. J.}\ }\textbf {\bibinfo {volume} {652}},\ \bibinfo {pages}
  {1028--1043} (\bibinfo {year} {2006})},\ \Eprint
  {http://arxiv.org/abs/astro-ph/0608502} {arXiv:astro-ph/0608502 [astro-ph]}
  \BibitemShut {NoStop}%
\bibitem [{\citenamefont {Patrick}\ \emph {et~al.}(2011)\citenamefont
  {Patrick}, \citenamefont {Reeves}, \citenamefont {Lobban}, \citenamefont
  {Porquet},\ and\ \citenamefont
  {Markowitz}}]{doi:10.1111/j.1365-2966.2011.19224.x}%
  \BibitemOpen
  \bibfield  {author} {\bibinfo {author} {\bibfnamefont {A.~R.}\ \bibnamefont
  {Patrick}}, \bibinfo {author} {\bibfnamefont {J.~N.}\ \bibnamefont {Reeves}},
  \bibinfo {author} {\bibfnamefont {A.~P.}\ \bibnamefont {Lobban}}, \bibinfo
  {author} {\bibfnamefont {D.}~\bibnamefont {Porquet}}, \ and\ \bibinfo
  {author} {\bibfnamefont {A.~G.}\ \bibnamefont {Markowitz}},\ }\bibfield
  {title} {\enquote {\bibinfo {title} {Assessing black hole spin in deep suzaku
  observations of seyfert 1 agn},}\ }\href {\doibase
  10.1111/j.1365-2966.2011.19224.x} {\bibfield  {journal} {\bibinfo  {journal}
  {Monthly Notices of the Royal Astronomical Society}\ }\textbf {\bibinfo
  {volume} {416}},\ \bibinfo {pages} {2725--2747} (\bibinfo {year}
  {2011})}\BibitemShut {NoStop}%
\bibitem [{\citenamefont {Patrick}\ \emph {et~al.}(2012)\citenamefont
  {Patrick}, \citenamefont {Reeves}, \citenamefont {Porquet}, \citenamefont
  {Markowitz}, \citenamefont {Braito},\ and\ \citenamefont {Lobban}}]{8175999}%
  \BibitemOpen
  \bibfield  {author} {\bibinfo {author} {\bibfnamefont {A.~R.}\ \bibnamefont
  {Patrick}}, \bibinfo {author} {\bibfnamefont {J.~N.}\ \bibnamefont {Reeves}},
  \bibinfo {author} {\bibfnamefont {D.}~\bibnamefont {Porquet}}, \bibinfo
  {author} {\bibfnamefont {A.~G.}\ \bibnamefont {Markowitz}}, \bibinfo {author}
  {\bibfnamefont {V.}~\bibnamefont {Braito}}, \ and\ \bibinfo {author}
  {\bibfnamefont {A.~P.}\ \bibnamefont {Lobban}},\ }\bibfield  {title}
  {\enquote {\bibinfo {title} {A suzaku survey of fe \#x2009;k lines in seyfert
  1 active galactic nuclei},}\ }\href {\doibase
  10.1111/j.1365-2966.2012.21868.x} {\bibfield  {journal} {\bibinfo  {journal}
  {Monthly Notices of the Royal Astronomical Society}\ }\textbf {\bibinfo
  {volume} {426}},\ \bibinfo {pages} {2522--2565} (\bibinfo {year}
  {2012})}\BibitemShut {NoStop}%
\bibitem [{\citenamefont {Johnstone}(2001)}]{johnstone2001}%
  \BibitemOpen
  \bibfield  {author} {\bibinfo {author} {\bibfnamefont {Iain~M.}\ \bibnamefont
  {Johnstone}},\ }\bibfield  {title} {\enquote {\bibinfo {title} {On the
  distribution of the largest eigenvalue in principal components analysis},}\
  }\href {\doibase 10.1214/aos/1009210544} {\bibfield  {journal} {\bibinfo
  {journal} {Ann. Statist.}\ }\textbf {\bibinfo {volume} {29}},\ \bibinfo
  {pages} {295--327} (\bibinfo {year} {2001})}\BibitemShut {NoStop}%
\bibitem [{\citenamefont {{Kachru}}\ \emph {et~al.}(2003)\citenamefont
  {{Kachru}}, \citenamefont {{Kallosh}}, \citenamefont {{Linde}},\ and\
  \citenamefont {{Trivedi}}}]{2003PhRvD..68d6005K}%
  \BibitemOpen
  \bibfield  {author} {\bibinfo {author} {\bibfnamefont {S.}~\bibnamefont
  {{Kachru}}}, \bibinfo {author} {\bibfnamefont {R.}~\bibnamefont {{Kallosh}}},
  \bibinfo {author} {\bibfnamefont {A.}~\bibnamefont {{Linde}}}, \ and\
  \bibinfo {author} {\bibfnamefont {S.~P.}\ \bibnamefont {{Trivedi}}},\
  }\bibfield  {title} {\enquote {\bibinfo {title} {{de Sitter vacua in string
  theory}},}\ }\href {\doibase 10.1103/PhysRevD.68.046005} {\bibfield
  {journal} {\bibinfo  {journal} {\prd}\ }\textbf {\bibinfo {volume} {68}},\
  \bibinfo {eid} {046005} (\bibinfo {year} {2003})},\ \Eprint
  {http://arxiv.org/abs/hep-th/0301240} {hep-th/0301240} \BibitemShut {NoStop}%
\bibitem [{\citenamefont {{Bowman}}\ \emph {et~al.}(2018)\citenamefont
  {{Bowman}}, \citenamefont {{Rogers}}, \citenamefont {{Monsalve}},
  \citenamefont {{Mozdzen}},\ and\ \citenamefont
  {{Mahesh}}}]{2018Natur.555...67B}%
  \BibitemOpen
  \bibfield  {author} {\bibinfo {author} {\bibfnamefont {J.~D.}\ \bibnamefont
  {{Bowman}}}, \bibinfo {author} {\bibfnamefont {A.~E.~E.}\ \bibnamefont
  {{Rogers}}}, \bibinfo {author} {\bibfnamefont {R.~A.}\ \bibnamefont
  {{Monsalve}}}, \bibinfo {author} {\bibfnamefont {T.~J.}\ \bibnamefont
  {{Mozdzen}}}, \ and\ \bibinfo {author} {\bibfnamefont {N.}~\bibnamefont
  {{Mahesh}}},\ }\bibfield  {title} {\enquote {\bibinfo {title} {{An absorption
  profile centred at 78 megahertz in the sky-averaged spectrum}},}\ }\href
  {\doibase 10.1038/nature25792} {\bibfield  {journal} {\bibinfo  {journal}
  {\nat}\ }\textbf {\bibinfo {volume} {555}},\ \bibinfo {pages} {67--70}
  (\bibinfo {year} {2018})}\BibitemShut {NoStop}%
\bibitem [{\citenamefont {Lidz}\ and\ \citenamefont
  {Hui}(2018)}]{Lidz:2018fqo}%
  \BibitemOpen
  \bibfield  {author} {\bibinfo {author} {\bibfnamefont {Adam}\ \bibnamefont
  {Lidz}}\ and\ \bibinfo {author} {\bibfnamefont {Lam}\ \bibnamefont {Hui}},\
  }\bibfield  {title} {\enquote {\bibinfo {title} {{The Implications of a
  Pre-reionization 21 cm Absorption Signal for Fuzzy Dark Matter}},}\
  }\href@noop {} {\  (\bibinfo {year} {2018})},\ \Eprint
  {http://arxiv.org/abs/1805.01253} {arXiv:1805.01253 [astro-ph.CO]}
  \BibitemShut {NoStop}%
\bibitem [{\citenamefont {Schneider}(2018)}]{Schneider:2018xba}%
  \BibitemOpen
  \bibfield  {author} {\bibinfo {author} {\bibfnamefont {Aurel}\ \bibnamefont
  {Schneider}},\ }\bibfield  {title} {\enquote {\bibinfo {title} {{Constraining
  Non-Cold Dark Matter Models with the Global 21-cm Signal}},}\ }\href@noop {}
  {\  (\bibinfo {year} {2018})},\ \Eprint {http://arxiv.org/abs/1805.00021}
  {arXiv:1805.00021 [astro-ph.CO]} \BibitemShut {NoStop}%
\bibitem [{\citenamefont {{Helfer}}\ \emph {et~al.}(2017)\citenamefont
  {{Helfer}}, \citenamefont {{Marsh}}, \citenamefont {{Clough}}, \citenamefont
  {{Fairbairn}}, \citenamefont {{Lim}},\ and\ \citenamefont
  {{Becerril}}}]{2017JCAP...03..055H}%
  \BibitemOpen
  \bibfield  {author} {\bibinfo {author} {\bibfnamefont {T.}~\bibnamefont
  {{Helfer}}}, \bibinfo {author} {\bibfnamefont {D.~J.~E.}\ \bibnamefont
  {{Marsh}}}, \bibinfo {author} {\bibfnamefont {K.}~\bibnamefont {{Clough}}},
  \bibinfo {author} {\bibfnamefont {M.}~\bibnamefont {{Fairbairn}}}, \bibinfo
  {author} {\bibfnamefont {E.~A.}\ \bibnamefont {{Lim}}}, \ and\ \bibinfo
  {author} {\bibfnamefont {R.}~\bibnamefont {{Becerril}}},\ }\bibfield  {title}
  {\enquote {\bibinfo {title} {{Black hole formation from axion stars}},}\
  }\href {\doibase 10.1088/1475-7516/2017/03/055} {\bibfield  {journal}
  {\bibinfo  {journal} {\jcap}\ }\textbf {\bibinfo {volume} {3}},\ \bibinfo
  {eid} {055} (\bibinfo {year} {2017})},\ \Eprint
  {http://arxiv.org/abs/1609.04724} {arXiv:1609.04724} \BibitemShut {NoStop}%
\bibitem [{\citenamefont {{Levkov}}\ \emph {et~al.}(2017)\citenamefont
  {{Levkov}}, \citenamefont {{Panin}},\ and\ \citenamefont
  {{Tkachev}}}]{2017PhRvL.118a1301L}%
  \BibitemOpen
  \bibfield  {author} {\bibinfo {author} {\bibfnamefont {D.~G.}\ \bibnamefont
  {{Levkov}}}, \bibinfo {author} {\bibfnamefont {A.~G.}\ \bibnamefont
  {{Panin}}}, \ and\ \bibinfo {author} {\bibfnamefont {I.~I.}\ \bibnamefont
  {{Tkachev}}},\ }\bibfield  {title} {\enquote {\bibinfo {title} {{Relativistic
  Axions from Collapsing Bose Stars}},}\ }\href {\doibase
  10.1103/PhysRevLett.118.011301} {\bibfield  {journal} {\bibinfo  {journal}
  {Physical Review Letters}\ }\textbf {\bibinfo {volume} {118}},\ \bibinfo
  {eid} {011301} (\bibinfo {year} {2017})},\ \Eprint
  {http://arxiv.org/abs/1609.03611} {arXiv:1609.03611} \BibitemShut {NoStop}%
\bibitem [{\citenamefont {{Desjacques}}\ \emph {et~al.}(2018)\citenamefont
  {{Desjacques}}, \citenamefont {{Kehagias}},\ and\ \citenamefont
  {{Riotto}}}]{2018PhRvD..97b3529D}%
  \BibitemOpen
  \bibfield  {author} {\bibinfo {author} {\bibfnamefont {V.}~\bibnamefont
  {{Desjacques}}}, \bibinfo {author} {\bibfnamefont {A.}~\bibnamefont
  {{Kehagias}}}, \ and\ \bibinfo {author} {\bibfnamefont {A.}~\bibnamefont
  {{Riotto}}},\ }\bibfield  {title} {\enquote {\bibinfo {title} {{Impact of
  ultralight axion self-interactions on the large scale structure of the
  Universe}},}\ }\href {\doibase 10.1103/PhysRevD.97.023529} {\bibfield
  {journal} {\bibinfo  {journal} {\prd}\ }\textbf {\bibinfo {volume} {97}},\
  \bibinfo {eid} {023529} (\bibinfo {year} {2018})},\ \Eprint
  {http://arxiv.org/abs/1709.07946} {arXiv:1709.07946} \BibitemShut {NoStop}%
\bibitem [{\citenamefont {Teukolsky}(1972)}]{PhysRevLett.29.1114}%
  \BibitemOpen
  \bibfield  {author} {\bibinfo {author} {\bibfnamefont {Saul~A.}\ \bibnamefont
  {Teukolsky}},\ }\bibfield  {title} {\enquote {\bibinfo {title} {Rotating
  black holes: Separable wave equations for gravitational and electromagnetic
  perturbations},}\ }\href {\doibase 10.1103/PhysRevLett.29.1114} {\bibfield
  {journal} {\bibinfo  {journal} {Phys. Rev. Lett.}\ }\textbf {\bibinfo
  {volume} {29}},\ \bibinfo {pages} {1114--1118} (\bibinfo {year}
  {1972})}\BibitemShut {NoStop}%
\bibitem [{\citenamefont {Leaver}(1985)}]{10.2307/2397876}%
  \BibitemOpen
  \bibfield  {author} {\bibinfo {author} {\bibfnamefont {E.~W.}\ \bibnamefont
  {Leaver}},\ }\bibfield  {title} {\enquote {\bibinfo {title} {An analytic
  representation for the quasi-normal modes of kerr black holes},}\ }\href
  {http://www.jstor.org/stable/2397876} {\bibfield  {journal} {\bibinfo
  {journal} {Proceedings of the Royal Society of London. Series A, Mathematical
  and Physical Sciences}\ }\textbf {\bibinfo {volume} {402}},\ \bibinfo {pages}
  {285--298} (\bibinfo {year} {1985})}\BibitemShut {NoStop}%
\bibitem [{\citenamefont {Marsh}\ and\ \citenamefont
  {Pop}(2015)}]{Marsh:2015wka}%
  \BibitemOpen
  \bibfield  {author} {\bibinfo {author} {\bibfnamefont {David J.~E.}\
  \bibnamefont {Marsh}}\ and\ \bibinfo {author} {\bibfnamefont {Ana-Roxana}\
  \bibnamefont {Pop}},\ }\bibfield  {title} {\enquote {\bibinfo {title} {{Axion
  dark matter, solitons and the cusp-core problem}},}\ }\href {\doibase
  10.1093/mnras/stv1050} {\bibfield  {journal} {\bibinfo  {journal} {Mon. Not.
  Roy. Astron. Soc.}\ }\textbf {\bibinfo {volume} {451}},\ \bibinfo {pages}
  {2479--2492} (\bibinfo {year} {2015})},\ \Eprint
  {http://arxiv.org/abs/1502.03456} {arXiv:1502.03456 [astro-ph.CO]}
  \BibitemShut {NoStop}%
\bibitem [{\citenamefont {Ma}\ \emph {et~al.}(2013)\citenamefont {Ma},
  \citenamefont {Hinshaw},\ and\ \citenamefont {Scott}}]{Ma:2013kun}%
  \BibitemOpen
  \bibfield  {author} {\bibinfo {author} {\bibfnamefont {Yin-Zhe}\ \bibnamefont
  {Ma}}, \bibinfo {author} {\bibfnamefont {Gary}\ \bibnamefont {Hinshaw}}, \
  and\ \bibinfo {author} {\bibfnamefont {Douglas}\ \bibnamefont {Scott}},\
  }\bibfield  {title} {\enquote {\bibinfo {title} {{WMAP Observations of Planck
  ESZ Clusters}},}\ }\href {\doibase 10.1088/0004-637X/771/2/137} {\bibfield
  {journal} {\bibinfo  {journal} {Astrophys. J.}\ }\textbf {\bibinfo {volume}
  {771}},\ \bibinfo {pages} {137} (\bibinfo {year} {2013})},\ \Eprint
  {http://arxiv.org/abs/1303.4728} {arXiv:1303.4728 [astro-ph.CO]} \BibitemShut
  {NoStop}%
\bibitem [{\citenamefont {{Long}}\ \emph {et~al.}(2017)\citenamefont {{Long}},
  \citenamefont {{McAllister}},\ and\ \citenamefont
  {{Stout}}}]{2017JHEP...02..014L}%
  \BibitemOpen
  \bibfield  {author} {\bibinfo {author} {\bibfnamefont {C.}~\bibnamefont
  {{Long}}}, \bibinfo {author} {\bibfnamefont {L.}~\bibnamefont
  {{McAllister}}}, \ and\ \bibinfo {author} {\bibfnamefont {J.}~\bibnamefont
  {{Stout}}},\ }\bibfield  {title} {\enquote {\bibinfo {title} {{Systematics of
  axion inflation in Calabi-Yau hypersurfaces}},}\ }\href {\doibase
  10.1007/JHEP02(2017)014} {\bibfield  {journal} {\bibinfo  {journal} {Journal
  of High Energy Physics}\ }\textbf {\bibinfo {volume} {2}},\ \bibinfo {eid}
  {14} (\bibinfo {year} {2017})},\ \Eprint {http://arxiv.org/abs/1603.01259}
  {arXiv:1603.01259 [hep-th]} \BibitemShut {NoStop}%
\bibitem [{\citenamefont {Bachlechner}\ \emph {et~al.}(2016)\citenamefont
  {Bachlechner}, \citenamefont {Long},\ and\ \citenamefont
  {McAllister}}]{Bachlechner:2015qja}%
  \BibitemOpen
  \bibfield  {author} {\bibinfo {author} {\bibfnamefont {Thomas~C.}\
  \bibnamefont {Bachlechner}}, \bibinfo {author} {\bibfnamefont {Cody}\
  \bibnamefont {Long}}, \ and\ \bibinfo {author} {\bibfnamefont {Liam}\
  \bibnamefont {McAllister}},\ }\bibfield  {title} {\enquote {\bibinfo {title}
  {{Planckian Axions and the Weak Gravity Conjecture}},}\ }\href {\doibase
  10.1007/JHEP01(2016)091} {\bibfield  {journal} {\bibinfo  {journal} {JHEP}\
  }\textbf {\bibinfo {volume} {01}},\ \bibinfo {pages} {091} (\bibinfo {year}
  {2016})},\ \Eprint {http://arxiv.org/abs/1503.07853} {arXiv:1503.07853
  [hep-th]} \BibitemShut {NoStop}%
\bibitem [{\citenamefont {Bachlechner}\ \emph
  {et~al.}(2015{\natexlab{a}})\citenamefont {Bachlechner}, \citenamefont
  {Long},\ and\ \citenamefont {McAllister}}]{Bachlechner:2014gfa}%
  \BibitemOpen
  \bibfield  {author} {\bibinfo {author} {\bibfnamefont {Thomas~C.}\
  \bibnamefont {Bachlechner}}, \bibinfo {author} {\bibfnamefont {Cody}\
  \bibnamefont {Long}}, \ and\ \bibinfo {author} {\bibfnamefont {Liam}\
  \bibnamefont {McAllister}},\ }\bibfield  {title} {\enquote {\bibinfo {title}
  {{Planckian Axions in String Theory}},}\ }\href {\doibase
  10.1007/JHEP12(2015)042} {\bibfield  {journal} {\bibinfo  {journal} {JHEP}\
  }\textbf {\bibinfo {volume} {12}},\ \bibinfo {pages} {042} (\bibinfo {year}
  {2015}{\natexlab{a}})},\ \Eprint {http://arxiv.org/abs/1412.1093}
  {arXiv:1412.1093 [hep-th]} \BibitemShut {NoStop}%
\bibitem [{\citenamefont {Halverson}\ and\ \citenamefont
  {Langacker}(2018)}]{Halverson:2018xge}%
  \BibitemOpen
  \bibfield  {author} {\bibinfo {author} {\bibfnamefont {James}\ \bibnamefont
  {Halverson}}\ and\ \bibinfo {author} {\bibfnamefont {Paul}\ \bibnamefont
  {Langacker}},\ }\bibfield  {title} {\enquote {\bibinfo {title} {{TASI
  Lectures on Remnants from the String Landscape}},}\ }in\ \href
  {https://inspirehep.net/record/1647588/files/arXiv:1801.03503.pdf} {\emph
  {\bibinfo {booktitle} {{Theoretical Advanced Study Institute in Elementary
  Particle Physics: Physics at the Fundamental Frontier (TASI 2017) Boulder,
  CO, USA, June 5-30, 2017}}}}\ (\bibinfo {year} {2018})\ \Eprint
  {http://arxiv.org/abs/1801.03503} {arXiv:1801.03503 [hep-th]} \BibitemShut
  {NoStop}%
\bibitem [{\citenamefont {Bachlechner}\ \emph
  {et~al.}(2015{\natexlab{b}})\citenamefont {Bachlechner}, \citenamefont
  {Dias}, \citenamefont {Frazer},\ and\ \citenamefont
  {McAllister}}]{Bachlechner:2014hsa}%
  \BibitemOpen
  \bibfield  {author} {\bibinfo {author} {\bibfnamefont {Thomas~C.}\
  \bibnamefont {Bachlechner}}, \bibinfo {author} {\bibfnamefont {Mafalda}\
  \bibnamefont {Dias}}, \bibinfo {author} {\bibfnamefont {Jonathan}\
  \bibnamefont {Frazer}}, \ and\ \bibinfo {author} {\bibfnamefont {Liam}\
  \bibnamefont {McAllister}},\ }\bibfield  {title} {\enquote {\bibinfo {title}
  {{Chaotic inflation with kinetic alignment of axion fields}},}\ }\href
  {\doibase 10.1103/PhysRevD.91.023520} {\bibfield  {journal} {\bibinfo
  {journal} {Phys. Rev.}\ }\textbf {\bibinfo {volume} {D91}},\ \bibinfo {pages}
  {023520} (\bibinfo {year} {2015}{\natexlab{b}})},\ \Eprint
  {http://arxiv.org/abs/1404.7496} {arXiv:1404.7496 [hep-th]} \BibitemShut
  {NoStop}%
\bibitem [{\citenamefont {{Baik}}\ \emph {et~al.}(2004)\citenamefont {{Baik}},
  \citenamefont {{Ben Arous}},\ and\ \citenamefont
  {{Peche}}}]{2004math......3022B}%
  \BibitemOpen
  \bibfield  {author} {\bibinfo {author} {\bibfnamefont {J.}~\bibnamefont
  {{Baik}}}, \bibinfo {author} {\bibfnamefont {G.}~\bibnamefont {{Ben Arous}}},
  \ and\ \bibinfo {author} {\bibfnamefont {S.}~\bibnamefont {{Peche}}},\
  }\bibfield  {title} {\enquote {\bibinfo {title} {{Phase transition of the
  largest eigenvalue for non-null complex sample covariance matrices}},}\
  }\href@noop {} {\bibfield  {journal} {\bibinfo  {journal} {ArXiv Mathematics
  e-prints}\ } (\bibinfo {year} {2004})},\ \Eprint
  {http://arxiv.org/abs/math/0403022} {math/0403022} \BibitemShut {NoStop}%
\bibitem [{\citenamefont {{Bloemendal}}\ and\ \citenamefont
  {{Vir{\'a}g}}(2010)}]{2010arXiv1011.1877B}%
  \BibitemOpen
  \bibfield  {author} {\bibinfo {author} {\bibfnamefont {A.}~\bibnamefont
  {{Bloemendal}}}\ and\ \bibinfo {author} {\bibfnamefont {B.}~\bibnamefont
  {{Vir{\'a}g}}},\ }\bibfield  {title} {\enquote {\bibinfo {title} {{Limits of
  spiked random matrices I}},}\ }\href@noop {} {\bibfield  {journal} {\bibinfo
  {journal} {ArXiv e-prints}\ } (\bibinfo {year} {2010})},\ \Eprint
  {http://arxiv.org/abs/1011.1877} {arXiv:1011.1877 [math.PR]} \BibitemShut
  {NoStop}%
\bibitem [{\citenamefont {{Mo}}(2011)}]{2011arXiv1101.5144M}%
  \BibitemOpen
  \bibfield  {author} {\bibinfo {author} {\bibfnamefont {M.~Y.}\ \bibnamefont
  {{Mo}}},\ }\bibfield  {title} {\enquote {\bibinfo {title} {{The rank 1 real
  Wishart spiked model}},}\ }\href@noop {} {\bibfield  {journal} {\bibinfo
  {journal} {ArXiv e-prints}\ } (\bibinfo {year} {2011})},\ \Eprint
  {http://arxiv.org/abs/1101.5144} {arXiv:1101.5144 [math.PR]} \BibitemShut
  {NoStop}%
\bibitem [{\citenamefont {Acharya}\ \emph {et~al.}(2006)\citenamefont
  {Acharya}, \citenamefont {Bobkov}, \citenamefont {Kane}, \citenamefont
  {Kumar},\ and\ \citenamefont {Vaman}}]{Acharya:2006ia}%
  \BibitemOpen
  \bibfield  {author} {\bibinfo {author} {\bibfnamefont {Bobby~Samir}\
  \bibnamefont {Acharya}}, \bibinfo {author} {\bibfnamefont {Konstantin}\
  \bibnamefont {Bobkov}}, \bibinfo {author} {\bibfnamefont {Gordon}\
  \bibnamefont {Kane}}, \bibinfo {author} {\bibfnamefont {Piyush}\ \bibnamefont
  {Kumar}}, \ and\ \bibinfo {author} {\bibfnamefont {Diana}\ \bibnamefont
  {Vaman}},\ }\bibfield  {title} {\enquote {\bibinfo {title} {{An M theory
  Solution to the Hierarchy Problem}},}\ }\href {\doibase
  10.1103/PhysRevLett.97.191601} {\bibfield  {journal} {\bibinfo  {journal}
  {Phys. Rev. Lett.}\ }\textbf {\bibinfo {volume} {97}},\ \bibinfo {pages}
  {191601} (\bibinfo {year} {2006})},\ \Eprint
  {http://arxiv.org/abs/hep-th/0606262} {arXiv:hep-th/0606262 [hep-th]}
  \BibitemShut {NoStop}%
\bibitem [{\citenamefont {Acharya}\ \emph {et~al.}(2007)\citenamefont
  {Acharya}, \citenamefont {Bobkov}, \citenamefont {Kane}, \citenamefont
  {Kumar},\ and\ \citenamefont {Shao}}]{Acharya:2007rc}%
  \BibitemOpen
  \bibfield  {author} {\bibinfo {author} {\bibfnamefont {Bobby~Samir}\
  \bibnamefont {Acharya}}, \bibinfo {author} {\bibfnamefont {Konstantin}\
  \bibnamefont {Bobkov}}, \bibinfo {author} {\bibfnamefont {Gordon~L.}\
  \bibnamefont {Kane}}, \bibinfo {author} {\bibfnamefont {Piyush}\ \bibnamefont
  {Kumar}}, \ and\ \bibinfo {author} {\bibfnamefont {Jing}\ \bibnamefont
  {Shao}},\ }\bibfield  {title} {\enquote {\bibinfo {title} {{Explaining the
  Electroweak Scale and Stabilizing Moduli in M Theory}},}\ }\href {\doibase
  10.1103/PhysRevD.76.126010} {\bibfield  {journal} {\bibinfo  {journal} {Phys.
  Rev.}\ }\textbf {\bibinfo {volume} {D76}},\ \bibinfo {pages} {126010}
  (\bibinfo {year} {2007})},\ \Eprint {http://arxiv.org/abs/hep-th/0701034}
  {arXiv:hep-th/0701034 [hep-th]} \BibitemShut {NoStop}%
\end{thebibliography}%


\end{document}